\documentclass[a4paper]{amsart}
\usepackage{latexsym}
\usepackage{amsmath}
\usepackage{amssymb}
\usepackage{amsfonts}
\usepackage{esint}
\usepackage{graphicx}
\usepackage{color}
\definecolor{Blue}{rgb}{0.,0.,1.}
\definecolor{Red}{rgb}{1.,0.,0.}
\newcounter{smallarabics}
\newenvironment{arabicenumerate}
{\begin{list}{{\normalfont\textrm{(\arabic{smallarabics})}}}
  {\usecounter{smallarabics}\setlength{\itemindent}{0cm}
   \setlength{\leftmargin}{5ex}\setlength{\labelwidth}{4ex}
   \setlength{\topsep}{0.75\parsep}\setlength{\partopsep}{0ex}
   \setlength{\itemsep}{0ex}}}
{\end{list}}

\newcounter{smallroman}

\newcommand{\ben}{\begin{arabicenumerate}}  
\newcommand{\een}{\end{arabicenumerate}}

\def\init{\setcounter{equation}{0}}

\newtheorem{theoreme}{Theorem }[section]
\newtheorem{proposition}[theoreme]{Proposition}
\newtheorem{lemma}[theoreme]{Lemma}
\newtheorem{definition}[theoreme]{Definition}
\newtheorem{corollary}[theoreme]{Corollary}
\newtheorem{remark}[theoreme]{Remark}
\newtheorem{example}[theoreme]{Example}
\newcommand{\beq}{\begin{equation}}
\newcommand{\eeq}{\end{equation}}

\newcommand{\bex}{\begin{example}}
\newcommand{\eex}{\end{example}}
\def\bel{\begin{lemma}}
\def\eel{\end{lemma}}
\def\bet{\begin{theoreme}}
\def\eet{\end{theoreme}}
\def\bed{\begin{definition}}
\def\eed{\end{definition}}
\def\ber{\begin{remark}}
\def\eer{\end{remark}}

\def\rr{{\mathbb R}}

\def\cc{{\mathbb C}}
\def\nn{{\mathbb N}}

\def\part{{\rm par}}

\def\Im{{\rm Im}}

\def\slim{{\rm s-}\lim}

\def\bar{\overline}

\def\l{l}

\def\cinf{C^\infty}
\def\c0inf{C_0^\infty}

\def\s{s}

\def\proof{
\noindent{\bf Proof.}\ \ }

\def\ch{{\mathfrak h}}

\def\cY{{\mathcal Y}}

\def\cS{{\mathcal S}}

\def\i{{\rm i}}

\def\Dom{{\rm Dom}}

\def\qed{$\Box$\medskip}

\def \p{ \partial}

\def\12{\frac{1}{2}}
\def\14{\frac{1}{4}}
\def\x{\langle x \rangle}
\def\xt{\frac{x}{t}}
\def\supp{{\rm supp}}

\def\e{{\rm e}}

\def\d{{\rm d}}

\def\bbbone{{\mathchoice {\rm 1\mskip-4mu l} {\rm 1\mskip-4mu l}
{\rm 1\mskip-4.5mu l} {\rm 1\mskip-5mu l}}}
\def\one{\bbbone}
\def\cH{{\mathcal H}}

\def\l{l}

\def\coinf{C_0^\infty}

\def\cK{{\mathcal K}}

\def \p{ \partial}

\def\12{\frac{1}{2}}
\def\x{\langle x \rangle}
\def\xt{\frac{x}{t}}
\def\supp{{\rm supp}}

\def\e{{\rm e}}

\def\d{{\rm d}}

\def\cH{{\mathcal H}}

\def\bep{\begin{proposition}}
\def\eep{\end{proposition}}

\def\Op{{\rm Op}^{\rm w}}
\def\s{s}

\newcommand{\mat}[4]{\left(\begin{array}{cc}#1 &#2  \\ #3 &#4 \end{array}\right)}
\newcommand{\lin}[2]{\left(\begin{array}{c}#1 \\#2\end{array}\right)}

\def\CARal{{\rm C\hskip 0.25 em \hbox{\raise 1.72 ex 
\hbox{$\scriptscriptstyle\rm al$}\kern -0.57 em A}R}}

\def\t{{\scriptscriptstyle\#}}
\def\otimesal{\mathop{\hbox{\raise 1.5 ex
  \hbox{$\scriptscriptstyle\rm al$}
\kern -0.92 em \hbox{$\otimes$}}}}
\def\oplusal{\mathop{\hbox{\raise 1.5 ex
  \hbox{$\scriptscriptstyle\rm al$}
\kern -0.92 em \hbox{$\oplus$}}}}
\def\Gammal{\hbox{\raise 1.68 ex 
\hbox{$\scriptscriptstyle\rm al$}\kern -0.50 em $\Gamma$}}
\def\Bal{\hbox{\raise 1.68 ex 
\hbox{$\scriptscriptstyle\rm  al$}\kern -0.50 em $B$}}
\def\CARal{{\rm C\hskip 0.25 em \hbox{\raise 1.72 ex 
\hbox{$\scriptscriptstyle\rm al$}\kern -0.57 em A}R}}
\def\t{{\scriptscriptstyle\#}}

\def\cE{{\mathcal E}}

\begin{document}

\title[Scattering theory for Klein-Gordon equations]{Scattering theory for Klein-Gordon equations with non-positive energy}
\author{C. G\'erard}
\thanks{We  would like to warmly thank Vladimir Georgescu for several helpful discussions on Krein spaces and functional calculus for definitizable operators}
\address{D\'epartement de Math\'ematiques, Universit\'e de Paris XI, 91405 Orsay Cedex France}
\email{christian.gerard@math.u-psud.fr}
\begin{abstract}
 We study  the scattering theory for  charged Klein-Gordon equations:
 \[
\left\{
\begin{array}{l}
(\p_{t}- \i v(x))^{2}\phi(t,x)+\epsilon^{2}(x, D_{x})\phi(t,x)=0,\\[2mm]
\phi(0, x)= f_{0}, \\[2mm]
 \i^{-1} \p_{t}\phi(0, x)= f_{1},
 \end{array}\right.
\]
where: 
\[
\epsilon^{2}(x, D_{x})=  -\sum_{1\leq j, k\leq n}\left(\p_{x_{j}}-\i b_{j}(x)\right)a^{jk}(x)\left(\p_{x_{k}}-\i b_{k}(x)\right)+ m^{2}(x),
\]
 describing a Klein-Gordon field minimally coupled to an external electromagnetic field described by the
electric potential $v(x)$  and magnetic potential $\vec{b}(x)$.   The flow of the Klein-Gordon equation  
preserves the   energy:
\[
h[f, f]:= \int_{\rr^{n}}\overline{f}_{1}(x) f_{1}(x)+ \overline{f}_{0}(x)\epsilon^{2}(x, D_{x})f_{0}(x) - \overline{f}_{0}(x) v^{2}(x) f_{0}(x) \d x.
\]
We consider the situation when the energy is not positive. In this case the flow cannot be written as a unitary group on a Hilbert space, and the Klein-Gordon equation may have complex eigenfrequencies. 

Using the theory of definitizable operators on Krein spaces and time-dep\-endent methods, we prove the existence and completeness of wave operators, both in the short- and long-range cases.  The range of the wave operators are characterized in terms of the spectral theory of the generator, as in the usual Hilbert space case.  
\end{abstract}
\maketitle

\section{Introduction}\label{seci}
\subsection{Klein-Gordon equations with non-positive energy}\label{seci.1}

Klein-Gordon field equations coupled with an external electromagnetic field appear in several problems of mathematical physics. 
It was realized since the forties by  Schiff, Snyder and Weinberg \cite{SSW} that for the Klein-Gordon equation on Minkowski space:
\beq\label{kgkg}
(\p_{t}- \i v(x) )^{2}\phi(t, x)- \Delta_{x} \phi(t,x)+ m^{2}\phi(t,x)=0,
\eeq
 complex eigenfrequencies appear if the electrostatic potential becomes too large, which causes difficulties with the quantization of this  field equation.
This phenomenon is usually called the {\em Klein paradox}.  It can be traced back to the fact that the conserved energy
\[
\int_{\rr^{d}}|\p_{t}\phi(t,x)|^{2} \d x + \int_{\rr^{d}} |\nabla_{x} \phi(t, x)|^{2}+ (m^{2}- v^{2}(x))|\phi(t,x)|^{2}\d x
\]
is not positive definite if $\|v\|_{\infty}$ is too large. 

A related problem appears when one considers the Klein-Gordon equation on some curved space-times of general relativity, like the Kerr space-time describing a rotating black hole. Again the conserved energy is not positive definite.  A nice reference describing these problems is the appendix of the book by Fulling \cite{Fu}.

The aim of  this paper  is to study in details the scattering theory for  a class of Klein-Gordon equations generalizing (\ref{kgkg}).  We consider 
the charged Klein-Gordon equation:
\beq\label{ei.1}
\left\{
\begin{array}{l}
(\p_{t}- \i v(x))^{2}\phi(t,x)+\epsilon^{2}(x, D_{x})\phi(t,x)=0,\\[2mm]
\phi(0, x)= f_{0}, \\[2mm]
 \i^{-1} \p_{t}\phi(0, x)= f_{1},
\end{array}\right.
\eeq
in $\rr_{t}\times \rr^{n}_{x}$ where
\[
\epsilon^{2}(x, D_{x})= - \sum_{1\leq j, k\leq n}\left(\p_{x_{j}}+\i b_{j}(x)\right)a^{jk}(x)\left(\p_{x_{k}}-\i b_{k}(x)\right)+ m^{2}(x),\]
 describing a Klein-Gordon field minimally coupled to an external electromagnetic field described by the
(real) electric potential $v(x)$  and magnetic potential $\vec{b}(x)$.  The function $x\mapsto m(x)$ corresponds to a variable mass term, incorporating for example a scalar curvature term .

The Cauchy problem (\ref{ei.1})  can  be rewritten as
\[
 f^{t}= \e^{-\i  t B}f, \ B= -\mat{0}{\one}{\epsilon^{2}- v^{2}}{2v}, \hbox{ for }f^{t}=\lin{\phi(t)}{-\i \p_{t}\phi(t)}.
\]
The evolution $\e^{- \i tB}$ preserves the  {\em energy}:
\[
h[f, f]:= \int_{\rr^{n}}\overline{f}_{1}(x) f_{1}(x)+ \overline{f}_{0}(x)\epsilon^{2}(x, D_{x})f_{0}(x) - \overline{f}_{0}(x) v^{2}(x) f_{0}(x) \d x.
\]
We are interested in this paper in the scattering theory, i.e. in the complete classification of the asymptotic behavior of $\e^{- \i tB}f$ for all initial data $f$, when $t\to \pm\infty$.

Typical assumptions  that we will impose in this paper (see Subsect. \ref{sec10.1} for details) are:
\[
\p_{x}^{\alpha}v(x), \ \p_{x}^{\alpha}b_{j}(x), \ \p_{x}^{\alpha}([a^{jk}](x)- \one)\hbox{ and }\p_{x}^{\alpha}( m(x)-m)\in O(\langle x\rangle^{- \mu-|\alpha|}),
\]
for some $\mu>0$, although a	 singular short-range part of $v$ can also be accommodated. 
The asymptotic mass $m$ is assumed to be strictly positive.
In analogy to the scattering theory for Schr\"{o}dinger operators, the case $1<\mu$ (resp. $0<\mu\leq 1$) will be called the {\em short-range} (resp. {\em long-range}) case.

\subsection{Scattering theory}\label{seci.2}
If the energy  $h$ is {\em positive definite}, i.e. the electric potential is not too large,   one can use it to  equip the space of initial data with a Hilbert space structure.

Under typical assumptions one obtains the {\em energy space}  $\cE= H^{1}(\rr^{n})\oplus L^{2}(\rr^{n})$, and the 
 group $\e^{- \i t B}$
becomes a  strongly continuous unitary group on $\cE$, whose scattering theory can be studied by  Hilbert space methods. We mention among many others the 
papers \cite{E, Lu, N,S,V, W, Wi}.  
In this paper we are interested in the situation when  the energy is {\em not} positive.  In this case the generator $B$ may have complex eigenvalues, or real eigenvalues with non trivial Jordan blocks. 
It follows that in  general the energy norm $\|\e^{-\i tB}f\|_{\cE}$  may be polynomially or exponentially  growing in $t$.  

To our knowledge the only result about scattering theory in this situation is due to Kako \cite{K} 
where the  case 
\[
\epsilon^{2}= -\Delta +m^{2}, \ v(x)\in O(\langle x\rangle^{-\mu}), \ \mu>2,
\] 
is treated. In \cite{K}, spectral projections $\one_{I}(B)$  for  bounded intervals $I$ such that $\pm m\not\in I^{\rm cl}$ are constructed by stationary arguments, and local wave operators
\[
\slim_{t\to \pm\infty}\e^{\ i t B}\e^{- \i t B_{\infty}}\one_{I}(B_{\infty})= W^{\pm}_{I}
\]
are shown to exist, for $B_{\infty}$ being the generator of the free Klein-Gordon equation obtained for $ \epsilon^{2}= -\Delta +m^{2}$ and $v(x)\equiv 0$.

 Their ranges are shown to be equal to the  range of  $\one_{I}(B)$, which is a result of local asymptotic completeness of wave operators. 

However the results of \cite{K} do not yield a complete classification of $\e^{- \i tB}$ on the whole energy space $\cE$, because the 
closure of the ranges of $W^{\pm}_{I}$ when $I$ runs over all allowed intervals $I$ is not characterized in terms of the spectral theory of $B$.
For example it is not known if the limits 
\[
\slim_{\epsilon\to 0,  R\to +\infty}\one_{[m+\epsilon, R]}(\pm B)= P_{\pm}\hbox{ exist}.
\]
Besides the stationary method used in \cite{K} does not adapt easily to long-range potentials.

In this paper we reconsider this problem using two tools: 

 the first tool is the theory of selfadjoint operators on {\em Krein spaces} (see  Appendices \ref{sec0} and \ref{secIk.} for a brief summary). Krein spaces are   complete, hilbertizable  vector spaces equipped with a bounded, non-degenerate 
but {\em non-positive} hermitian sesquilinear form $h[\cdot, \cdot]$,  the adjoint of a densely defined linear operator being defined with respect to $h$.
Among selfadjoint operators on a Krein space, the class of {\em definitizable} selfadjoint operators is of primary importance. 
Definitizable operators  on a Krein space are quite close to selfadjoint operators on a Hilbert space:  although they may have complex eigenvalues, they admit a (smooth) functional calculus on the real line (see Subsect. \ref{secIk.4} for  a self-contained presentation) and also spectral projections $\one_{I}(B)$ for a class of intervals $I\subset \rr$.

The idea of using Krein space theory to study the Klein-Gordon equation with a non-positive energy is of course not new. Equations coming from classical mechanics (like the Klein-Gordon equation)  are actually  typical applications of Krein space theory. 
We mention among others the  papers  \cite{J2, J3, LNT1, LNT2}. However in these works the Krein space theory is mainly used to prove the existence of the dynamics $\e^{- \i t B}$,  to show that the number of complex eigenvalues of $B$ is finite or to get a priori estimates on the spectrum of $B$.
The scattering theory is  not considered.

Our second tool is an adaptation to the framework of definitizable selfadjoint operators on Krein spaces of the  {\em time-dependent} approach to  Hilbert space
scattering theory, in the version initiated by Sigal and Soffer \cite{SS}, based on {\em propagation estimates}. 
The method of propagation estimates proved very powerful and flexible to study scattering theory for Schr\"{o}dinger operators, in particular for the problem of asymptotic completeness of wave operators.    For wave or Klein-Gordon equations on some stationary manifolds it has been developed by H\"{a}fner \cite{H}.

Its adaptation to the Krein space setup requires some care, because one needs to work with two sesquilinear forms, the non-positive one defining the Krein scalar product, and a positive one defining the hilbertizable topology, the dynamics $\e^{- \i tB}$ preserving the first, but of course not the second. 
\subsection{Description of the results}\label{seci.3b}
Let us now briefly describe the results of this paper, summarized in Sect.  \ref{mainresults}.  We assume decay hypotheses like those outlined in Subsect. \ref{seci.1}, although the electric potential may have a singular short-range part (see  Subsect. \ref{sec10.1} for precise statements). 

 If $B$ is a definitizable operator on a Krein space, there is a finite subset of $\sigma(B)\cap \rr$, called the set of {\em critical points} of $B$ which plays a special role.   We have to assume that the points $\pm m$, which are boundaries of the essential spectrum of $B$, 	are not critical points.  
 A sufficient condition for this to hold is that there are no eigenstates of   $B$ for the eigenvalues $\pm m$ with {\em negative} energy.

As in Hilbert space scattering theory we split the energy space $\cE$ into 
\beq\label{direct}
\cE= \cE_{\rm pp}(B)\oplus^{\perp} \cE_{\rm scatt}(B),
\eeq
where  $\cE_{\rm pp}(B)$ is  the closure of the sum of the (generalized) eigenspaces of $B$ for all (real and complex) eigenvalues of $B$, and $\cE_{\rm scatt}(B)$ is its orthogonal complement for $h[\cdot, \cdot]$. 

We first discuss the short-range case $\mu>1$. Let us denote by $\cE_{\infty}$ the energy space for the free Klein-Gordon equation, whose generator is denoted by $B_{\infty}$. 
We prove in Thm. \ref{youpla}  that for all $f\in \cE_{\infty}$ there exist $f^{\pm}\in \cE_{\rm scatt}(B)$ such that
\[
\e^{- \i t B_{\infty}}f= \e^{- \i t B}f^{\pm}+ o(1), \ t\to \pm \infty.
\]
The maps $ \Omega_{s}^{\pm}: f\mapsto f^{\pm}$ are called the {\em short-range wave operators}.  As in the Schr\"{o}dinger case the wave operators intertwine the free and interacting dynamics. Moreover we prove that 
\[
{\rm Ran}\Omega_{s}^{\pm}= \cE_{\rm scatt}(B),
\]
hence the wave operators $\Omega^{\pm}_{s}$ are {\em complete}.

 In the long-range case $0<\mu\leq 1$,  we need to assume that the potential $v$  is of constant sign near infinity, which is not a serious restriction from the point of view of physical applications.  
 Similarly to the Schr\"{o}dinger case, we introduce  a {\em time-independent modifier} $T$ (which is a Fourier integral operator), and   show  for all $f\in \cE_{\infty}$ the existence of $f^{\pm}\in \cE_{\rm scatt}(B)$ such that
 \[
T\e^{- \i t B_{\infty}}f= \e^{- \i t B}f^{\pm}+ o(1), \ t\to \pm \infty.
\]
 The {\em long-range wave operators} $\Omega_{l}^{\pm}: f\mapsto f^{\pm}$ have the same properties as in the short-range case, in particular they are complete.

 \subsection{Symplectic point of view}\label{seci.3}
  Introducing the conjugate variables:
 \[
\varphi_{t}(x)= \phi(t, x), \ \pi_{t}(x)= \p_{t}\phi(t, x)- \i v(x)\phi(t,x),
\]
one can consider the evolution for (\ref{ei.1}) as a (complex) symplectic flow, obtained from the symplectic form
\[
\overline{(\varphi_{1}, \pi_{1})}\omega(\varphi_{2}, \pi_{2}):= \int_{\rr^{n}} \overline{\pi_{1}}(x) \varphi_{2}(x)-\overline{\varphi_{1}}(x)\pi_{2}(x) \d x,
\]
and the  {\em classical Hamiltonian}:
\[
h(\varphi, \pi):=\int_{\rr^{n}} \overline{\pi}(x) \pi(x)+ \overline{\varphi}(x) \epsilon^{2}\varphi (x)
-\i\overline{\varphi}(x) v(x)\pi(x)+\i \overline{\pi}(x) v(x)\varphi(x)\d x.
\] 
This point of view is of course important for the quantization of the Klein-Gordon equation (see e.g. \cite{DG2}).

From the results explained above, it is easy to show that the two subspaces $\cE_{\rm pp}(B)$ and $\cE_{\rm scatt}(B)$ are symplectic spaces and that the direct sum in (\ref{direct}) is orthogonal for the symplectic form.  
The wave operators are invertible symplectic transformations.  On $\cE_{\rm scatt}(B)$
 the symplectic flow $\e^{- \i tB}$ is hence symplectically equivalent to the free symplectic flow $\e^{- \i tB_{\infty}}$.
Therefore the scattering theory for the Klein-Gordon equation
(\ref{ei.1}) is parallel to the scattering theory for two-body
Schroedinger operators, replacing unitary groups by symplectic flows.
\subsection{Some open problems}
Let us mention some open problems.  The first problem would be to accommodate potentials with local Coulomb singularities ( for $d=3$). The coupling constant should be sufficiently small so that $\epsilon^{2}- v^{2}$ is well defined.  
Another  problem would be to consider the {\em wave equation} i.e. put  $m=0$.    

A related problem which could  be treated by the methods of this paper is the modified wave equation:
\[
(\p_{t}+ \p_{x}v)(\p_{t}+ v\p_{x})\phi(t, x) -\p_{x}^{2}\phi(t, x)+ \p_{x}^{4}\phi(t,x)=0, \hbox{ in }\rr_{t}\times \rr_{x},
\]
where  $v(x)\to v_{\infty}$ with $|v_{\infty}|<1$ at $\pm \infty$. This field equation appears in {\em the black hole laser effect}, see eg Coutant-Parentani 
\cite{CP}.

A more   difficult problem would be to extend the methods of this paper to the  one-dimensional Klein-Gordon or wave equations  with step-like potentials  considered by Bachelot \cite{Ba}. 
If the electric potential $v(x)$ has two limits  $v_{\pm}$ at $\pm\infty$ with $|v_{+}- v_{-}|\geq 2m$ then 
the energy space is not a Pontryagin space (see Def. \ref{pontryag}) and hence the generator of the dynamics is not necessarily definitizable.

\subsection{Plan of the paper}
In Sect. \ref{sec1} we discuss the abstract charged Klein-Gordon equation,  recalling its symplectic and Krein space aspects. Sect. \ref{sec10} is devoted to the concrete Klein-Gordon equation that we consider in this paper. We collect the various hypotheses and describe properties of the essential and point spectrum of the generator $B$. 

In Sect. \ref{mainresults} we state the main results of the paper, on existence and completeness of wave operators, both in the short- and long-range case. In Sect. \ref{sec6} we describe an approximate diagonalization of the generator $B$ and collect various technical estimates, obtained from standard pseudo-differential calculus. The most important is  Prop. \ref{commut}, which is the analog of the Mourre estimate in our setup.
In Sect. \ref{sec3} we prove various propagation estimates. Their intuitive content is similar to the positive energy case (see e.g. \cite{H}) but  the abstract setup is different and explained in Appendix \ref{sec4}.

 In Sect. \ref{srwave} we prove the existence and completeness of intermediate wave operators, allowing to remove a short-range part of the potential. The resulting dynamics has a positive energy, and its scattering theory is studied in Sect. \ref{sec11}, using rather standard arguments. The proofs of the main results of Sect. \ref{mainresults} are given in Sect.  \ref{youlou}.  

 Appendix \ref{sec0} and Appendix \ref{secIk.}  are devoted to a rather self-contained introduction to Krein spaces and  definitizable operators. For Krein spaces a classic reference is the book by Bognar \cite{B}. The theory of definitizable operators is covered in a survey article by Langer \cite{La}. We include in Appendix \ref{secIk.} a presentation of smooth and Borel functional calculus for definitizable operators based on almost analytic extensions. 
 
 Appendix \ref{sec4} is devoted to propagation estimates for unitary groups on a Krein space with a definitizable generator.  Finally we recall in Appendix \ref{pdocalc} various facts on pseudo-differential and Fourier integral operators. 
\subsection{Notation}\label{titi}
We now collect some notation which will be use throughout this paper.

The domain of a linear operator $a$ on a Hilbert space $\ch$ will be denoted by $\Dom a$. If $a$ is 
selfadjoint operator, we will write $a>0$ if $a\geq 0$
and ${\rm Ker}a=\{0\}$.
Note that if $a>0$ and
$s\in \rr$, $\|h\|_{s}= \|a^{-s}h\|_{\ch}$ is a norm on
$\Dom  a^{-s}$.
We denote then by $a^{s}\ch$ the completion of $\Dom  a^{-s}$
for the norm $\|\ \|_{s}$.
The map $a^{s}$ extends as a
unitary operator from $a^{t}\ch$ to $a^{s+t}\ch$.
One example of this notation are the familiar  Sobolev spaces, where
$H^{s}(\rr^{d})$ is equal to $(-\Delta +1)^{-s/2}L^{2}(\rr^{d})$.

For a map $ \rr\ni \mapsto B(t)\in B(\ch)$ we write $B(t)\in O(t^{\mu})$ if $\|B(t)\|\leq C t^{\mu}$ for $ |t|\geq 1$.

A complex symplectic space will be denoted by the letter $\cY$. A Krein space (see Appendix \ref{sec0}) will be denoted by the letter $\cK$. The energy space associated to a Klein-Gordon equation (which is an example of a Krein space) will be denoted by the letter $\cE$.

As usual we set $D_{x}= \i^{-1}\p_{x}$,  and $\langle x\rangle= (1+x^{2})^{\12}$.


For  $-\infty <a<b\leq +\infty $ we denote by $F(a\leq \lambda\leq b)$ a function in $\cinf(\rr)$ supported in $[a, b]$ and equal to $1$ in $[a/2, b/2]$.
\section{Abstract Klein-Gordon equation}\label{sec1}\init
In this section we consider an abstract Klein-Gordon equation
corresponding to the Klein-Gordon field minimally coupled to an
external electro-magnetic field. We discuss the various ways of
writing it as a one-parameter evolution group and the associated
conserved quantities. Our approach in most of this section will be formal, ie
we will not discuss the problem of existence and uniqueness of
solutions.

Let $\ch$ be a complex Hilbert space, $\epsilon, v$ two selfadjoint operators on $\ch$. The scalar product on $\ch$ will be denoted by $(h_{1}| h_{2})$ or sometimes by $\overline{h_{1}}\cdot h_{2}$.

The abstract Klein-Gordon equation is:
\begin{equation}
\label{e1.1}
(\p_{t}- \i v)^{2}\phi(t)+ \epsilon^{2}\phi(t)=0,
\end{equation}
where $\phi: \rr\to \ch$.

\subsection{Symplectic setup}\label{sec1.1}

Let $\cY=\ch\oplus \ch$ whose elements are denoted by $(\varphi,
\pi)$. We consider  $\cY$ as a complex symplectic space by equipping it
with  the complex symplectic form ( i.e. sesquilinear, non-degenerate,
anti-hermitian):
\[
\overline{(\varphi_{1}, \pi_{1})}\omega(\varphi_{2}, \pi_{2}):= \overline{\pi_{1}}\cdot \varphi_{2}-\overline{\varphi_{1}}\cdot\pi_{2}  .
\]
The {\em classical Hamiltonian} is:
\[
h(\varphi, \pi):= \overline{\pi}\cdot \pi+ \overline{\varphi}\cdot \epsilon^{2}\varphi 
-\i\overline{\varphi} \cdot v\pi+\i \overline{\pi}\cdot v\varphi.
\]
We consider
$\omega, h$ as maps from $\cY$ to $\cY^{*}$, where $\cY^{*}$ is the
space of anti-linear forms on $\cY$ and set:
\beq\label{e0.002}
A:=-\i \omega^{-1}h=\mat{v}{-\i}{\i\epsilon^{2}}{v}.
\eeq

If we set
\[
\lin{\varphi_{t}}{\pi_{t}}:= \e^{\i t A}\lin{\varphi}{\pi}
\]
then
\beq\label{e0.001}
\begin{array}{l}
\frac{\d }{\d t}\varphi_{t}= \pi_{t}+\i v \varphi_{t},\\[2mm]
\frac{\d }{\d t}\pi_{t}= - \epsilon^{2}\varphi_{t} + \i v \pi_{t},
\end{array}
\eeq
hence $\phi(t)= \varphi_{t}$ solves the following Cauchy problem for
the Klein-Gordon equation:
\[\left\{
\begin{array}{l}
(\p_{t}- \i v)^{2}\phi(t)+ \epsilon^{2}\phi(t)=0,\\[2mm]
 \phi(0)= \varphi, \ \p_{t} \phi(0)= \pi+\i v \varphi.
\end{array}\right.
\]
Clearly  the group $\e^{\i t A}$ preserves  $\omega$ ($\e^{\i  tA}$ is a
complex symplectic flow) and $h$.
\subsection{PDE setup}\label{sec1.3}
Since in this paper we will mostly  use the conserved
quantity $h$, it is more convenient to adopt the standard setup from
partial differential equations.  We 
consider the Cauchy problem:
\beq\label{asma}
\left\{
\begin{array}{l}
(\p_{t}- \i v)^{2}\phi(t)+ \epsilon^{2}\phi(t)=0,\\[2mm]
 \phi(0)= f_{0}, \ -\i\p_{t} \phi(0)= f_{1}.
\end{array}\right.
\eeq
Setting
\[
f^{t}= \lin{\phi(t)}{-\i \p_{t}\phi(t)},
\]
(\ref{asma}) is rewritten as:
\beq\label{e0.003}
 f^{t}= \e^{-\i  t B}f, \ B= -\mat{0}{\one}{\epsilon^{2}- v^{2}}{2v}.
\eeq
 The symplectic form and classical Hamiltonian become:
 \[\begin{array}{rl}
 \overline{f}\omega  g=& -\i \big(\overline{f_{0}}\cdot g_{1}+ \overline{f_{1}}\cdot g_{0}- 2 \overline{f_{0}}\cdot v g_{0}\big),\\[2mm]
 h[f, f]=& \|f_{1}\|^{2}+ (f_{0}|(\epsilon^{2}- v^{2})f_{0}).
 \end{array}
\]
The sesquilinear form   $q= \i \omega$ is  hermitian  and often called the {\em charge}.
From (\ref{e0.002}) we  obtain that:
\beq\label{atchoum}
\overline{f}\cdot \omega g= \i h[f, B^{-1}g].
\eeq
\subsection{Krein space approach to the abstract Klein-Gordon equation}\label{sec1.4}
We now recall some results  due to \cite{LNT1}  about
the Krein space approach to the abstract Klein-Gordon equation and  existence of the dynamics
$\e^{-\i tB}$. We refer the reader to Appendices \ref{sec0}, \ref{secIk.} for terminology concerning Krein spaces.
We first introduce some assumptions:
\[
\begin{array}{rl}
\hbox{(E1)}&
\epsilon^{2}\geq m^{2}, \ m>0,\\[2mm]
 \hbox{(E2)}& v\hbox{ is }\epsilon-\hbox{ bounded with relative bound }<1,\\[2mm]
\hbox{(E3)}&\  0\not\in \sigma_{\rm p}(\epsilon^{2}- v^{2}),\\[2mm]
\hbox{(E4)} &\ {\rm Tr}\one_{]-\infty,
0]}(\epsilon^{2}- v^{2})<\infty.
\end{array}
\]
\begin{proposition}[\cite{LNT1}]
\label{1.1}  Assume (E1), (E2), (E3). Then:
\ben
\item  the Hilbert space $\cE=\epsilon^{-1}\ch\oplus \ch$ equipped with the
sesquilinear form $h[\cdot, \cdot]$ is a Krein space;
\item the operator $B$  on $\cE$ defined in (\ref{e0.003}) with domain $\Dom
(\epsilon^{2}- v^{2})\oplus \epsilon^{-1}\ch$ is closed, selfadjoint
on  $(\cE, h[\cdot, \cdot])$ with $0\in \rho(B)$.  

 Assume  (E1),  (E2), (E3), (E4).  Then:
\item $(\cE, h[\cdot, \cdot])$ is a
Pontryagin space; 
\item $B$ generates a strongly continuous group $(\e^{-\i tB})_{t\in
\rr}$ of bounded unitary operators on $(\cE, h[\cdot, \cdot])$ and
hence the Cauchy problem (\ref{asma}) has a unique
solution.
\een
\end{proposition}

\section{Concrete Klein-Gordon equation}\label{sec10}\init
In this section we fix some hypotheses that we will impose for the rest of the paper. We also discuss some properties of the  eigenvalues  of $B$. For $m\in \rr$ we denote by $S^{m}(\rr^{d})$ the space of functions $f\in C^{\infty}(\rr^{d})$ such that $\p^{\alpha}_{x}f\in O(\langle x\rangle^{m-|\alpha|})$, for all $\alpha\in \nn$. 
\subsection{Assumptions}\label{sec10.1}
We set $\ch:=L^{2}(\rr^{d})$ and consider the Klein-Gordon equation (\ref{e1.1}) for:
\[
\epsilon^{2}= c(x)^{-1}\langle (D_{x}- b(x)), A(x)(D_{x}- b(x))\rangle c(x)^{-1}+ m^{2}(x),
\]
where $A(x)=[a^{jk}](x)$, $b(x)= (b_{1}(x), \dots, b_{d}(x))$,
$a^{jk}(x), \ b_{j}(x), \ c(x) \ m(x)$ are  real functions satisfying:
\[
\hbox{(A1)}\ \begin{array}{l}
c_{0}\one \leq [a^{jk}](x)\leq c_{1}\one, \ c_{0}\leq c(x)\leq c_{1}, \ c_{0}\leq m(x)\leq c_{1}, \\[2mm]
\hbox{ for some }c_{0}>0,\\[2mm]
[a^{jk}](x)-\one,  \ b(x), \ c(x)-1, \ m(x)-m\in S^{
-\mu_{0}}(\rr^{d}), \\[2mm]
\hbox{ for some }m>0, \ \mu_{0}>0.
\end{array}
\]Clearly $\epsilon^{2}$ is selfadjoint on $H^{2}(\rr^{d})= \epsilon^{-2}\ch$ and there
exists $m_{0}>0$ such that $\epsilon^{2}\geq m_{0}^{2}$.

Concerning the scalar potential $v$ we assume that $v= v(x)$ is a
multiplication operator with
\[
\hbox{(A2)}\ v^{k}\epsilon^{-k}:\ch\to \ch\hbox{ is compact for }k=1,2.
\]
It follows that $\epsilon^{2}- v^{2}$ with domain $\epsilon^{-2}\ch$ is
selfadjoint and bounded below. It follows also from (A1), (A2) that:
\[
\sigma_{\rm ess}(\epsilon^{2}- v^{2})= \sigma_{\rm ess}(\epsilon^{2})=
[m^{2}, +\infty[.
\]
We assume also the analog of (E3):
\[
\hbox{(A3)}\ 0\not \in \sigma_{\rm p}(\epsilon^{2}-v^{2}).
\]
By Prop. \ref{1.1} the Krein space
\[
\cE:=\epsilon^{-1}\ch\oplus \ch= H^{1}(\rr^{d})\oplus L^{2}(\rr^{d}),
\]
equipped with
\[
h[f, f]= \|f_{1}\|^{2}+ (f_{0}| (\epsilon^{2}- v^{2})f_{0})
\]
is a Pontryagin space.  It follows from Lemma \ref{arsouille} that $B$ is definitizable with a definitizing polynomial $p$ of even degree.  We denote by $c_{p}(B)$ the set of associated critical points.
 
It follows also from (A2) that $\Dom B= \epsilon^{-2}\ch\oplus \epsilon^{-1}\ch$ and we can write
\[
B= B_{0}+ V, \ B_{0}=-\mat{0}{\one}{\epsilon^{2}}{0}, V= \mat
{0}{0}{v^{2}}{-2v},
\]
where $V$ is $B_{0}-$compact. We have then
\beq
\sigma_{\rm ess}(B)= \sigma_{\rm ess}(B_{0})= ]-\infty, -m]\cup[m,
+\infty[,
\label{e3.01b}
\eeq
where the second equality follows by
using the  transformation $U$ defined in Subsect. \ref{reduc} for
$v\equiv 0$.

It will be useful to be able to split the potential $v$ as the sum of a short-range part and a smooth
long-range part. Therefore we assume:
\[
\hbox{(A4)}\ v(x)= v_{s}(x)+ v_{l}(x),
\] 
where:
\beq\label{e3.01}
v_{\l}(x)\in S^{-\mu_{l}}(\rr^{d})\ \mu_{l}>0,
\eeq
 \beq\label{e3.02}
\langle x\rangle^{\mu_{s}}v_{s}^{k}\epsilon^{-k}\hbox{ is bounded for }k=1,2, \ \mu_{s}>1.
\eeq
\subsection{Properties of  eigenvalues and critical points}\label{absence}
In this subsection we discuss the possible location of critical points (see Def. \ref{criti}) and  real eigenvalues of $B$. 
For the scattering theory it will be important to know that there are no critical points of  $B$ embedded in the essential spectrum. \begin{proposition}\label{noemb}
Assume that  $v= v_{1}+ v_{2}$ where:
\[
\hbox{\rm (B1)}\ \left\{\begin{array}{l}
\p^{\alpha}_{x}v_{1}\in O(\langle x\rangle^{-\mu -|\alpha|}), \
|\alpha|\leq 2,\\[2mm]
v_{2}\hbox{ has compact support}, \ v_{2}\in L^{d}(\rr^{d}).
\end{array}\right.
\]
Then $\sigma_{\rm p}(B)\cap \rr\subset [-m, m]$.
\end{proposition}
\proof
An easy computation shows that:
\[
\epsilon^{2}= h+ m^{2},
\]
for
\[
h= \sum_{jk}D_{j}c^{jk}(x)D_{k}+  \sum_{j}d_{j}(x)D_{j}+
D_{j}d_{j}(x)+ r(x),
\]
with $[c^{jk}](x)-\one$, $d_{j}(x)$, $r(x)$ in
$S^{-\mu_{0}}(\rr^{d})$. 
We see that $Bf= \lambda f$ iff  $h(\lambda)f_{0}= E f_{0}$ and $ f_{1}=- \lambda f_{0}$ for 
\[
h(\lambda)=h-v^{2} - 2\lambda v,\ E= \lambda^{2}-m^{2}.
\]
Therefore to prove the proposition it suffices to show that $h(\lambda)$ has no strictly positive eigenvalues.  According to 
\cite[Thm. 11]{KT}, this is the case if we can write the 
potential $2\lambda v-v^{2}$  as $V_{l}+ V_{s}$ where $V_{\rm
s}\in L^{d/2}(\rr^{d})$ and:
\[
\p^{\alpha}V_{l}\in o(\langle x\rangle^{-|\alpha|}), \
|\alpha|\leq 1, \ \p^{\alpha}V_{l}\in O(\langle
x\rangle^{-|\alpha|}), \ |\alpha|=2.
\]
We take $V_{l}= 2\lambda v_{1}-v_{1}^{2}$, $V_{s}= 2\lambda
v_{2}-v_{2}^{2}-2 v_{1}v_{2}$. \qed
 
 \medskip
 
 We introduce now an important implicit condition, stating that $\pm m$ are not critical points:
 \[
\hbox{ \rm (B2)}\   m, - m \not \in c_{p}(B).
\]

We now give some sufficient conditions for (B2). We recall that $\epsilon^{2}= h+ m^{2}$.
\begin{lemma}
 \ben
 \item Assume that
 \[\left.
\begin{array}{l}
hu +(\pm 2mv -v^{2})u=0, \ u\in H^{2}(\rr^{d}),\\[2mm]
m^{2}\|u\|^{2}_{L^{2}}\mp m (u| vu)\leq 0,
\end{array}\right\}
\Rightarrow u=0.
\]
Then {\rm (B2)} holds;
\item Assume that $\epsilon^{2}= -\Delta + m^{2}$. If either
\[
\|v\|_{\infty}< \sqrt{2}m,
\]
or
\[
v\hbox{ has constant sign, } \|v\|_{\infty}< 2m,
\]
then {\rm (B2)} holds.
 \een
\end{lemma}
\proof The condition in (1) is equivalent to
\[
Bf= \pm mf, \ h[f, f]\leq 0\Rightarrow f=0.
\]
By   the proof of Thm. \ref{arsouille}, we see that critical points are eigenvalues of the restriction of $B$ to a subspace on which $h$ is non positive. Hence  the condition in (1) implies that  $\pm m\not\in c_{p}(B)$. To prove the first statement of (2), we note that using the first condition in (1), the second becomes $m^{2}\|u\|^{2}+\12 \|\nabla u\|^{2}-\12 \|vu\|^{2}\leq 0$. To  prove the second statement of (2), we note that if $v\leq 0$ the second condition of (1)  for  the $-$ sign implies  that $u=0$. If $0\leq v< 2m$ then $2m v- v^{2}> 0$ hence the first condition of (1)  for the $+$ sign implies that $u=0$. The argument for the reversed signs is similar. \qed

\section{Main results}\label{mainresults}
In this section we describe the results of this paper.

 \subsection{Spectrum of $B$}\label{spica}
We first  summarize what we know about the spectrum of $B$. We set $\sigma_{\rm pp}^{\cc}(B)= \sigma_{\rm pp}(B)\backslash \rr$, $\sigma_{\rm pp}^{\rr}(B)= \sigma_{\rm pp}(B)\cap \rr$.
\begin{proposition}
Assume hypotheses (A), (B). Then:
\ben
\item $\sigma_{\rm ess}(B)= ]-\infty, -m]\cup [m, +\infty[$;
\item $\sigma_{\rm pp}^{\cc}(B)=\bigcup_{j=1}^{N}\{z_{j}, \overline{z}_{j}\}$, where $z_{j}$, $\overline{z}_{j}$ are eigenvalues of finite algebraic multiplicities;
\item $\sigma_{\rm pp}^{\rr}(B)\subset [-m, m]$ is a (finite or infinite) sequence  $(\lambda_{i})_{i\in \nn}$ of eigenvalues which can accumulate only at $\pm m$,  the eigenvalues in $]-m, m[$ have finite algebraic multiplicities;
\item $\sigma_{\rm pp}^{\rr}\backslash c_{p}(B)$ have trivial Jordan blocks.
\een
\end{proposition}
 \begin{figure}[!ht]
\includegraphics*[scale=1]{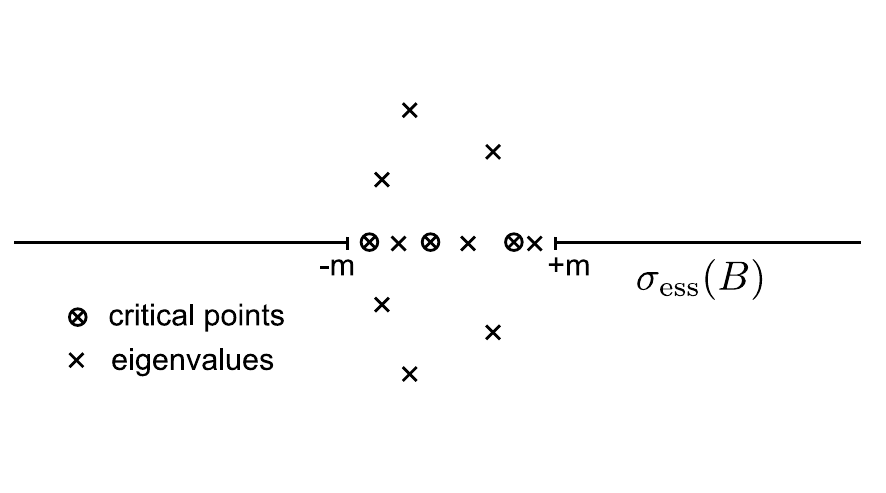}
\caption{The spectrum of $B$}
\label{fig1}
\end{figure}
\subsection{ Bound and scattering states}
We set
\[
\begin{array}{rl}
\one_{\rm pp}^{\cc}(B):=& \sum_{z\in \sigma_{\rm pp}^{\cc}(B)}E(z, B), \\[2mm]
 \one_{\rm pp}^{\rr}(B):= &\sum_{\lambda\in \sigma_{\rm pp}^{\rr}(B)}\one_{\{\lambda\}}(B),\\[2mm]
 \one_{\rm pp}(B):= &\one_{\rm pp}^{\cc}(B)+  \one_{\rm pp}^{\rr}(B).
 \end{array}
\]
Here $E(z, B)$ for $z\in \sigma_{\rm pp}^{\cc}(B)$ is the Riesz spectral projection on $z$ (see Appendix \ref{secIk.}). If $\lambda\in \sigma_{\rm pp}^{\rr}(B)\backslash c_{p}(B)$, then  $\one_{\{\lambda\}}(B)$ is defined in Subsect. \ref{secIk.5}. If $\lambda\in c_{p}(B)$ then $\one_{\{\lambda\}}(B)= \one_{[\lambda-\epsilon, \lambda+\epsilon]}(B)$ for all $\epsilon>0$ small enough.

The first sum is finite, the second strongly convergent. In fact, since $\sigma_{\rm ess}(B)\cap[-m , m]=\emptyset$, there  exist $ \delta>0$  such that all  real eigenvalues except for a finite number of them belong to $I= [-m-\delta, -M + \delta]\cup [m-\delta, m+\delta]$.  Since $\pm m$ are not critical points of $B$, the subspace $\one_{I}(B)\cE$ is positive. By Prop. \ref{pr:kp} $h[\cdot, \cdot]^{\12}$ is a Hilbertian norm on  $\one_{I}(B)\cE$, equivalent to the energy norm. This proves the strong convergence by the usual Hilbert space argument. 
We set:
\[
\cE_{\rm pp}(B):=  \one_{\rm pp}(B)\cE, \ \cE=: \cE_{\rm pp}(B)\oplus^{\perp}\cE_{\rm scatt}(B).
\]
The properties of $\cE_{\rm pp}(B)$ and $\cE_{\rm scatt}(B)$ are summarized in the following proposition:
\begin{proposition}\label{alfie}
 \ben
 \item $\cE_{\rm pp}(B)$ and $ \cE_{\rm scatt}(B)$ are Krein subspaces of $\cE$, invariant under $(\e^{-\i tB})_{t\in \rr}$;
 \item $\cE_{\rm pp}(B)$ and $ \cE_{\rm scatt}(B)$  are closed symplectic subspaces of $\cE$ and are symplectically orthogonal;
  \item Let $u\in \cE_{\rm pp}(B)$. Then
 \[
\e^{- \i t B}u= \sum_{z\in \sigma_{\rm pp}^{\cc}(B)}\e^{- \i tB}E(z, B)u + \sum_{\lambda\in \sigma_{\rm pp}^{\rr}(B)}\e^{-\i tB}\one_{\{\lambda\}}(B)u,
\]
where the first sum is finite, the second strongly convergent, uniformly for $t\in \rr$;
 \item one has
 \[
 \cE_{\rm scatt}(B)=\one_{]-\infty, -m[\cup]m, +\infty[}(B)\cE.
 \]

 \een

\end{proposition}
The space $\cE_{\rm scatt}(B)$ will be called the space of {\em scattering states} for $B$.

\proof  (1) follows from Appendices  \ref{sec0}, \ref{secIk.}.  (2) follows from (1) and (\ref{atchoum}).  To prove (3) we use  that the real eigenvalues of $B$ can accumulate  only at $\pm m$, that $\pm m\not\in c_{p}(B)$ and that the eigenvalues of $B$ in $\rr\backslash c_{p}(B)$ have trivial Jordan blocks (see  Prop. \ref{coco}). (4)
 follows from Prop. \ref{lati} (4)  and the fact that $\pm m\not\in c_{p}(B)$.  \qed
\subsection{Existence and completeness of short-range wave operators}
In this subsection we assume hypotheses (A1) for $\mu_{0}>1$, (A2), (A3), (A4) for $v_{l}=0$, and  (B). In other words we are in the short-range case.
 We set
$\cE_{\infty}:= H^{1}(\rr^{d})\oplus L^{2}(\rr^{d})$, equipped with the usual energy scalar product:
\[
h_{\infty}[f, f]= (f_{1}| f_{1})+ (f_{0}|\epsilon_{\infty}^{2}f_{0}), \  \epsilon_{\infty}^{2}:= (-\Delta +m^{2}),
\]
so that $\cE_{\infty}= \cE$ as topological spaces. We set also
 \[
B_{\infty}:=-\mat{0}{\one}{\epsilon_{\infty}^{2}}{0}, 
\]
which is the generator of the free Klein-Gordon evolution with mass $m$. 
\bet\label{youpla}
Assume  hypotheses (A1) for $\mu_{0}>1$, (A2), (A3), (A4) for $v_{l}=0$, and  (B). Then: 
 \ben
 \item for all $f\in \cE_{\infty}$ there exist  unique $f^{\pm}\in \cE_{\rm scatt}(B)$ such that
 \[
 \e^{- \i t B}f^{\pm}- \e^{-\i tB_{\infty}}f\to 0, \ t\to \pm \infty.
\]
\item  Let us define  the {\em short-range wave operators} $\Omega^{\pm}_{s}$:
\[
\Omega^{\pm}_{s}: 
\begin{array}{rl}
\cE_{\infty}&\to \cE_{\rm scatt}(B),\\
 f&\mapsto f^{\pm}.
\end{array}
\]
Then:
\[
\begin{array}{rl}
(i)&\Omega^{\pm}_{s}\hbox{ are bounded symplectic transformations}, \\[2mm]
(ii)&\Omega^{\pm}_{s}\e^{-\i tB_{\infty}}= \e^{-\i t B}\Omega^{\pm}_{s}, \ t\in \rr,\\[2mm]
 (iii)&\Omega^{\pm}_{s}\hbox{ are unitary   from }(\cE_{\infty}, h_{\infty}[\cdot, \cdot])\hbox{ to }(\cE_{\rm scatt}(B), h[\cdot,\cdot]).
\end{array}
\]
\een
\eet
The proof will be given in Subsect. \ref{you1}.
\subsection{Existence and completeness of long-range wave operators}
We assume now hypotheses (A), (B), i.e. we are in the long-range case. As in the case of Schr\"{o}dinger operators, it is necessary to introduce a {\em modified free dynamics} to define the wave operators.  We choose to use {\em time-independent modifiers} analogous to those introduced by Isozaki-Kitada for Schr\"{o}dinger operators \cite{IK}.  
 It turns out that it is  necessary to assume that the long-range potential $v_{l}$ is of constant sign near infinity. 
 This is  not a serious restriction from the point of view of physical applications.
Hence we introduce the hypothesis
 \[
{\rm (C)}\ \pm v_{\l}(x)\geq  0\hbox{ for }|x|\gg 1.
\]
Let us now define the time-independent modifiers.  We recall from  Subsect. \ref{absence} that  $\epsilon^{2}=  \Op\tilde{\epsilon}^{2}$ for 
\[
\tilde{\epsilon}^{2}(x, \xi)= \sum_{jk}\xi_{j}c^{jk}(x)\xi_{k}+ 2\sum_{j} d_{j}(x)\xi_{j}+ r(x)+ m^{2},
\]
where 
\[
[c^{jk}](x)-\one,  d_{j}(x), r(x)\in S^{0, -\mu_{0}}(\rr^{2d}).
\]
We denote by $\varphi_{\pm}( x, \xi)$   the functions introduced in Lemma \ref{kitada}. We recall that they solve  the eikonal equations (see Appendix \ref{astruc}):
\[
\pm \tilde{\epsilon}(x, \p_{x}\varphi_{\pm}(x, \xi)) -v_{l}(x)= \pm(\xi^{2}+m^{2})^{\12}, 
\]
in some outgoing/incoming regions.  We denote by $j_{\pm}$ the associated Fourier integral operators defined as:
\[
j_{\pm}u(x)= (2\pi)^{-d}\int \e^{\i \varphi_{\pm}(x, \xi)- \i y\cdot \xi}u(y) \d y \d \xi,
\]
which are bounded operators on $L^{2}(\rr^{d})$ and $H^{1}(\rr^{d})$.
\begin{definition}\label{def-de-T}
The time-independent modifier $T$ is defined as
\[
 T:=\pm\12\mat{j_{+}- j_{-}}{-(j_{+}+ j_{-})\epsilon_{\infty}^{-1}}{-(j_{+}+ j_{-})\epsilon_{\infty}}{j_{+}- j_{-}},
\]
where we use the $\pm$ sign according to the sign of $v_{l}$ in {\rm (C)}.
\end{definition}
\bet\label{youpli}
Assume  hypotheses  {\rm(A)},  {\rm(B)} and  {\rm(C)}. Then: 
 \ben
 \item for all $f\in \cE_{\infty}$ there exist  unique $f^{\pm}\in \cE_{\rm scatt}(B)$ such that
 \[
 \e^{- \i t B}f^{\pm}- T\e^{-\i tB_{\infty}}f\to 0, \ t\to \pm \infty.
\]
\item   Let us define  the {\em long-range wave operators} $\Omega^{\pm}_{l}$:
\[
\Omega^{\pm}_{l}: 
\begin{array}{rl}
\cE_{\infty}&\to \cE_{\rm scatt}(B),\\
 f&\mapsto f^{\pm}.
\end{array}
\]
Then:
\[
\begin{array}{rl}
(i)&\Omega^{\pm}_{l}\hbox{ are bounded, symplectic transformations}, \\[2mm]
(ii)&\Omega^{\pm}_{l}\e^{-\i tB_{\infty}}= \e^{-\i t B}\Omega^{\pm}_{l}, \ t\in \rr,\\[2mm]
 (iii)&\Omega^{\pm}_{l}\hbox{ are unitary   from }(\cE_{\infty}, h_{\infty}[\cdot, \cdot])\hbox{ to }(\cE_{\rm scatt}(B), h[\cdot,\cdot]).
\end{array}
\]
\een
\eet
The proof will be given in Subsect. \ref{you2}.
\section{Technical  estimates}\init \label{sec6}

In this section we   describe  an approximate diagonalization of $B$. We also collect some technical estimates which will be used later. We  assume hypotheses (A) and (B1).
\subsection{Approximate diagonalization}\label{reduc}
We use the notation $F(a\leq \lambda\leq b)$ explained in Subsect. \ref{titi}.

Replacing $v_{l}(x)$  by $F(|x|\geq c)v_{l}(x)$ and $v_{\rm
s}(x)$ by $v_{s}(x)+ F(|x|\leq c)v_{l}(x)$ for $c\gg 1$ we can
assume that:
\[
\epsilon^{2}- v_{l}^{2}\geq m^{2}/4.
\]
We set:
\[
r:= v^{2}- v_{\l}^{2}= v_{\s}^{2}+ 2 v_{\s}v_{\l}, \  b:=
(\epsilon^{2}- v_{\l}^{2})^{\12},
\]
and write $B$ as the operator sum:
\[
B=B_{l}+R, \ B_{l}:=-\mat{0}{\one}{ (\epsilon^{2}- v_{\l}^{2})}{2 v_{\l}}, \ R= \mat{0}{0}{r}{-2 v_{\s}}.
\]
We equip the Krein space $\cE$ with the norm
\[
\|f\|^{2}= \|f_{1}\|^{2}+ (f_{0}| (\epsilon^{2}- v_{\l}^{2})f_{0}),
\]
which yields the same topology.
 
It is convenient to unitarily (as Hilbert spaces) map $\cE$ to $\ch\oplus\ch$ using the map:
\beq\label{def-de-u}
U:=\frac{1}{\sqrt{2}}\mat{b}{-\one}{b}{\one},\ 
U^{-1}=\frac{1}{\sqrt{2}}\mat{b^{-1}}{b^{-1}}{-\one}{\one}.
\eeq
Then  $U:\ \cE\to\ch\oplus\ch$ is unitary and
\[
U B_{\l}U^{*}= \mat{b-v_{\l}}{ v_{\l}}{v_{\l}}{-b- v_{\l}}, \ URU^{*}= -\frac{1}{2}\mat{ rb^{-1}+ 2 v_{\s}}{rb^{-1}- 2 v_{\s}}{ rb^{-1}- 2 v_{\s}}{ rb^{-1}+ 2 v_{\s}}.
\]
We set now:
\[
L_{0}:=\mat{b}{0}{0}{-b},  
\]
\[
V_{l}:= \mat{-v_{\l}}{v_{\l}}{v_{\l}}{-v_{\l}}, \ V_{s}=-\frac{1}{2}\mat{ rb^{-1}+ 2 v_{\s}}{rb^{-1}- 2 v_{\s}}{ rb^{-1}- 2 v_{\s}}{ rb^{-1}+ 2 v_{\s}}.
\]
and
\beq\label{def-de-L}
L:= UBU^{-1}= L_{0}+V,\ V:= V_{l}+ V_{s},
\eeq
which is closed with $\Dom L= \epsilon^{-1}\ch\oplus \epsilon^{-1}\ch$.

After conjugation by $U$, the Krein bilinear form $h[\cdot, \cdot]$ becomes:
\[
[u, u]:=h[U^{-1}u, U^{-1 }u]=(u|(\one + K)u),
\]
for
\begin{equation}
\label{urlu}
K= -\12\mat{b^{-1}rb^{-1}}{b^{-1}rb^{-1}}{b^{-1}rb^{-1}}{b^{-1}rb^{-1}}.
\end{equation}
Let us summarize the results of this reduction: 

-we work on the space  $\cK= \ch\oplus \ch$, $\ch= L^{2}(\rr^{d})$ with its natural Hilbert space topology. 

-the space $\cK$ is equipped with the Krein bilinear form:
\[
[u, u]= (u|(\one +K)u), 
\]
where $K$ is defined in (\ref{urlu}), so that $(\cK, [\cdot, \cdot])$ is a Pontryagin space.  

-we consider the operator $L$ defined in (\ref{def-de-L}), which is closed with domain $\epsilon^{-1}\ch\oplus \epsilon^{-1}\ch= H^{1}(\rr^{d})\oplus H^{1}(\rr^{d})$.  The operator $L$ is selfadjoint on the Pontryagin space
$(\cK, [\cdot, \cdot])$ and hence definitizable with a definitizing polynomial of even degree.

 \subsection{Estimates on scalar operators}
In this subsection we collect various estimates involving $b$, $v_{\l}$ and $v_{\s}$. 
We set
  \beq
  \label{conjug}
\epsilon_{\infty}:= (D^{2}+ m^{2})^{\12},\ a:= \12(x\cdot \frac{D}{D^{2}+ m^{2}}+ \frac{D}{D^{2}+ m^{2}}\cdot x).
\eeq
Note that $a$ is selfadjoint on $L^{2}(\rr^{d})$ with domain $\Dom a= \{u\in L^{2}\ :\ au\in L^{2}\}$.  The symbol classes $S^{m,p}(\rr^{2d})$
 are defined in Appendix  \ref{pdocalc}.

\begin{proposition}\label{3.1}
 \ben
 \item $[b, \x]$ is bounded;
 \item $b^{2}[b^{-1}, \x]$ is bounded;
 \item
 \[
[b, \i a]= b^{-3}(b^{2}-m^{2})+ T, \hbox{ where }T\hbox{ is compact};
\]
\item let $F(x, \xi)\in S^{0, 0}(\rr^{2d})$ with $\supp F\subset \{|x|\geq c \}$ for some $c>0$. Then
\[
\begin{array}{rl}
[b, \i F(\xt, D_{x})]=& \frac{1}{2t}\epsilon_{\infty}^{-\12}(D_{x}\cdot \nabla_{x}F(\xt, D_{x})+ D_{x}\cdot \nabla_{x}F(\xt, D_{x}))\epsilon_{\infty}^{-\12}\\[2mm]
&+ O(t^{-1- \delta}), \ \delta>0;
\end{array}
\]
\item Let $G, G_{1}\in \cinf(\rr^{d})$ bounded with all derivatives with  $0\not\in \supp G, G_{1}$ and $G_{1}G=G$. Then:
\[
b^{-1}G(\xt)= G_{1}(\xt) b^{-1}G(\xt)+ R(t),
\]
where
\[
bR(t), \ b^{2}R(t)b^{-1}\in O(t^{-2}).
\]
\een
\end{proposition}
\begin{proposition}\label{3.1bis} Let $F(x, \xi)$ as in Prop. \ref{3.1}. Then
 \ben
 \item 
 \[
[v_{\l}(x), F(\xt, D_{x}) ]\in O(t^{-1- \mu_{l}});
\]
\item $v_{\s}F(\xt, D_{x}) b^{-1}, \ F(\xt, D_{x})v_{\s} b^{-1}\in O(t^{-\mu_{s}})$;
 \item $rb^{-1} F(\xt, D_{x})b^{-1}\  F(\xt, D_{x})rb^{-2}\in O((t^{-\mu_{s}})+ O(t^{-2})$.
\een
It follows that
\[
[v_{\s},F(\xt, D_{x})] b^{-1}\in O(t^{-\mu_{s}}), \ [rb^{-1}, F(\xt, D_{x})]b^{-1}\in O(t^{-\mu_{s}})+ O(t^{-2}).
\]
\end{proposition}
\proof 
The proofs will be given in Subsect.  \ref{ap1}. \qed

\subsection{Estimates on matrix operators}\label{matmat}

If $c$ is an operator on $\ch$, we still denote by $c$ the operator $\mat{c}{0}{0}{c}$.

\begin{lemma}\label{3.0}
 Let $\chi\in \coinf(\rr)$ and $F(x, \xi)$ as in Prop. \ref{3.1}. Then
 \[
 \begin{array}{rl}
(1)& \chi(L)- \chi(L_{0}) \hbox{ is compact},\\[2mm]
(2) & \left(\chi(L)- \chi(L_{0})\right)F(\xt, D_{x})\in O(t^{- \delta}), \ \delta>0,\\[2mm]
(3) & \langle x\rangle^{1+\delta}V_{s}\epsilon^{-1}\hbox{ is bounded},\ \delta>0,\\[2mm]
(4)& \x \chi(L)\x^{-1}, \ \x^{-1}\chi(L)\x\hbox{ are bounded},\\[2mm]
(5)& [\chi(L), F(\xt, D_{x})]\in O(t^{-\delta}), \ \delta>0.
\end{array}
\]
\end{lemma}
\proof we will use the functional calculus defined in Subsect. \ref{secIk.4}, based on almost-analytic extensions.

(1):   We have:
\[
\chi(L)- \chi(L_{0})= \frac{\i}{2\pi}\int_{\cc}\frac{\p\tilde{\chi}(z)}{\p\bar{z}} (z-L)^{-1}V(z-L_{0})^{-1}\: d z\wedge d\bar{z}.
\]
\def\Im{{\rm Im}}
We use the estimates (see Lemma \ref{besson} for the first one):
\[
\|(z- L)^{-1}\|\leq C|{\rm Im}z |^{-m-1},  \ \|L_{0}(z- L_{0})^{-1}\|\leq C|\Im z|^{-1},\ z\in \supp \tilde{\chi},
\]
and  the fact that $VL_{0}^{-1}$ is compact by (A2). This proves (1).

(2): We write
\[
\begin{array}{rl}
&(z-L)^{-1}V(z-L_{0})^{-1}F(\xt, D_{x})\\[2mm]
=&(z-L)^{-1}VF(\xt, D_{x})(z-L_{0})^{-1}+ (z-L)^{-1}V(z-L_{0})^{-1}[L_{0}, F(\xt, D_{x})](z-L_{0})^{-1}.
\end{array}
\]
It follows from Prop.  \ref{3.1bis} (2), (3) that $V_{s}F(\xt, D_{x})L_{0}^{-1}\in O(t^{-1- \delta})$. By (\ref{e3.01}) $V_{l}F(\xt, D_{x})\in O(t^{-\mu_{l}})$. Therefore the first term in the r.h.s. is $O(t^{-\delta})\langle z\rangle|\Im z|^{-m-2}$. The second term is $O(t^{-1})\langle z\rangle|\Im z|^{-m-3}$.  This gives (2). 

(3): $\langle x\rangle^{1+\delta}v_{\s}\epsilon^{-1}$ and $\langle x\rangle^{1+\delta}r\epsilon^{-2}$ are bounded by (\ref{e3.02}). This implies (3). 

(4): We first claim that $[L_{0}, \x]$  and $[V, \x]= [V_{s}, \x]$ are bounded.  The first claim follows from Prop. \ref{3.1} (1).  To prove the second we note that 
\[
[rb^{-1}, \x]= r[b^{-1}, \x]=rb^{-2}b^{2}[b^{-1}, \x]
\]
is bounded, since $rb^{-2}$ is bounded by (\ref{e3.02}) and $b^{2}[b^{-1}, \x]$ is bounded by Prop. \ref{3.1} (2). Next we write
 \[
\x(z-L)^{-1}\x^{-1}= (z-L)^{-1}+ (z-L)^{-1}[L, \x](z-L)^{-1}\x^{-1},
\]
and use (\ref{e3b}).

(5): by (2) it suffices to estimate $[\chi(L_{0}), F]$. The result follows then from pseudo-differential  calculus, see e.g. \cite[Appendix D]{DG}. 
\qed

\begin{proposition}\label{3.3}
Let $F$ be as above and $\chi\in \coinf(\rr)$. Then:
\begin{equation}
\label{e3.2}
\chi(L)[V, \i F(\xt, D_{x})]\chi(L)\in O(t^{-1- \delta}), \ \delta>0.
\end{equation}
\end{proposition}
\proof we apply Prop. \ref{3.1bis} (1) and Prop. \ref{3.1bis} (2), (3), using that $b\chi(L)$ is bounded since $\chi(L)$ maps $\ch\oplus \ch$ into $\Dom L= \epsilon^{-1}\ch\oplus \epsilon^{-1}\ch$. \qed

\begin{proposition}\label{3.5}
 Let $G\in \cinf(\rr^{d})$ bounded with all derivatives with  $0\not\in \supp G$ and $K$  be defined in (\ref{urlu}). Then:
 \[
KG(\xt)\in O(t^{-\mu_{s}})+ O(t^{-2}).
\]
\end{proposition}
\proof Let $G_{1}(\xt)$ as in Prop. \ref{3.1} (5).We have 
\[
b^{-1}rb^{-1}G(\xt)= b^{-1}r G_{1}(\xt)b^{-1}G(\xt)+ b^{-1}r R(t)= b^{-1}r G_{1}(\xt)b^{-1}G(\xt)+ b^{-1}r b^{-1}bR(t)
\]
Now $b^{-1}rG_{1}(\xt)b^{-1}\in O(t^{-\mu_{s}})$,  $b^{-1}rb^{-1}$ is bounded and $b R(t)\in O(t^{-2})$ by Prop. \ref{3.1} (5). \qed

\begin{proposition}\label{3.00}
\[
\begin{array}{rl}
(1)&[L_{0}, \ i a]= \mat{b^{-3}(b^{2}- m^{2})}{0}{0}{-b^{-3}(b^{2}- m^{2})}+ R,  \hbox{ where }R\hbox{ is compact},\\[2mm]
(2)& \chi(L)[V, \i a]\chi(L)\hbox{ is compact}.
 \end{array}
\]
\end{proposition}

\proof:
(1) follows from Prop. \ref{3.1} (3).
Since $[v_{\l}, \i a]$ is compact so is $[V_{l}, \i a]$. From Lemma \ref{3.0} (3) we see that $\chi(L)V_{s}a$ and $ a V_{s}\chi(L)$ are compact. This implies (2). \qed

\subsection{Positive commutators}
In this subsection we prove a key result on local positivity of the commutator $[L, \i a]$. Its form is  different from the usual selfadjoint case, because the two scalar products $[\cdot, \cdot]$ and $(\cdot | \cdot)$ play a role (see Rem. \ref{ferdine}).
\begin{proposition}\label{commut}
 Let $a$ be defined in (\ref{conjug}) and let $\lambda_{0}\in \rr$ with $\pm \lambda_{0} >m$.  Set 
 $\chi_{\delta}(\lambda)= F(\delta^{-1}\lambda)$ where $F\in \coinf(\rr)$, $F(\lambda)\equiv 1$ for  $|\lambda-\lambda_{0}|\leq 1$, $F(\lambda)\equiv 0$ for $|\lambda-\lambda_{0}|\geq 2$.
 
 Then there exists $c_{0}>0$, $g\in \coinf(\rr)$ and $R_{\delta}\in B(\cK)$ such that:
 \beq\label{irlta}
\begin{array}{rl}
&\pm\chi_{\delta}(L)[L, \i a]\chi_{\delta}(L)\\[2mm]
=& c_{0}\chi_{\delta}(L)(\one + K)^{-1}\chi_{\delta}(L)+ \chi_{\delta}^{2}(L)g^{2}(L)+ \chi_{\delta}(L)(\one +K)^{-1}R_{\delta}\chi_{\delta}(L), 
\end{array}
\eeq
with
\[
\lim_{\delta\to 0}R_{\delta}=0.
\]
\end{proposition}

\begin{remark}\label{ferdine}
 Taking expectation values of (\ref{irlta}) for the Krein scalar product, we obtain
\[
\begin{array}{rl}
&\pm [u,\chi_{\delta}(L)[L, \i a]\chi_{\delta}(L) u]\\[2mm]
=&c_{0}(\chi_{\delta}(L)u| \chi_{\delta}(L)u)+ [g(L)\chi_{\delta}(L)u, g(L)\chi_{\delta}(L)u]\\[2mm]
&+ (\chi_{\delta}(L)u| R_{\delta}\chi_{\delta}(L)u), \ u \in \cK.
\end{array}
\]
\end{remark}
\proof We treat only the case $\lambda_{0}>m$, the case $\lambda_{0}<-m$ being similar except for the change of signs.
Let $\chi\in \coinf(\rr)$ supported in $]m, +\infty[$. Then:
\[
\begin{array}{rl}
\chi(L)[L, \i a]\chi(L)=& \chi(L) \mat{b^{-3}(b^{2}- m^{2})}{0}{0}{-b^{-3}(b^{2}- m^{2})}\chi(L)+ R_{1}\\[2mm]
= &\chi(L_{0}) \mat{b^{-3}(b^{2}- m^{2})}{0}{0}{-b^{-3}(b^{2}- m^{2})}\chi(L_{0})+ R_{1}+ R_{2}\\[2mm]
=&\chi(L_{0}) \mat{b^{-3}(b^{2}- m^{2})}{0}{0}{b^{-3}(b^{2}- m^{2})}\chi(L_{0})+ R_{1}+ R_{2}\\[2mm]
=&\chi(L_{0})\left(|L_{0}|^{-3}(L_{0}^{2}- m^{2})\right)\chi(L_{0})+ R_{1}+ R_{2}
\end{array}
\]
where  $R_{1}$, $R_{2}$ are compact
and we used successively Prop. \ref{3.00}, Lemma \ref{3.0} (1) and the fact that $\supp \chi\subset ]m, +\infty[$. 

Clearly there exists $c_{0}>0$ and $g\in \coinf(\rr)$ such that:
\[
\chi^{2}(\lambda) \left(|\lambda|^{-3}(\lambda^{2}- m^{2})\right)=  c_{0}\chi^{2}(\lambda)+ \chi^{2}(\lambda)g^{2}(\lambda).
\]
Using once more Lemma \ref{3.0} (1), we obtain:
\[
\chi(L)[L, \i a]\chi(L)= c_{0}\chi^{2}(L)+ \chi^{2}(L)g^{2}(L)+ R, 
\]
where $R$ is compact. Since $K$ is compact this yields:
\begin{equation}
\label{e3.11}
\begin{array}{rl}
&\chi(L)[L, \i a]\chi(L)\\[2mm]
=& c_{0}\chi(L)(\one + K)^{-1}\chi(L)+c_{0}\chi(L)K(\one + K)^{-1}\chi(L)+ \chi^{2}(L)g^{2}(L)+ R\\[2mm]
=&c_{0}\chi(L)(\one + K)^{-1}\chi(L)+ \chi^{2}(L)g^{2}(L)+ R_{1},
\end{array}
\end{equation}
where $R_{1}$ is compact.

Since $\lambda_{0}>m$, we know  using (B1) and Prop. \ref{noemb} that $\lambda_{0}$ is not an eigenvalue of $L$ and hence by Prop. \ref{lati} (2): 
\beq\label{bebert}
\slim_{\delta\to 0}\chi_{\delta}(L)=0, \hbox{ and hence }\lim_{\delta\to 0} \chi_{\delta}(L)R_{1}=0.
\eeq
Multiplying to the left and right both sides of (\ref{e3.11}) by $\chi_{\delta}(L)$ we obtain:
\[
\begin{array}{rl}
&\chi_{\delta}(L)[L, \i a]\chi_{\delta}(L)\\[2mm]
=& c_{0}\chi_{\delta}(L)(\one + K)^{-1}\chi_{\delta}(L)+ \chi_{\delta}^{2}(L)g^{2}(L)+ \chi_{\delta}(L)\chi_{2\delta}(L)R_{1}\chi_{2\delta}(L)\chi_{\delta}(L)\\[2mm]
=&c_{0}\chi_{\delta}(L)(\one + K)^{-1}\chi_{\delta}(L)+ \chi_{\delta}^{2}(L)g^{2}(L)+ \chi_{\delta}(L)(\one +K)^{-1}R_{\delta}\chi_{\delta}(L),
\end{array}
\]
for
\[
R_{\delta}= (\one +K)\chi_{2\delta}(L)R_{1}\chi_{2\delta}(L).
\]
Using (\ref{bebert}) this completes the proof of the proposition. \qed

\section{Propagation estimates}\init\label{sec3}
In all this section we assume hypotheses (A), (B).
We will prove propagation estimates for $\e^{-\i tL}$. Their content is parallel to the well-known selfadjoint case.  The abstract setup for propagation estimates on Krein spaces is however different, and developed in Appendix \ref{sec4}.
We denote as usual by
\[
\begin{array}{l}
 {\bf D}_{0}\Phi(t)=  \p_{t}\Phi(t)+ [L_{0}, \i \Phi(t)],\\[2mm]
 {\bf D}\Phi(t)= \p_{t}\Phi(t)+ [L, \i \Phi(t)],
\end{array}
\]
the  free and interacting Heisenberg derivatives. We recall that we use the notation explained at the beginning of Subsect. \ref{matmat} for scalar operators. We recall also that all operator inequalities are to be understood in Hilbert sense, it $A\geq 0$ means $(u|Au)\geq 0$.

\subsection{Large velocity estimates}\label{sec3.2}
\begin{proposition}\label{3.4}
 Let $\chi\in \coinf(\rr)$  with $\supp \chi\cap c_{p}(L)= \emptyset$. Then: 
 
 (i) if $f\in \coinf(]1, +\infty[)$:
 \[
\int_{1}^{+\infty}\| f(\frac{|x|}{t})\e^{-\i tL}\chi(L)u\|^{2}\frac{dt }{t}\leq C\|u\|^{2}.
\]
(ii) if $f\in \cinf(\rr)$ is bounded and $\supp f\subset ]1, +\infty[$ then: 
\[
\slim_{t\to +\infty} f(\frac{|x|}{t})\e^{-\i tL}\chi(L)=0.
\]
\end{proposition}
\proof

{\it Proof of (i)}: 

\medskip

Fix $1<\theta_{0}<\theta_{1}<\theta_{2}$, $f\in \coinf([\theta_{0}, +\infty[)$ with $f\equiv 1$ on $[\theta_{1}, \theta_{2}]$ and $F(s)= -\int_{-\infty}^{s} f^{2}(s_{1})ds_{1}$. Set:
\[
\Phi(t)=  F(\frac{|x|}{t}).
\]
Let us fix $\chi_{1}\in \coinf(\rr)$ supported away from critical points with $\chi_{1}\chi= \chi$. Then:
\[
 \chi_{1}(L){\bf D}F(\frac{|x|}{t}) \chi_{1}(L)= \chi_{1}(L){\bf D}_{0}F(\frac{|x|}{t})\chi_{1}(L)+ \chi(L)[V, \i F(\frac{|x|}{t})]\chi_{1}(L),
\]
where
\beq\label{e3.3a}
\|\chi_{1}(L)[V, \i F(\frac{|x|}{t})]\chi_{1}(L)\|\in O(t^{-1-\delta}),
\eeq
  by Prop. \ref{3.3}.
We have by Prop. \ref{3.1} (4):
\[
\begin{array}{rl}
&\p_{t}F(\frac{|x|}{t})\pm [b, \i F(\frac{|x|}{t})]\\[2mm]
=&-\frac{|x|}{t^{2}}F'(\frac{|x|}{t})\pm \frac{1}{2t}\epsilon_{\infty}^{-\12}\left(D_{x}\cdot \frac{x}{|x|}F'(\frac{|x|}{t})+ F'(\frac{|x|}{t})\frac{x}{|x|}\cdot D_{x}\right)\epsilon_{\infty}^{-\12}+ O(t^{-1-\delta})\\[2mm]
=&\frac{1}{t}f(\frac{|x|}{t})\left(\frac{|x|}{t}\mp\12\left(\epsilon_{\infty}^{-1}D_{x}\cdot \frac{x}{|x|}+
 \frac{x}{|x|}\cdot D_{x}\epsilon_{\infty}^{-1}\right)\right)f(\frac{|x|}{t})+ O(t^{-1-\delta})+ O(t^{-2})\\[2mm]
\geq &  \frac{C}{t}f^{2}(\frac{|x|}{t}) + O(t^{-1-\delta}),
\end{array}
\]
using that $\|\epsilon_{\infty}^{-1}D_{x}\|\leq 1$ and $\theta_{0}>1$.  Hence
\begin{equation}
\label{e3.3b}
{\bf D}_{0}F(\frac{|x|}{t})\geq  \frac{C}{t}f^{2}(\frac{|x|}{t}) + O(t^{-1-\delta}).
\end{equation}
We claim that we can  apply Corollary \ref{abs4}  for
\[
\begin{array}{rl}
R_{0}(t)=& \chi_{1}(L)[V, \i F(\frac{|x|}{t})]\chi_{1}(L),\\[2mm]
D(t)= &{\bf D}_{0}F(\frac{|x|}{t}).
\end{array}
\]
In fact applying Prop. \ref{3.5} we  see that $KD(t)\in O(t^{-1- \delta})$ hence condition {\it ii)} 
of Corollary \ref{abs4} is satisfied. Condition {\it i)} is satisfied by (\ref{e3.3a}), conditions {\it iii)} and {\it iv)} 
follow from (\ref{e3.3b}).

{\it  Proof of (ii)}:  

\medskip	
We take $F$, $f$ as above and 
set $F_{Rt}= F(\frac{|x|}{Rt})$, $f_{Rt}= f(\frac{|x|}{Rt})$ for $R\geq 1$. Using the  computations above with $t$ replaced by $Rt$ we get that:
\begin{equation}
\label{connard.e1}
\begin{array}{l}
KF_{Rt}\in O((Rt)^{-\delta}), \ \delta>0\\[2mm]
K{\bf D}_{0}F_{Rt}\in O((Rt)^{-1- \delta}), \\[2mm]
\end{array}
\end{equation}
\begin{equation}
\label{connard.e2}
\begin{array}{l}
 \chi(L)[V, \i F_{Rt}]\chi(L)\in O((Rt)^{-1- \delta}), \ \chi\in \coinf(\rr),\\[2mm]
{\bf D}_{0}F_{Rt}= \frac{1}{t}f_{Rt}T(R, t)f_{Rt}+ O( (Rt)^{-1- \delta}), \\[2mm]
c_{0}\one \leq T(R, t)\leq c_{1}\one,  \ c_{0}>0.
\end{array}
\end{equation}
Setting now $u_{t}= \e^{- \ i tL}\chi(L)u$ we deduce from (\ref{connard.e1}), (\ref{connard.e2}) that:
\beq\label{e3.4}
\begin{array}{rl}
\frac{d}{dt}[u_{t}, F_{Rt}u_{t}]=&[u_{t}, {\bf D}_{0}F_{Rt}u_{t}]+ O((Rt)^{-1 -\delta})\\[2mm]
=&(u_{t}|  {\bf D}_{0}F_{Rt}u_{t})+  O((Rt)^{-1 -\delta})\\[2mm]
=&(f_{Rt}u_{t}| T(R, t)f_{Rt}u_{t})+ O((Rt)^{-1 -\delta}).
\end{array}
\eeq
Using (i) and Cauchy-Schwarz inequality we see that
 \[
\lim_{t\to +\infty}[u_{t}, F_{Rt}u_{t}]\hbox{ exists}.
\]
Moreover  by Prop. \ref{3.5} and (\ref{connard.e1}) we get:
\[
[u_{t}, F_{Rt}u_{t}]= (u_{t}| F_{Rt}u_{t})+ (u_{t}| K F_{Rt}u_{t})= (u_{t}| F_{Rt}u_{t})+ 0((Rt)^{-\delta}),
\]
hence:
\beq\label{connard.e3}
\lim_{t\to +\infty}[u_{t}, F_{Rt}u_{t}]=\lim_{t\to+\infty}(u_{t}| F_{Rt}u_{t})\leq 0.
\eeq
\[
\lim_{t\to+\infty}\||F_{Rt}|^{\12}u_{t}\|\hbox{ exists}.
\]
Integrating  (\ref{e3.4})  between $t=1$ and $t=+\infty$ we obtain
\[
\begin{array}{rl}
&\lim_{t\to +\infty}[u_{t}, F_{Rt}u_{t}]=\lim_{t\to +\infty}(u_{t}| F_{Rt}u_{t})\\[2mm]
=& [u_{1}, F_{R}u_{1}]+ \int_{1}^{+\infty}(f_{Rt}u_{t}| T(R, t)f_{Rt}u_{t})dt+ O(R^{-1- \delta}).
\end{array}
\]
Since $T(R,t)\geq 0$ by (\ref{connard.e2}), this implies using also (\ref{connard.e3}) that
\[
0\geq \lim_{t\to +\infty}[u_{t}, F_{Rt}u_{t}]\geq |[u_{1}, F_{R}u_{1}]| + O(R^{-1-\delta}).
\]

The terms on the right  tend to $0$ when $R\to +\infty$ hence:
\beq\label{e3.6}
\lim_{R\to +\infty}\lim_{t\to +\infty}[u_{t}, F_{Rt}u_{t}]=0.
\eeq
Since $F_{1}- F_{R}\in \coinf(]1, + \infty[)$ we have also  using (i):
\[
\lim_{t\to +\infty}[u_{t}, (F_{t}- F_{Rt})u_{t}]=0,
\] 
and letting $R\to \infty$ in the identity above using also (\ref{e3.6}) we obtain
\[
\lim_{t\to +\infty}[u_{t}, F_{t}u_{t}]=0.
\]
By (\ref{connard.e3}) this yields
\[
\lim_{t\to +\infty}\| |F_{t}|^{\12}u_{t}\|=0,
\]
which  completes the proof of the proposition. \qed

\subsection{Phase space propagation estimates}
Set 
\beq\label{e3.7}
{\bf v}_{\s}= \epsilon_{\infty}^{-1}D_{x}, \ {\bf v}:=\mat{\one}{0}{0}{- \one}{\bf v}_{\s},
\eeq
considered as a  vector of commuting operators on $\ch\oplus \ch$.
Note that ${\bf v}= [L_{\infty}, \i x]$, where $L_\infty$ is defined in
(\ref{shitique}),  hence  has the meaning of the velocity operator.
\begin{proposition}\label{3.6}
 Let $\chi\in \coinf(\rr)$ with $\supp \chi\cap[-m,m]=\emptyset$. Then: 
 
 (i) if $f\in \coinf(]0, +\infty[)$:
 \[
\int_{1}^{+\infty}\| f(\frac{|x|}{t})({\bf v}-\xt)\e^{-\i tL}\chi(L)u\|^{2}\frac{dt }{t}\leq C\|u\|^{2}.
\]
(ii) moreover:
\[
\slim_{t\to +\infty} f(\frac{|x|}{t})({\bf v}-\xt)\chi(L)\e^{-\i tL}=0.
\]
\end{proposition}

\proof  

{\it Proof of (i)}:  
\medskip	

We assume first that $\supp \chi\subset ]m, +\infty[$.  The case $\supp \chi\subset ]-\infty, -m[$ can be treated similarly replacing ${\bf v}_{\s}$ by $-{\bf v}_{\s}$ in the arguments below.

We fix $0<\theta_{1}<\theta_{2}<\theta_{3}$, $\theta_{3}>1$ and choose as usual a function $R\in\cinf(\rr^{d})$ such that $R\equiv 0$ near $0$, $\nabla^{2}R(x)\geq 0$ and $R(x)= \12 x^{2}-c$ for  $|x|>\theta_{1}$. We fix also $J\in\coinf(\rr)$ with $J\equiv 1$ on $[0, \theta_{3}]$. 

Set
\[
M(t)= \12 \langle {\bf v}_{\s}-\xt, \nabla R(\xt)\rangle + \12 \langle \nabla R(\xt), {\bf v}_{\s}-\xt\rangle + R(\xt),
\]
which is  a $2\times 2$ matrix of operators on  $\ch$, following   the notation introduced at the beginning of Subsect. \ref{matmat}.
\[
\Phi(t)=J(\frac{|x|}{t})M(t)J(\frac{|x|}{t}),
\]
which is uniformly bounded. Let us pick $\chi_{1}\in \coinf (]m, +\infty[)$with  $\chi_{1}\chi=\chi$. Then 
\[
\begin{array}{rl}
 &\chi_{1}(L){\bf D}\Phi(t) \chi_{1}(L)\\[2mm]
 =& \chi_{1}(L){\bf D}_{0}\left(J(\frac{|x|}{t})M(t)J(\frac{|x|}{t})\right)\chi_{1}(L)+ \chi_{1}(L)[V, \i \Phi(t)]\chi_{1}(L)\\[2mm]
=& \chi_{1}(L){\bf D}_{0}\left(J(\frac{|x|}{t})M(t)J(\frac{|x|}{t})\right)\chi_{1}(L)+ O(t^{-1-\delta}),
\end{array}
\]
by Prop. \ref{3.3}. Next
\[
\begin{array}{rl}
&{\bf D}_{0}\left(J(\frac{|x|}{t})M(t)J(\frac{|x|}{t})\right)\\[2mm]
=&{\bf D}_{0}\left(J(\frac{|x|}{t})\right)M(t)J(\frac{|x|}{t})+J(\frac{|x|}{t}) M(t){\bf D}_{0}\left(J(\frac{|x|}{t})\right)\\[2mm]
&+J(\frac{|x|}{t}) \left({\bf D}_{0}M(t)\right)J(\frac{|x|}{t}).
\end{array}
\]
As in the proof of Prop. \ref{3.5} we have
\[
{\bf D}_{0}J(\frac{|x|}{t})= \frac{1}{t}f(\frac{|x|}{t}) T(t)f(\frac{|x|}{t})+ O(t^{-1-\delta}),
\]
where $f\in \coinf(]\theta_{0}, +\infty[)$ and $\|T(t)\|\in O(1)$. Commuting $f(\frac{|x|}{t})$ to the left or right we obtain:
\[
\begin{array}{rl}
&{\bf D}_{0}\left(J(\frac{|x|}{t})\right)M(t)J(\frac{|x|}{t})+J(\frac{|x|}{t}) M(t){\bf D}_{0}\left(J(\frac{|x|}{t})\right)\\[2mm]
=& \frac{1}{t}f(\frac{|x|}{t})\left(T(t)M(t)+ M(t)T(t)\right) f(\frac{|x|}{t})+ O(t^{-1-\delta})\\[2mm]
=&\frac{1}{t}f(\frac{|x|}{t})N(t) f(\frac{|x|}{t})+ R_{1}(t),
\end{array}
\]
where 
\beq\label{connard.e4}
N(t):=T(t)M(t)+ M(t)T(t) \in O(1), \ R_{1}(t)\in O(t^{-1- \delta}).
\eeq 

We now compute ${\bf D}_{0}M(t)$. We consider the (scalar) operator:
\[
m_{0}(t)= \12 \langle {\bf v}_{\s}-\xt, \nabla R(\xt)\rangle + \12 \langle \nabla R(\xt), {\bf v}_{\s}-\xt\rangle + R(\xt),
\]
so that $M(t)= m_{0}(t)\otimes\one_{2}$,  
and compute
\[
\begin{array}{rl}
\p_{t}m_{0}(t)=&-\frac{1}{2t}\left(\langle {\bf v}_{\s}- \xt, \nabla^{2}_{x}R(\xt) \xt \rangle +\langle\xt, \nabla^{2}R(\xt) ({\bf v}_{\s}-\xt )\rangle\right),\\[2mm]
[b, \i m_{0}(t)]
=&\frac{1}{2t}\left(\langle {\bf v}_{\s}- \xt, \nabla^{2}_{x}R(\xt) {\bf v}_{\s} \rangle +\langle {\bf v}_{\s}, \nabla^{2}R(\xt) ({\bf v}_{\s}-\xt) \rangle\right)\\[2mm]
& +\ O(t^{-1-\delta}),
\end{array}
\]
\beq\label{irlot}
\begin{array}{rl}
\p_{t}m_{0}(t)+ [b, \i m_{0}(t)]
= &\frac{1}{2t}\langle {\bf v}_{\s}-\xt, \nabla^{2}R(\xt) ({\bf v}_{\s}-\xt)\rangle +O(t^{-1-\delta})\\[2mm]
\geq & \frac{1}{2t}\langle {\bf v}_{\s}-\xt,\one_{[\theta_{1}, \theta_{3}]}(\frac{|x|}{t}) ({\bf v}_{\s}-\xt)\rangle + O(t^{-1-\delta}).
\end{array}
\eeq
using Prop. \ref{3.1} (4). Hence
\[
{\bf D}_{0}M(t)=\mat{\p_{t}m_{0}(t)+ [b, \i m_{0}(t)]}{0}{0}{\p_{t}m_{0}(t)- [b, \i m_{0}(t)]}\in O(t^{-1}).
\]
Applying  Lemma \ref{3.0} (2) we get that
\beq\label{irloto}
\begin{array}{rl}
\chi_{1}(L){\bf D}_{0}M(t)\chi_{1}(L)= &\chi_{1}(L_{0})){\bf D}_{0}M(t)\chi_{1}(L)+ O(t^{-1- \delta})\\[2mm]
=&\chi_{1}(L_{0})\left(\p_{t}m_{0}(t)+ [b, \i m_{0}(t)]\right)\chi_{1}(L)+ O(t^{-1- \delta})\\[2mm]
=&\chi_{1}(L)\left(\p_{t}m_{0}(t)+ [b, \i m_{0}(t)]\right)\chi_{1}(L)+ O(t^{-1- \delta}),
\end{array}
\eeq
where we used that $\supp \chi_{1}\subset ]m, +\infty[$ in the second line.

We can apply Corollary \ref{abs4} for
\[
\begin{array}{rl}
R_{0}(t)=& \chi_{1}(L)[V, \i \Phi(t)]\chi_{1}(L),\\[2mm]
D(t)=& J(\frac{|x|}{t})\left(\p_{t}m_{0}(t)+ [b, \i m_{0}(t)]\right)J(\frac{|x|}{t})+ \frac{1}{t}f(\frac{|x|}{t})N(t) f(\frac{|x|}{t})+ R_{1}(t),\\[2mm]
B^{*}(t)B(t)= & \frac{1}{2t}J(\frac{|x|}{t})\langle {\bf v}_{\s}-\xt,\one_{[\theta_{1}, \theta_{3}]}(\frac{|x|}{t}) ({\bf v}_{\s}-\xt)\rangle J(\frac{|x|}{t}),
\end{array}
\]
where $N(t)$, $ R_{1}(t)$ are defined in (\ref{connard.e4}), with $R_{1}(t)\in O(t^{-1- \delta})$. 

By Prop. \ref{3.3}  we know that $R_{0}(t)\in O(t^{-1- \delta})$ hence condition {\it i)} of  Corollary \ref{abs4} is satisfied. Applying Prop. \ref{3.5} we see that $\|K D(t)\|\in O(t^{-1- \delta})$, hence condition  {\it ii)} of  Corollary \ref{abs4} are satisfied. 
Condition {\it iii)}  follows from  the last inequality in (\ref{irlot}), condition {\it iv)} from Prop. \ref{3.4}.

\medskip

{\it Proof of (ii)}: 

\medskip	

 set $u_{t}= \e^{-\i tL}u$, $\Phi(t)= \langle( {\bf v}-\xt),f^{2}(\xt)({\bf v}- \xt)\rangle$. Then 
\beq\label{e3.8}
[\chi(L)u_{t}, \Phi(t)\chi(L)u_{t}]= (\chi(L)u_{t}| \Phi(t)\chi(L)u_{t})+ o(1),
\eeq
by Prop.  \ref{3.5}. We compute
\[
\begin{array}{rl}
\chi(L){\bf D}\Phi(t)\chi(L)&= \chi(L){\bf D}_{0}\Phi(t)\chi(L)+ O(t^{-1-\delta})\\[2mm]
\end{array}
\]
We have
\[
\Phi(t)= \mat{\Phi_{+}(t)}{0}{0}{\Phi_{-}(t)}, \ \Phi_{\pm}(t)= \langle \pm {\bf v}_{\s}- \xt,f^{2}(\xt)(\pm {\bf v}_{\s}- \xt)\rangle,
\]
hence
\[
{\bf D}_{0}\Phi(t)=\mat{\p_{t}\Phi_{+}(t)+ [b, \i \Phi_{+}(t)]}{0}{0}{\p_{t}\Phi_{-}(t)- [b, \i \Phi_{-}(t)]}.
\]
Using  Prop. \ref{3.1} (4), we get that:
\[
\begin{array}{rl}
&\p_{t}\Phi_{\pm}(t)\pm [b, \i \Phi_{\pm}(t)]\\[2mm]
=& -\frac{2}{t}\langle\pm {\bf v}_{\s}- \xt,f^{2}(\xt)(\pm {\bf v}_{\s}- \xt)\rangle + \frac{1}{t}\langle\pm {\bf v}_{\s}- \xt,
m_{\pm}(\xt, D_{x})(\pm {\bf v}_{\s}- \xt)\rangle,\\[2mm]
&+ O(t^{-1- \delta}),
\end{array}
\]
for 
\[
 m_{\pm}(\xt, D_{x})=\12\left(\langle\nabla f^{2}(\xt), (\pm {\bf v}_{\s}- \xt)\rangle + {\rm h.c.}\right).
\]
Setting
\[
T(\xt, D_{x})= \mat{m_{+}(\xt, D_{x})}{0}{0}{m_{-}(\xt, D_{x})}, 
\]
this yields
\[
\begin{array}{rl}
&{\bf D}_{0}\Phi(t)\\[2mm]
=& -\frac{2}{t}\langle{\bf v}- \xt,f^{2}(\xt)({\bf v}- \xt)\rangle + \frac{1}{t}\langle{\bf v}- \xt,
T(\xt, D_{x})({\bf v}- \xt)\rangle+ O(t^{-1- \delta}).
\end{array}
\]
It follows that
\[
\begin{array}{rl}
&\frac{\d }{\d t}[\chi(L)u_{t}, \Phi(t)\chi(L)u_{t}]\\[2mm]
=&(g(\xt)({\bf v}- \xt)\chi(L)u_{t}| T(t)(g(\xt)({\bf v}- \xt)\chi(L)u_{t})+ O(t^{-1-\delta})\|u\|^{2},
\end{array}
\]
 where $\|T(t)\|\in O(t^{-1})$, $g\in \coinf(\rr^{d}\backslash \{0\})$.

By Corollary \ref{abs4} (2),  {\it (i)} above  and (\ref{e3.8}) we know that 
\[
\lim_{t\to +\infty}[\chi_{1}(L)u_{t},\Phi(t)\chi_{1}(L)u_{t} ]= \lim_{t\to +\infty}\|f(\xt) ({\bf v}- \xt) \chi_{1}(L)u_{t}\|^{2}\hbox{ exists}.
\]
Using again {\it (i)} this limit is equal to $0$.
\qed
\subsection{Minimal velocity estimates}
\begin{proposition}\label{minimal}
 Let $\chi\in \coinf(\rr\backslash [-m, m])$. Then there exists $\theta_{0}>0$ such that
 \[
\int_{1}^{+\infty}\|\one_{[0, \theta_{0}]}(\xt)\chi(L)\e^{- \i tL}u\|^{2}\frac{\d t}{t}\leq C\|u\|^{2}.
\]
 \end{proposition}
\proof
Let $J\in \coinf(\rr^{d})$ with $J(x)\equiv 1$ for $|x|\leq \theta_{0}$, $J(x)\equiv 0$ for $|x|\geq 2\theta_{0}$, where $ \theta_{0}$ will be chosen small enough later. 
We set
\[
M(t)= J(\xt)+ \12\langle {\bf v}_{\s}- \xt, \nabla J(\xt)\rangle+ \12 \langle \nabla J(\xt),{\bf v}_{\s}- \xt\rangle. 
\]
For later use let us first note some properties of $M(t)$. 

We set $M^{c}(t)= \one - M(t)$ and note that $M^{c}(t)$ is given by the same formula 
with $J(x)$ replaced by $J^{c}(x)= \one - J(x)$,  where $0\not\in \supp J^{c}$. Therefore we obtain from Prop. \ref{3.3} that
\begin{equation}
\label{e3.20}
\chi_{1}(L)[V, \i M(t)]\chi(L)\in O(t^{-1- \delta}), \ \delta>0.
\end{equation}
 The same argument using Lemma \ref{3.0} (5) shows that
 \begin{equation}
\label{lavigue}
[\chi(L), M(t)]\in O(t^{-\mu}), \ \mu>0, \ \chi\in \coinf(\rr).
\end{equation}
Using formula (\ref{link}) and Prop. \ref{3.5},  we obtain also
\beq\label{e3.25}
\begin{array}{rl}
M^{\dag}(t)=&M(t)+(\one +K)^{-1}[M(t), K]\\[2mm]
=&M(t)- (\one +K)^{-1}[M^{c}(t), K]\\[2mm]
=&M(t)+ O(t^{- \mu}), \ \mu>0.
\end{array}
\eeq
 
 By a compactness argument it suffices to prove the proposition for $\chi$ supported in a small neighborhood of some 
$\lambda_{0}\in \rr\backslash [-m, m]$. Without loss of generality we can  assume that $\lambda_{0}>m$, the case $\lambda_{0}<-m$ being similar, replacing ${\bf v}_{\s}$ by $-{\bf v}_{\s}$ and $a$ by $-a$ . 

We choose the cutoff functions $\chi_{\delta}(\lambda)$ as in  Prop.  \ref{commut} and fix another cutoff function $\chi_{1}$ such that
$\supp \chi_{1}\subset ]m, +\infty [$ and $\chi_{1}\chi_{\delta}= \chi_{\delta}$ for $\delta\ll 1$.
We set:
\[
\Phi(t)= \chi_{1}(L)M(t)\chi_{\delta}(L)\frac{a}{t}\chi_{\delta}(L) M(t)\chi_{1}(L).
\]
Note that using Lemma \ref{3.0} (4) we know that 
\begin{equation}
\label{e3.21}
\frac{a}{t}\chi_{\delta}(L)M(t)\in O(1),
\end{equation}  and hence $\Phi(t)$ is uniformly bounded. We have
\[
\begin{array}{rl}
{\bf D}\Phi(t)
=  &\chi_{1}(L){\bf D}M(t)\chi_{\delta}(L)\frac{a}{t}\chi_{\delta}(L) M(t)\chi_{1}(L)\\[2mm]
&+ \chi_{1}(L)M(t)\chi_{\delta}(L)\frac{a}{t}\chi_{\delta}(L){\bf D}M(t)\chi_{1}(L)\\[2mm]
&+\chi_{1}(L)M(t)\chi_{\delta}(L){\bf D}\left(\frac{a}{t}\right)\chi_{\delta}(L) M(t)\chi_{1}(L)\\[2mm]
=:& R_{1}(t)+ R_{2}(t),
\end{array}
\]
where $R_{2}(t)$ is the term in the third line above.
\medskip

{\it Estimates on $R_{1}(t)$.}

\medskip	

  We claim that:
\beq\label{tristan}
[u, R_{1}(t)u]= \frac{1}{t}(T(t)\chi_{1}(L)u|B(t)T(t)\chi_{1}(L)u)+ O(t^{-1- \mu})\|u\|^{2},
\eeq
where:
\beq\label{e3.22b}
T(t):= G(\xt)({\bf v}_{\s}- \xt), \ \|B(t)\|\in O(1).
\eeq
Let us prove our claim.
We have
\[
{\bf D}M(t)= {\bf D}_{0}M(t)+ [V, \i M(t)],
\]
As we saw above we have:
\[
\chi_{1}(L)[V, \i M(t)]\chi_{\delta}(L)\in O(t^{-1- \mu}), \ \mu>0.
\]
Next using (\ref{irloto}) with $J$ instead of $R$ we obtain that:
\[
\chi_{1}(L){\bf D}_{0}M(t)\chi_{\delta}(L)=\chi_{1}(L)\mat{m(t)}{0}{0}{m(t)}\chi_{\delta}(L)+ O(t^{-1- \mu}), 
\]
for
\[
m(t)= \frac{1}{2t}\langle {\bf v}_{\s}- \xt, \nabla^{2}J(\xt)( {\bf v}_{\s}-\xt)\rangle= \frac{1}{2t}\langle {\bf v}_{\s}- \xt, G(\xt)\nabla^{2}J(\xt)G(\xt) ({\bf v}_{\s}-\xt)\rangle,
\]
for $G\in\coinf(\rr^{d}\backslash\{0\})$ with $G\equiv 1$ on $\supp \nabla J$. 
 We claim now that
\[
 R_{1}(t)
= \frac{1}{t}\chi_{1}(L)\langle ({\bf v}_{\s}- \xt)G(\xt), B(t)G(\xt)({\bf v}_{\s}-\xt)\rangle\chi_{1}(L) +O(t^{-1- \mu}),
\]
for $B(t)\in O(1)$. This follows from the identities above by commuting $G(\xt)({\bf v}_{\s}- \xt)$ to the left or to the right, using that
\beq\label{alphonse}
[G(\xt)({\bf v}_{\s}- \xt), \ i \chi_{\delta}(L)]\in O(t^{-\mu}), \ [G(\xt)({\bf v}_{\s}- \xt), a]\in O(1),
\eeq
the first estimate in (\ref{alphonse}) follows from  Lemma \ref{3.0} (5), the second from pseudo-differential calculus.   
Recall that  $T(t)= G(\xt)({\bf v}_{\s}- \xt)$. Then using Prop. \ref{3.5} and the fact that $0\not\in \supp G$ we obtain that
\[
\begin{array}{rl}
[u, R_{1}(t)u]=& \frac{1}{t}[\chi_{1}(L)u, T^{*}(t)B(t)T(t)\chi_{1}(L)u]+ O(t^{-1- \mu})\|u\|^{2}\\[2mm]
=& \frac{1}{t}(\chi_{1}(L)u|(\one +K) T^{*}(t)B(t)T(t)\chi_{1}(L)u)+ O(t^{-1- \mu})\|u\|^{2}\\[2mm]
=& \frac{1}{t}(T(t)\chi_{1}(L)u|B(t)T(t)\chi_{1}(L)u)+ O(t^{-1- \mu})\|u\|^{2},
\end{array}
\]
which implies  (\ref{tristan}).

\medskip	

{\it Estimates on $R_{2}(t)$.}

\medskip	

  We claim that:

\beq\label{moorburg}
\begin{array}{rl}
&[u, R_{2}(t)u]\\[2mm]
=&\frac{c_{0}}{t}(M(t)\chi_{\delta}(L)u|M(t) \chi_{\delta}(L)u)\\[2mm]
&+ \frac{1}{t}[g(L)M(t)\chi_{\delta}(L)u, g(L)M(t)\chi_{\delta}(L)u]\\[2mm]
&+\frac{1}{t}(M(t)\chi_{\delta}(L)u|B_{2}(t)M(t) \chi_{\delta}(L)u)+ O(t^{-1-\mu})\|u\|^{2},
\end{array}
\eeq
for  $\|B_{2}(t)\|\leq c_{0}/2$.

We have:
\[
\chi_{\delta}(L){\bf D}\left(\frac{a}{t}\right)\chi_{\delta}(L)= \frac{1}{t}\chi_{\delta}(L)[L, \i a]\chi_{\delta}(L)- \frac{1}{t}\chi_{\delta}(L)\frac{a}{t}\chi_{\delta}(L).
\]
Applying (\ref{e3.25}) we obtain:
\[
\begin{array}{rl}
[u, R_{2}(t)u]=&[\chi_{1}(L)u, M(t)\chi_{\delta}(L){\bf D}\left(\frac{a}{t}\right)\chi_{\delta}(L)M(t)\chi_{1}(L)u]\\[2mm]
=&[M(t)\chi_{1}(L)u, \chi_{\delta}(L){\bf D}\left(\frac{a}{t}\right)\chi_{\delta}(L)M(t)\chi_{1}(L)u]+ O(t^{-1- \mu})\|u\|^{2}\\[2mm]
=&\frac{1}{t}[M(t)\chi_{1}(L)u, \chi_{\delta}(L)[L, \i a]\chi_{\delta}(L)M(t)\chi_{1}(L)u]\\[2mm]
&-  \frac{1}{t}
[M(t)\chi_{1}(L)u, \chi_{\delta}(L)\frac{a}{t}\chi_{\delta}(L)M(t)\chi_{1}(L)u]+O(t^{-1- \mu})\|u\|^{2}\\[2mm]
=:& [u, R_{3}(t)u]+ [u,R_{4}(t)u]+ O(t^{-1- \mu})\|u\|^{2}.
\end{array}
\]
We consider  first $R_{3}(t)$.
Applying  Prop. \ref{commut} (see Remark \ref{ferdine}) we obtain that
\[
\begin{array}{rl}
&[u, R_{3}(t)u]\\[2mm]

=&\frac{1}{t}[M(t)\chi_{1}(L)u, \chi_{\delta}(L)[L, \i a]\chi_{\delta}(L)M(t)\chi_{1}(L)u]\\[2mm]
=&\frac{c_{0}}{t}(\chi_{\delta}(L)M(t)\chi_{1}(L)u| \chi_{\delta}(L)M(t)\chi_{1}(L)u)\\[2mm]
&+ \frac{1}{t}[\chi_{\delta}(L)g(L)M(t)\chi_{1}(L)u, \chi_{\delta}(L)g(L)M(t)\chi_{1}(L)u]\\[2mm]
&+\frac{1}{t}(\chi_{\delta}(L)M(t)\chi_{1}(L)u| R_{\delta} \chi_{\delta}(L)M(t)\chi_{1}(L)u).
\end{array}
\]
Since  $[\chi_{\delta}(L), M(t)]\in O(t^{-\mu})$ by (\ref{lavigue}), we finally get:
\begin{equation}
\label{harras}
\begin{array}{rl}
&[u, R_{3}(t)u]\\[2mm]
=&\frac{c_{0}}{t}(M(t)\chi_{\delta}(L)u|M(t) \chi_{\delta}(L)u)\\[2mm]
&+ \frac{1}{t}[g(L)M(t)\chi_{\delta}(L)u, g(L)M(t)\chi_{\delta}(L)u]\\[2mm]
&+\frac{1}{t}(M(t)\chi_{\delta}(L)u|R_{\delta}M(t) \chi_{\delta}(L)u)+ O(t^{-1-\mu})\|u\|^{2}.
\end{array}
\end{equation}
From Prop. \ref{commut}, we know that  $\|R_{\delta}\|\leq c_{0}/4$, if $\delta$ is small enough.  We fix such a $\delta$ and consider $R_{4}(t)$. We fix a cutoff function $G$ with 
$G(x)\equiv 1$ for $|x|\leq 2 \theta_{0}$, $G(x)\equiv 0$ for $|x|\geq 3\theta_{0}$.  Clearly
\begin{equation}
\label{lili}
M(t)= G(\xt)M(t)+ O(t^{-1}), \ [\chi_{\delta}(L), G(\xt)]\in O(t^{-\mu}), \ \mu>0,
\end{equation}
the first identity follows from pdo calculus, the second from Lemma \ref{3.0} (5),  using that $0\not \in \supp (\one - G)$. Using (\ref{lili})  and (\ref{lavigue}) we get that:
\beq\label{kracht}
\begin{array}{rl}
&[u, R_{4}(t)u]\\[2mm]
=&-  \frac{1}{t}[M(t)\chi_{1}(L)u, \chi_{\delta}(L)\frac{a}{t}\chi_{\delta}(L)G(\xt)M(t)\chi_{1}(L)u]+ O(t^{-2})\|u\|^{2}\\[2mm]
=&-  \frac{1}{t}[M(t)\chi_{1}(L)u, \chi_{\delta}(L)\frac{a}{t}G(\xt)\chi_{\delta}(L)M(t)\chi_{1}(L)u]+ O(t^{-1-\mu})\|u\|^{2}\\[2mm]
=&-  \frac{1}{t}[\chi_{\delta}(L)M(t)\chi_{1}(L)u, \frac{a}{t}G(\xt)\chi_{\delta}(L)M(t)\chi_{1}(L)u]+ O(t^{-1-\mu})\|u\|^{2}\\[2mm]
=&-\frac{1}{t}(\chi_{\delta}(L)M(t)\chi_{1}(L)u| (\one +K)\frac{a}{t}G(\xt)\chi_{\delta}(L)M(t)\chi_{1}(L)u)+ O(t^{-1- \mu})\|u\|^{2}\\[2mm]
=&\frac{1}{t}(\chi_{\delta}(L)M(t)\chi_{1}(L)u| R_{5}(t)\chi_{\delta}(L)M(t)\chi_{1}(L)u)+ O(t^{-1-\mu})\|u\|^{2}\\[2mm]
=&\frac{1}{t}(M(t)\chi_{\delta}(L)u| R_{5}(t)M(t)\chi_{\delta}(L)u)+ O(t^{-1-\mu})\|u\|^{2},
\end{array}
\eeq
for $R_{5}(t)= (\one +K)\frac{a}{t}G(\xt)$. Picking $\theta_{0}$ small enough we can be sure that
\[
\|R_{5}(t)\|\leq c_{0}/4.
\]
Adding (\ref{harras}) and (\ref{kracht}), we obtain (\ref{moorburg})
for $B_2(t)= R_{\delta}(t)+ R_5(t))$.

We can now complete the proof of the proposition. We apply Prop. \ref{abs2} for:
\[
\begin{array}{rl}
B(t)= &(\frac{c_{0}}{t})^{\12}M(t)\chi_{\delta}(L),\\[2mm]
C(t)= &B^{*}(t)B(t)+ \frac{1}{t}(\one +K) \chi^{2}_{\delta}(L)g^{2}(L),\\[2mm]
C_{1}(t)=& \frac{1}{t}T^{*}(t)B_{1}(t) T(t) + O(t^{-1- \mu}),\\[2mm]
C_{2}(t)=&\frac{1}{t}\chi_{\delta}(L)M(t)B_{2}(t)M(t)\chi_{\delta}(L),
\end{array}
\]
where $B_{1}(t)$ is defined in (\ref{tristan}), $B_{2}(t)$ in (\ref{moorburg}).
We note that $C(t)\geq B^{*}(t)B(t)$ since
\[
(u|(\one +K) g^{2}(L)u)= [g(L)u, g(L)u]\geq 0,
\]
using that $\supp g\subset ]m, +\infty[$.  We obtain that:
\[
\int_{1}^{+\infty}\|M(t)\chi_{\delta}(L)\e^{- \i tL}u\|^{2}\frac{\d t}{t}\leq C\|u\|^{2}.
\]
Using Prop. \ref{3.6} we deduce that
\[
\int_{1}^{+\infty}\|J(\xt)\chi_{\delta}(L)\e^{- \i tL}u\|^{2}\frac{\d t}{t}\leq C\|u\|^{2}.
\]
This completes the proof of the proposition.
\qed

\medskip

\begin{proposition}\label{minimalbis}
 Let $\chi\in \coinf(\rr\backslash [-m, m])$. Then there exists $\theta_{0}>0$ such that
 \[
\slim_{t\to +\infty}\one_{[0, \theta_{0}]}(\frac{|x|}{t})\chi(L)\e^{- \i tL}=0.
\]
\end{proposition}
\proof 
We can assume that $\supp \chi\subset ]m,+\infty[$.
Let $F\in \coinf(\rr^{d})$ with $F(x)\equiv 1$  for $|x|\leq \theta_{0}$,  $F(x)\equiv 0$ for $|x|\geq 2\theta_{0}$, where $ \theta_{0}$ will be chosen sufficiently small later. We fix also  $\chi_{1}\in \coinf(]m,+\infty[)$ with $\chi_{1}\chi=\chi$. Then by Lemma \ref{3.0} (5):
\[
F(\xt)\e^{- \i tL}\chi(L)u= \chi_{1}(L)F(\xt)\e^{- \i tL}\chi(L)u + O(t^{-\mu}).
\]
We have
\[
\begin{array}{rl}
& [ \chi_{1}(L)F(\xt)\e^{- \i tL}\chi(L)u,  \chi_{1}(L)F(\xt)\e^{- \i tL}\chi(L)u]\\[2mm]
=& [\e^{- \i tL}\chi(L)u,  F(\xt)^{\dag}\chi^{2}_{1}(L)F(\xt)\e^{- \i tL}\chi(L)u]\\[2mm]
=& [\e^{- \i tL}\chi(L)u,  F(\xt)\chi^{2}_{1}(L)F(\xt)\e^{- \i tL}\chi(L)u]+ O(t^{-\mu})\\[2mm]
=& [\e^{- \i tL}\chi(L)u,  \chi^{2}_{1}(L)F^{2}(\xt)\e^{- \i tL}\chi(L)u]+ O(t^{-\mu})\\[2mm]
=& [\e^{- \i tL}\chi(L)u, F^{2}(\xt)\e^{- \i tL}\chi(L)u]+ O(t^{-\mu})\\[2mm]
=& [\e^{- \i tL}\chi(L)u,  M(t)\e^{- \i tL}\chi(L)u]+ O(t^{-\mu}),
\end{array}
\]
for
\[
M(t)= F^{2}(\xt)+ \12 \langle {\bf v}_{\s}- \xt, \nabla F^{2}(\xt)\rangle + \12\langle \nabla F^{2}(\xt), {\bf v}_{\s}- \xt\rangle.
\]
In the third line we use  formula (\ref{link}), Prop. \ref{3.5} and the fact that $F(x)\equiv 1$ near $0$. In the fourth line we use again Lemma \ref{3.0} (5). In the sixth line we use Prop. \ref{3.6} ii). We have seen in the proof of Prop. \ref{minimal} that 
\[
\chi(L){\bf D}M(t)\chi(L)=\frac{1}{t}\chi(L)({\bf v}_{\s}- \xt)G(\xt)B(t)G(\xt)({\bf v}_{\s}- \xt)\chi(L)+ O(t^{-1- \mu}), 
\]
where $G\in \coinf(\rr^{d}\backslash \{0\})$, $B(t)\in O(1)$.  Moreover by Prop. \ref{3.5} 
\[
K({\bf v}_{\s}- \xt)G(\xt)B(t)G(\xt)({\bf v}_{\s}- \xt)\in O(t^{-\mu}).
\] Therefore by Corollary \ref{abs4} (2) the limit
\beq\label{percy}
\lim_{t\to +\infty} [ \chi_{1}(L)F(\xt)\e^{- \i tL}\chi(L)u,  \chi_{1}(L)F(\xt)\e^{- \i tL}\chi(L)u]\hbox{ exists}.
\eeq
We fix a bounded interval $I$ with $I^{\rm cl}\cap c_{p}(L)=\emptyset$ such that $\supp\chi, \supp \chi_{1}\subset I$. The subspace $\cK_{1}= \one_{I}(L)\cK$ is positive hence by Prop. \ref{pr:kp}  we obtain that
\beq\label{krefeld}
C\| \chi_{1}(L)v\|^{2}\leq  [ \chi_{1}(L)v,  \chi_{1}(L)v]\leq C^{-1}\|\chi_{1}(L)v\|^{2}
\eeq
If the limit in (\ref{percy}) is not equal to $0$ then applying (\ref{krefeld}) to $v= F(\xt)\e^{- \i tL}\chi(L)u$ we obtain that
\[
\liminf_{t\to +\infty}\| \chi_{1}(L)F(\xt)\e^{- \i tL}\chi(L)u\|^{2}>0.
\]
But  for $\theta_{0}$ small enough, this contradicts the convergence of the integral in Prop. \ref{minimal}. Therefore the limit in (\ref{percy}) is $0$ which implies the proposition using once more (\ref{krefeld}). \qed
\section{Existence and completeness of short-range wave operators}\label{srwave}

 \subsection{Short-range wave operators}
\def \Ll{L_{l}}
We set $L_{l}= L_{0}+ V_{l}$,  with domain $H^{1}(\rr^{d})\oplus H^{1}(\rr^{d})$ which is selfadjoint (in the usual Hilbert sense) on $(\cK, (\cdot| \cdot))$. Note that by Prop. \ref{noemb} we know that $\sigma_{\rm pp}(L_{l})\subset [-m, m]$. Note also that  $L_{l}$ satisfies all the estimates in Sects. \ref{sec6}, \ref{sec3}, since it equals $L$ if $v_{s}=0$. However $L_{l}$ is not selfadjoint on $(\cK, [\cdot, \cdot])$.
We denote by $\cK_{\rm c}(L_{l})$ the continuous spectral subspace of $L_{l}$, defined in the usual sense and set
\[
\cK_{\rm scatt}(L):= U \cE_{\rm scatt}(B),
\]
where the map $U$ is defined in (\ref{def-de-u}).

For later use we state a consequence of Subsect. \ref{secIk.5} and of the fact that $\pm m\not \in c_{p}(B)$, which will be important in
some strong convergence arguments later on. We use the notation explained in Subsect. \ref{titi} for $ F(a\leq \lambda\leq b)$.
\begin{lemma}\label{asf}
 There exists  $\epsilon>0$ such that
\[
\sup_{t\in \rr}\|\e^{-\i tL}F(m-\epsilon\leq \pm L \leq +\infty)\|<+\infty.
\]
\end{lemma}
\proof For the $+$ case we choose   $\epsilon>0$ such that  $[m-\epsilon, +\infty[\cap c_{p}(L)= \emptyset$ and apply   Prop. \ref{lati} (4).  The proof for the $-$ case is identical. \qed

\begin{theoreme}[Existence of short-range wave operators]\label{mainli}
 Assume hypotheses (A), (B). Then: 
 \ben
 \item for all $u\in \cK_{\rm c}(L_{l})$ there exist  unique $u^{\pm}\in \cK_{\rm scatt}(L)$ such that
 \[
 \e^{- \i t L}u^{\pm}- \e^{-\i tL_{l}}u\to 0, \ t\to \pm \infty.
\]
\item   Let us define  the {\em short-range wave operators} $W^{\pm}(L, L_{l})$ by:
\[
W^{\pm}(L, L_{l}): 
\begin{array}{rl}
\cK_{\rm c}(L_{l})&\to\cK_{\rm scatt}(L),\\
 u&\mapsto u^{\pm}.
\end{array}
\]
Then:
\[
\begin{array}{rl}
(i)&W^{\pm}(L, L_{l})\in B(\cK_{\rm c}(L_{l}), \cK_{\rm scatt}(L)), \\[2mm]
(ii)&W^{\pm}(L, L_{l})\e^{-\i tL_{l}}= \e^{-\i t L}W^{\pm}(L, L_{l}), \ t\in \rr,\\[2mm]
 (iii)&W^{\pm}(L, L_{l})\hbox{ are isometric  from }(\cK_{\rm c}(L_{l}), (\cdot | \cdot))\hbox{ to }(\cK_{\rm scatt}(L), [\cdot,\cdot]).
\end{array}
\]
\een
\end{theoreme}
\begin{theoreme}[Completeness of short-range wave operators]\label{mainil}
Assume hypotheses (A), (B). Then:
\ben
\item ${\rm Ran}W^{\pm}(L, \Ll)= \cK_{\rm scatt}(L)$, i.e. $W^{\pm}(L, \Ll)$ are unitary from $(\cK_{\rm c}(\Ll), (\cdot | \cdot))$ to $(\cK_{\rm scatt}(L), [\cdot, \cdot])$.
\item set $W^{\pm}(\Ll, L):=W^{\pm}(L, \Ll)^{-1}$. Then
\[
W^{\pm}(\Ll, L)u= \slim_{t\to \pm \infty}\e^{\i t\Ll}\e^{-\i tL}u, \ u \in \cK_{\rm scatt}(L).
\]
\een 
\end{theoreme}

{\bf Proof of Thm. \ref{mainli}.}

It suffices to prove the existence of $W^{+}(L, L_{l})\one_{]m,+\infty[}(L_{l})$ (the existence of $W^{+}(L, L_{l})\one_{]-\infty, m[}(L_{l})$ being similar). 

We fix a cutoff function $F\in \cinf(\rr)$ as in Lemma \ref{asf} (1), so that  $\e^{\i tL}F(L)$ is uniformly bounded for $t\in \rr$.

We first claim that the limit
\beq
\slim_{t\to +\infty}\e^{\i tL} F(L)\e^{-\i t \Ll}\one_{]m,
+\infty[}(\Ll)\hbox{ exists}.
\label{e5.a1}
\eeq
Let us prove (\ref{e5.a1}). By the  standard
density argument, it suffices to prove
the existence of 
\beq\label{e5.00}
\slim_{t\to +\infty}\e^{\i tL}F(L)\e^{-\i t\Ll}\chi^{2}(\Ll), 
\eeq
for $\chi\in \coinf(]m, +\infty[)$. Let us fix such a cutoff function
$\chi$. Applying  
Props. \ref{3.4}, \ref{3.6} and \ref{minimalbis} to 
$L_{l}$ we can find $J\in \coinf(\rr^{d}\backslash\{0\})$ such
that
\[
\slim_{t\to +\infty} (\one -M(t))\chi^{2}(\Ll)\e^{-\i tL_{l}}=0.
\]
for
\[
M(t)= J(\xt)+\12\langle \nabla J(\xt), {\bf  v}- \xt\rangle +\12\langle {\bf  v}- \xt,\nabla J(\xt)\rangle.
\]
Therefore to prove (\ref{e5.00}) it suffices to prove the existence of
\[
\slim_{t\to +\infty}\e^{\i tL}F(L)M(t)\chi^{2}(\Ll)\e^{-\i t\Ll}.
\]
Moreover using   Lemma 
\ref{3.0} (2)  and  Lemma \ref{3.0} (5) (which apply to $L_{l}$),
we have
\[
\begin{array}{rl}
&\e^{\i tL}F(L)M(t)\chi^{2}(\Ll)\e^{-\i t\Ll}u\\[2mm]
=&\e^{\i tL}F(L)\chi(\Ll)M(t)\chi(\Ll)\e^{-\i t\Ll}u+ o(1)\\[2mm]
=&\e^{\i tL}F(L)\chi(L)M(t)\chi(\Ll)\e^{-\i t\Ll}u+ o(1)\\[2mm]
=&\e^{\i tL}\chi(L)M(t)\chi(\Ll)\e^{-\i t\Ll}u+ o(1).\\[2mm]
\end{array}
\]
Therefore it suffices to prove the existence of
\[
\slim_{t\to +\infty}\e^{\i tL}\chi(L)M(t)\chi(\Ll)\e^{-\i t\Ll}.
\]
To prove it we apply Prop. \ref{abs3} (2). The asymmetric Heisenberg derivative 
equals:
\[
= \chi(L) {\bf D}_{0}M(t)\chi(L_{l})+ \chi(L)[V_{l}, \i M(t)]\chi(L_{l})+\i  \chi(L)V_{s}M(t)\chi(L_{l}).
\]
As in the proof of   Prop. \ref{3.6} (see eg (\ref{irloto})) we see that
\[
 \chi(L){\bf D}_{0}M(t)\chi(L_{l})=  \frac{1}{t}\chi(L)T^{*}(t)B(t)T(t)\chi(L_{l})+ O(t^{-1- \mu}),
\]
for 
\[
T(t):= G(\xt)({\bf  v}- \xt), \ \|B_{1}(t)\|\in O(1).
\]
By Prop. \ref{3.3} we know that  $[V_{l}, \i M(t)]\in O(t^{-1- \mu})$ and applying Prop. \ref{3.1bis} (2) and (3) we obtain that
$\chi(L)V_{s}M(t)\chi(L_{l})\in O(t^{-1-{ \mu}})$.

To obtain the estimates in Prop. \ref{abs3} (2), we use the propagation estimates of Prop. \ref{3.6} (which are also valid for $L_{l}$) and the fact 
that $T^{\dag}(t)= T(t)+ O(t^{-\mu})$, which follows from Prop. \ref{3.5} and (\ref{link}).

This completes the proof of (\ref{e5.a1}). Let hence
\beq\label{ista}
u^{+}:=\lim_{t\to +\infty}\e^{\i tL} F(L)\e^{-\i t \Ll}u, u\in \one_{]m, +\infty[}(L_{l})\cK.
\eeq
Note that $u^{+}\in \one_{[m-2\epsilon, +\infty[}(L)\cK$. Since  $[m-2\epsilon, +\infty[\cap c_{p}(L)=\emptyset$ it follows from Prop. \ref{lati} (4) that $\e^{-\i tL}$ is uniformly bounded on $\one_{[m-2\epsilon, +\infty[}(L)\cK$ and hence:
\[
\e^{-\i tL}u^{+}- F(L)\e^{-\i t \Ll}u\to 0, \hbox{ when }t\to +\infty.
\]
The same arguments used above  show that 
\[
\slim_{t\to \infty}(\one -F(L))\e^{-\i t\Ll}\one_{]m, +\infty[}(L_{l})= 0.
\]
Therefore
\[
\e^{-\i tL}u^{+}- \e^{-\i t \Ll}u\to 0, \hbox{ when }t\to +\infty,
\]
which completes the proof of (1).  Property (2) {\it (i)} follows from (\ref{ista}), property (2) {\it (ii)} is immediate. To prove (2) {\it (iii)} we write:
\[
\begin{array}{rl}
[u^{+}, u^{+}]=& [\e^{- \i tL}u^{+}, \e^{- \i t L}u^{+}]\\[2mm]
=& [ \e^{-\i t \Ll}u,  \e^{-\i t \Ll}u]+ o(1)\\[2mm]
=&( \e^{-\i t \Ll}u|(\one +K) \e^{-\i t \Ll}u)+ o(1)\\[2mm]
=& ( \e^{-\i t \Ll}u| \e^{-\i t \Ll}u)+ o(1)=(u| u),
\end{array}
\]
where in the last line we use that $K$ is compact and $ \e^{-\i t \Ll}u$ tends weakly to $0$ when $t\to +\infty$. \qed
\medskip

{\bf Proof of Thm. \ref{mainil}.}
It suffices to prove the existence of $W^{+}(L_{l}, L)\one_{]m, +\infty[}(L)$, ie that 
for any $u\in \one_{]m, +\infty[}(L)\cK$  there exists $u_{+}\in \cK$
 such that
 \[
\e^{- \i t L_{l}}u_{+}- \e^{-\i tL}u\to 0\hbox{ in norm when }t\to +\infty.
\]
Since $\e^{- \i t L_{l}}$ is unitary on $(\cK, (\cdot| \cdot))$ it suffices to prove the existence of 
\[
\slim_{t\to +\infty}\e^{\i tL_{1}}\e^{- \i tL}\one_{]m, +\infty[}(L),
\]
or by the usual density argument the existence of 
\[
\slim_{t\to +\infty}\e^{\i tL_{1}}\e^{- \i tL}\chi^{2}(L), \ \chi\in \coinf(]m, +\infty[).
\]
By the same argument as in the proof of Thm \ref{mainli} it suffices to prove the 
 existence
\[
\slim_{t\to +\infty}\e^{\i tL_{l}}\chi(L_{l})M(t)\chi(L)\e^{- \i tL}.
\]
We apply now Prop. \ref{abs3} (1), using the same propagation estimates as before. The details are left to the reader. \qed

\section{Long-range wave operators}\label{sec11}\init
 We have seen in Sect. \ref{srwave} that $W^{\pm}(L, \Ll)$ intertwine the dynamics 
 $\e^{- \i t \Ll}$ on $\cK_{\rm c}(\Ll)$ and $\e^{-\i tL}$  on $\cK_{\rm scatt}(L)$ . 
 
In this section we further simplify the dynamics $\e^{- \i t \Ll}$ using standard arguments of (Hilbert space) long-range scattering. Note that we are now in the familiar Hilbert space setting, and we are facing long-range scattering theory for matrix valued pseudo-differential operators.  
\subsection{Asymptotic diagonalization}\label{sec11.1}

Recall that 
$\Ll=\mat{b-v_{\l}}{v_{\l}}{v_{\l}}{-b -v_{\l}}$. We set
\[
L_{\rm diag}:=\mat{\epsilon-v_{\l}}{0}{0}{-\epsilon- v_{\l}} .
\]
We recall hypothesis (C):
 \[
{\rm (C)}\ \pm v_{\l}(x)\geq  0\hbox{ for }|x|\gg 1.
\]
which will be important in the sequel.

For simplicity we will denote  in the sequel  the symbol classes $S^{m,
p}(\rr^{2d})$ simply by $S^{m,p}$. We will also use the same
notation for matrix valued symbols and operators.

As in Subsect. \ref{absence} we have $\epsilon^{2}=\Op (\tilde{\epsilon}^{2})$ for:
\[
\tilde{\epsilon}^{2}= \sum_{jk}\xi_{j}c^{jk}(x)\xi_{k}+ 2\sum_{j} d_{j}(x)\xi_{j}+ r(x)+ m^{2},
\]
where 
\[
[c^{jk}](x)-\one,  d_{j}(x), r(x)\in S^{0, -\mu_{0}}(\rr^{2d}).
\]
We set $\mu=\inf(\mu_{0}, \mu_{\l})$.

Recall that we set $b=(\epsilon^{2}- v_{l}^{2})^{\12}$. It follows from Lemma \ref{urla} that:
\beq\label{e11.0}
\begin{array}{rl}
\epsilon= &\Op(\tilde{\epsilon})+ \Op S^{0, -1-\mu}, \\[2mm]
b=& \Op ((\tilde{\epsilon}^{2}-
v_{l}^{2})^{\12})+ \Op S^{0, -1-\mu},\\[2mm]
b^{-1}= &\Op ((\tilde{\epsilon}^{2}-
v_{l}^{2})^{-\12})+ \Op S^{-2, -1-\mu}.
\end{array}
\eeq

\begin{lemma}
Assume (A) and (C). Then there exists a matrix
valued symbol $\tilde{J}_{1}(x, \xi)$ with 
\[
\tilde{J}_{1}(x,\xi)= \pm\mat{1}{0}{0}{-1}+ S^{-1, -\mu},\hbox{ if }\pm v(x)>0\hbox{ for }|x|\gg 1
\]
such that if $J_{1}=\Op \tilde{J}_{1}$ one has:
\[
\begin{array}{l}
J_{1}= J_{1}^{*}+ \Op S^{0, -\mu}, J_{1}^{2}= \one + \Op S^{0,
-\mu},\\[2mm]
LJ_{1}- J_{1}L_{\rm diag}\in \Op S^{0, -1- \mu}.
\end{array}
\]
\end{lemma}
\proof Let $\tilde{\epsilon}, v\in \rr$ with $\tilde{\epsilon}\geq
m/2$, $|v|\leq m/2$ and $\tilde{b}= (\tilde{\epsilon}^{2}-
v^{2})^{\12}$. Set:
\[
\tilde{L}= \mat{\tilde{b}-v}{ v}{v}{-\tilde{b}-v}, \
\tilde{L}_{0}=\mat{\tilde{\epsilon}-v}{0}{0}{-\tilde{\epsilon}-v},
\]
\[
\tilde{J}_{1}=\frac{1}{(2\tilde{\epsilon})^{\12}}
\mat{(\tilde{\epsilon}-\tilde{b})^{-\12}v}{(\tilde{\epsilon}-\tilde{b})^{\12}}{(\tilde{\epsilon}-\tilde{b})^{\12}}{-(\tilde{\epsilon}-\tilde{b})^{-\12}v}.
\]
Then it is easy to see that
\beq
\tilde{J}_{1}=\tilde{J}_{1}^{-1}= \tilde{J}_{1}^{*},\
\tilde{J}_{1}\tilde{L}= \tilde{L}_{0}\tilde{J}_{1}.
\label{e11.1}
\eeq
We write
\[
\tilde{\epsilon}- \tilde{b}= \frac{v^{2}}{\tilde{\epsilon}+
\tilde{b}}, \ \tilde{\epsilon}+ \tilde{b}= \tilde{\epsilon}+
\tilde{\epsilon}(1-\frac{v^{2}}{\tilde{\epsilon}^{2}})^{\12},
\]
and substitute
\[
\tilde{\epsilon}= \tilde{\epsilon}(x, \xi), \ v= v_{l}(x), \tilde{b}(x,
\xi)= (\tilde{\epsilon}^{2}(x,\xi)- v_{l}^{2}(x))^{\12},
\]
which satisfy the above conditions for $|x|\gg 1$. We obtain that
\[
\tilde{\epsilon}+ \tilde{b}= 2\tilde{\epsilon}(1+ S^{-2, -2\mu}),
\]
hence
\beq\label{e11.1b}
\frac{(\tilde{\epsilon}-\tilde{b})^{-\12}v_{l}}{(2\tilde{\epsilon})^{\12}}=\frac{v_{l}}{|v_{l}|}\frac{(\tilde{\epsilon}+\tilde{b})^{\12}}{(2\tilde{\epsilon})^{\12}}=
\frac{v_{l}}{|v_{l}|}(1+ S^{-2, -2\mu}),
\eeq
\beq\label{e11.1c}
\frac{(\tilde{\epsilon}-\tilde{b})^{\12}}{(2\tilde{\epsilon})^{\12}}=\frac{|v_{l}|}{(\tilde{\epsilon}+
\tilde{b})^{\12}(2\tilde{\epsilon})^{\12}}=
\frac{|v_{l}|}{2\tilde{\epsilon}}(1+ S^{-2, - 2\mu}),
\eeq
for $|x|\gg 1$.
By (C)  $\frac{v_{l}}{|v_{l}|}= \pm 1 $  and $|v_{l}|\in S^{0,
-\mu}$ in $|x|\gg 1$. It follows from (\ref{e11.1b}),
(\ref{e11.1c}) that
\beq
\tilde{J}_{1}(x, \xi)= \pm\mat{1}{0}{0}{-1}+ S^{-1, -\mu}. 
\label{e11.2}
\eeq

We set now $J_{1}=\Op (\tilde{J}_{1})$ and get from (\ref{e11.0})
\[
L=\Op(\tilde{L})+ \Op S^{0, -1-\mu}, L_{\rm diag}= \Op(\tilde{L}_{0})+ \Op
S^{0, -1- \mu}.
\]
The lemma follows then  from (\ref{e11.1}) and pseudo-differential
calculus. \qed
\begin{proposition}\label{asti}
The limits
\[
W^{\pm}(L_{\l}, L_{\rm diag}):= \slim_{t\to \pm\infty}\e^{\i t
L_{\l}}J_{1}\e^{-\i t L_{\rm diag}}\one_{\rm c}(L_{\rm diag}),
\]
\[
W^{\pm}(L_{\l}, L_{\rm diag})^{*}:= \slim_{t\to \pm\infty}\e^{\i t
L_{\rm diag}}J_{1}^{*}\e^{-\i t L_{\l}}\one_{\rm c}(L_{\l})
\]
exist. One has
\[
\begin{array}{l}
W^{\pm}(L_{\l}, L_{\rm diag})^{*}W^{\pm}(L_{\l}, L_{\rm
diag})=\one_{\rm c}(L_{\rm diag}), \\[2mm]
W^{\pm}(L_{\l}, L_{\rm diag})W^{\pm}(L_{\l}, L_{\rm
diag})^{*}=\one_{\rm c}(L_{\l}),\\[2mm]
\e^{-\i tL}W^{\pm}(L_{\l}, L_{\rm diag})= W^{\pm}(L_{\l}, L_{\rm
diag})\e^{-\i tL_{\rm diag}}, \ t\in \rr.
\end{array}
\]
\end{proposition}
\proof  The proof of the proposition relies on standard arguments, therefore we will only sketch it. We first note that $L_{l}$ and $L_{\rm diag}$ satisfy the propagation estimates in Sect. \ref{sec3} (with easier proofs, since one can use Hilbert space arguments). 
For example to prove the existence of $W^{+}(\Ll, L_{\rm diag})$, it suffices to consider 
\[
\slim_{t\to +\infty}\e^{\i t
L_{\l}}\chi(\Ll)J_{1}M(t)\chi(L_{\rm diag})\e^{-\i t L_{\rm diag}},
\]
for $M(t)$ as in the proof of Thm. \ref{mainli}. In the Heisenberg derivatives, one obtain extra terms coming from $\Ll J_{1}- J_{1}L_{\rm diag}$. These terms yield integrable in time contributions, since $\Ll J_{1}- J_{1}L_{\rm diag}\in \Op S^{0, -1- \mu}(\rr^{2d})$,  hence is of short-range type.  The details are left to the reader.
\qed
\subsection{Isozaki-Kitada modifiers}\label{sec11.2}
By Prop. \ref{asti} we are reduced to the dynamics $\e^{-\i tL_{\rm diag}}$,  which we want to compare with $\e^{-\i tL_{\infty}}$ where
\beq\label{shitique}
L_{\infty}:=\mat{\epsilon_{\infty}}{0}{0}{\epsilon_{\infty}}, \ \epsilon_{\infty}= (D^{2}+ m^{2})^{\12}.
\eeq
Since both operators are diagonal, we 
 have to consider the scattering theory for long-range, scalar pseudo-differential operators. 
 As is well-known, it is necessary to introduce  {\em modified free dynamics} to define the wave operators. We choose to use  the time-independent modifiers introduced by Isozaki and Kitada \cite{IK} in the context of Schr\"{o}dinger operators. 

We start by stating  an easy extension of results of Robert \cite{Ro} on
solutions of eikonal equations. 
As in \cite{Ro} we define for $\alpha>0$,
$R\geq 1$ and $0<\sigma<1$ the incoming/outgoing regions:
\[
\Gamma(R, \alpha, \sigma)=\{(x, \xi)\ : \ |\xi|\geq \alpha, \
|x|\geq R, \ |x\cdot \xi|\geq \sigma|x||\xi|\}
\]

\begin{lemma}\label{kitada}
There exists functions $\varphi^{\pm}(x, \xi)$ such that:
\[
i) \ \varphi_{\pm}(x, \xi)= x\cdot \xi+ S^{-1, 1-\mu},
\] 
{\it ii)} for each $\alpha, \sigma$ there exists $R=R(\alpha, \sigma)$
such that
\[
\pm \tilde{\epsilon}(x, \p_{x}\varphi_{\pm}(x, \xi)) -v_{l}(x)= \pm(\xi^{2}+m^{2})^{\12} \hbox{ in }\Gamma(R, \alpha, \sigma).
\]
\end{lemma}
\proof We set
\[
p_{\pm}(x, \xi)=(\tilde{\epsilon}(x, \xi)\mp v_{l}(x))^{2}=
\tilde{\epsilon}^{2}(x, \xi) \pm 2v_{l}(x)\tilde{\epsilon}(x, \xi)+
v_{l}^{2}(x)= \tilde{\epsilon}^{2}+ S^{1, -\mu}.
\]
An easy modification of the arguments in \cite{Ro} shows that there
exist  functions $\varphi_{\pm}(x, \xi)$ satisfying {\it i)} solving
in $\Gamma(R,  \alpha, \sigma)$ the eikonal equations:
\[
\xi^{2}+ m^{2}= p_{\pm}(x, \p_{x}\varphi_{\pm}(x,, \xi))\hbox{ in }\Gamma(R,  \alpha,\sigma).
\]
Then {\it ii)} follows by taking the square roots. \qed

We {}define now the Isozaki-Kitada modifier:
\[
J_{2}:=\mat{j(\varphi_{+}, 1)}{0}{0}{j(\varphi_{-},1)},
\]
where the Fourier integral operators $j(\varphi_{\pm}, 1)$ are defined in Def.  \ref{urlc}.
\begin{proposition}\label{osti}
The limits
\[
W^{+}(L_{\rm diag}, L_{\infty}):= \slim_{t\to +\infty}\e^{\i t
L_{\rm diag}}J_{2}\e^{-\i t L_{\infty}},
\]
\[
W^{+}(L_{\rm diag}, L_{\infty})^{*}:= \slim_{t\to +\infty}\e^{\i t
L_{\infty}}J_{2}^{*}\e^{-\i t L_{\rm diag}}\one_{\rm c}(L_{\rm diag})
\]
exist. One has
\[
\begin{array}{l}
W^{+}(L_{\rm diag}, L_{\infty})^{*}W^{+}(L_{\rm diag},
L_{\infty})=\one, \\[2mm]
W^{+}(L_{\rm diag}, L_{\infty})W^{+}(L_{\rm diag},
L_{\infty})^{*}=\one_{\rm c}(L_{\rm diag}),\\[2mm]
\e^{-\i tL_{\rm diag}}W^{+}(L_{\rm diag},
L_{\infty})=W^{+}(L_{\rm diag}, L_{\infty})\e^{-\i tL_{\infty}}, \ t\in \rr.
\end{array}
\]
\end{proposition}
\proof
The proof is rather standard so we will again sketch it.  Since $L_{\rm diag}$, $J_{2}$ and $L_{\infty}$ are diagonal, it suffices to prove  the 
same result for the scalar operators  $\epsilon\mp v_{l}$, $j(\varphi_{\pm},1)$ and $\epsilon_{\infty}$.  Let us consider the case of $ \epsilon-v_{l}$.
We have to prove the existence of the limits:
\begin{equation}
\label{limat}
\slim_{t\to \pm \infty}\e^{\i t (\epsilon- v_{l})}j(\varphi_{+}, 1)\e^{- \i t \epsilon_{\infty}},
 \ \slim_{t\to \pm\infty}\e^{ \i t \epsilon_{\infty}}j(\varphi_{+}, 1)^{*}\e^{- \i t (\epsilon- v_{l})}\one_{\rm c}( \epsilon- v_{l}).
\end{equation}
Since $\tilde{\epsilon}(x, \xi)= \epsilon_{\infty}(\xi)+ S^{1, - \mu}(\rr^{2d})$ we see using (\ref{e11.0}) that  $\epsilon-v_{l}= \epsilon_{\infty}+ \Op r(x, D_{x})$ for 
$r\in S^{1, - \mu}(\rr^{2d})$. 
We can hence apply the results in \cite{Mu}, which we briefly recall. One first proves that there exists a function $S_{+}(t, \xi)$  solving the Hamilton-Jacobi equation:
\[
\p_{t}S_{+}(t, \xi)= \tilde{\epsilon}( \p_{\xi}S_{+}(t, \xi), \xi)- v_{l}( \p_{\xi}S_{+}(t, \xi)), \hbox{ in }|\xi|\geq \epsilon, |t|\geq T_{\epsilon}, \ \forall \ \epsilon>0
\]
and satisfying the estimates
\[
\p_{\xi}^{\alpha}\big( S_{\pm}(t, \xi)- t\epsilon_{\infty}(\xi)\big)\in O(t^{ 1- \mu}), \ \ \alpha\in \nn.
\]
Then it is shown in \cite{Mu} that the limits:
\[
\slim_{t\to \pm \infty}\e^{\i t (\epsilon- v_{l})}\e^{- \i S_{+}(t, D_{x})}, \ \slim_{t\to \pm\infty}\e^{\i S_{+}(t,D_{x})}\e^{- \i t (\epsilon- v_{l})}\one_{\rm c}( \epsilon- v_{l})
\]
exist and are inverse of each other.  To obtain (\ref{limat}) is suffices to prove the existence of the limits:
\[
\slim_{t\to \pm \infty}\e^{\i S_{+}(t, D_{x})} j(\varphi_{+}, 1)\e^{-\i t\epsilon_{\infty}}, \ \slim_{t\to \pm\infty}\e^{\i t \epsilon_{\infty}}j(\varphi_{+}, 1)^{*}\e^{- \i S_{+}( t, D_{x})}.
\]
This can easily be proved by the   Cook method, using stationary phase arguments and the eikonal and Hamilton-Jacobi equation satisfied by $\varphi_{+}$ and $S_{+}$. The details are left to the reader. \qed
\section{Proofs of Thms. \ref{youpla} and \ref{youpli}.}\label{youlou}
\subsection{Proof of Thm. \ref{youpla}}\label{you1}

If $v_{l}=0$ then $b= \epsilon$. By standard arguments we obtain the existence of the scalar wave operators:
\[
\begin{array}{rl}
W^{\pm}( \epsilon^{2}, \epsilon^{2}_{\infty})&=\slim_{t\to +\pm \infty}\e^{\i t\epsilon^{2}}\e^{-\i t \epsilon^{2}_{\infty}}, \\[2mm]
W^{\pm}( \epsilon^{2}_{\infty}, \epsilon^{2})&=\slim_{t\to +\pm \infty}\e^{\i t\epsilon_{\infty}^{2}}\e^{-\i t \epsilon^{2}}\one_{\rm c}( \epsilon^{2}).
\end{array}
\]
By the invariance principle for wave operators (see eg \cite{RS3}) the above limits are also equal to 
\[
\begin{array}{l}
\slim_{t\to +\pm \infty}\e^{\i t\epsilon}\e^{-\i t \epsilon_{\infty}}, \\[2mm]
\slim_{t\to +\pm \infty}\e^{\i t\epsilon_{\infty}}\e^{-\i t \epsilon}\one_{\rm c}( \epsilon).
\end{array}
\]
Using the chain rule of wave operators, it follows that Thms. \ref{mainli}, \ref{mainil} are still valid with $L_{l}$ replaced by $L_{\infty}$, yielding wave operators denoted by $W^{\pm}(L, L_{\infty})$. 
We denote by $U_{\infty}$ the analog of $U$ in Subsect.  \ref{reduc} for $v_{l}=0$, $\epsilon= \epsilon_{\infty}$, so that $U_{\infty}$ is unitary from $(\cE_{\infty}, h_{\infty}[\cdot , \cdot])$ to $\ch\oplus \ch$.  We set:
\[
\Omega^{\pm}_{s}:= U^{-1}W^{\pm}(L, L_{\infty})U_{\infty}.
\]
We obtain that 
\[
\e^{- \i t B}\Omega^{\pm}_{s}f-  U^{-1}U_{\infty}\e^{- \i tB_{\infty}}f\to 0, \ t\to \pm \infty.
\]
Now $U^{-1}U_{\infty}=\mat{b^{-1}\epsilon_{\infty}}{0}{0}{\one}$, and by stationary phase arguments we obtain that
\[
\lim_{t\to \pm \infty}(U^{-1}U_{\infty}-\one)\e^{-\i t B_{\infty}}f=0.
\]
Therefore 
\[
\e^{- \i t B}\Omega^{\pm}_{s}f-  \e^{- \i tB_{\infty}}f\to 0, \ t\to \pm \infty.
\]
The fact that $\Omega^{\pm}_{s}$ are bounded, 
unitary  from  $(\cE_{\infty}, h_{\infty}[\cdot, \cdot])$ to $(\cE_{\rm scatt}(B), h[\cdot, \cdot])$  and the intertwining property follow from the corresponding statements in Thms. \ref{mainli}, \ref{mainil}.  To prove that $\Omega^{\pm}_{s}$ are symplectic it suffices then to use  identity (\ref{atchoum}). 
\qed

\subsection{Proof of Thm. \ref{youpli}}\label{you2}

Set $J= J_{1}J_{2}$, where $J_{1}$, $J_{2}$ are defined in Subsects. \ref{sec11.1} and \ref{sec11.2}. Combining the results of Sects. \ref{srwave}, \ref{sec11} and the chain rule of wave operators, we obtain wave operators $W^{\pm}(L, L_{\infty})$, which are bounded and unitary from $\cK_{\infty}= \ch\oplus \ch$ to $(\cK_{\rm scatt}(L), [\cdot , \cdot])$  such that:
\[
\begin{array}{l}
\e^{-\i tL}W^{\pm}(L, L_{\infty})u- J\e^{-\i tL_{\infty}}u\to 0, \ t\to \pm \infty,\\[2mm]
\e^{-\i tL}W^{\pm}(L, L_{\infty})= W^{\pm}(L, L_{\infty}) \e^{-\i tL_{\infty}}.
\end{array}
\]
As above we set 
\[
\Omega^{\pm}_{l}:= U^{-1}W^{\pm}(L, L_{\infty})U_{\infty}, \ \hat{J}:= U^{-1}JU_{\infty},
\]
and obtain that $\Omega^{\pm}_{l}$ are bounded, 
unitary  from  $(\cE_{\infty}, h_{\infty}[\cdot, \cdot])$ to $(\cE_{\rm scatt}(B), h[\cdot, \cdot])$ and satisfy
\[
\begin{array}{l}
\e^{- \i tB}\Omega^{\pm}_{l}f-\hat{J}\e^{-\i tB_{\infty}}f\to 0, \ t\to \pm \infty,\\[2mm]
\e^{- \i t B}\Omega^{\pm}_{l}= \Omega^{\pm}_{l}\e^{-\i tB_{\infty}}. 
\end{array}
\]
The fact that $\Omega^{\pm}_{l}$ are symplectic follows as above from (\ref{atchoum}). To complete the proof of Thm. \ref{youpli}, it remains to use Lemma \ref{nerd} below, which shows that:
\[
\slim_{t\to \pm \infty}(T- \hat{J})\e^{- \i t B_{\infty}}=0.
\]
This completes the proof of the theorem. \qed

\medskip

\begin{lemma}\label{nerd}
 Let  $T$ be defined in (\ref{def-de-T}) and $\hat{J}:= U^{-1}J_{1}J_{2}U_{\infty}$. Then 
 \[
\slim_{t\to \pm \infty}(T- \hat{J})\e^{- \i t B_{\infty}}=0.
\]
\end{lemma}
\proof 
Recall first that
$J_{1}=\pm\mat{\one}{0}{0}{-\one}+ \Op S^{-1, -\mu_{l}}$, where
one chooses the $\pm$ sign according to the sign of $v_{l}$ near
infinity. For simplicity we consider the $+$ sign. Clearly it suffices to prove the lemma with  $J_{1}$ replaced by $\mat{\one}{0}{0}{-\one}$.

Denote the Fourier integral operators $j(\varphi_{\pm}, 1)$ 
simply by $j_{\pm}$. We obtain
\[
\hat{J}= \12\mat{b^{-1}(j_{+}-j_{-})\epsilon_{\infty}}{-b^{-1}(j_{+}+j_{-})}{-(j_{+}+ j_{-})\epsilon_{\infty}}{j_{+}- j_{-}},
\]
where $b= (\epsilon^{2}- v_{l}^{2})^{\12}$.

By (\ref{e11.0}) and Lemma \ref{urlb}, we have:
\[
b^{-1}j_{\pm}= j(\varphi_{\pm}, c_{\pm}),\hbox{ for }
c_{\pm}=(\tilde{\epsilon}^{2}- v_{l}^{2})^{-\12}(x, \p_{x}\varphi_{\pm})+ S^{-2, -1- \mu}.
\]
Since
\[
\tilde{\epsilon}(x, \p_{x}\varphi_{\pm}(x, \xi))= \epsilon_{\infty}(\xi)\pm v_{l}(x)\hbox{ in }\Gamma(R,  \alpha, \sigma),
\]
we obtain
\[
(\tilde{\epsilon}^{2}- v_{l}^{2})(x, \p_{x}\varphi(x, \xi))= \epsilon_{\infty}^{2}(\xi)\pm 2 \epsilon_{\infty}(\xi)v_{l}(x)\hbox{ in }\Gamma(R,  \alpha, \sigma),
\]
and hence:
\[
\begin{array}{rl}
c_{\pm}(x,\xi)=&\epsilon_{\infty}(\xi)^{-1}(1\pm 2\epsilon_{\infty}^{-1}(\xi)v_{l}(x))^{-\12}\\[2mm]
=&\epsilon_{\infty}^{-1}(\xi)+ S^{-1, -\mu}\hbox{ in }\Gamma(R, \alpha, \sigma).
\end{array}
\]
Let $S_{0}\subset S^{0, 0}$ be the space of symbols of the form:
\[
r=r_{0}+r_{1}, \ : \ r_{0}\in S^{0, -\mu}, \  \mu>0, \ r_{1}\in S^{0, 0}, \supp r_{1}\cap \Gamma(R, \alpha, \sigma)=\emptyset, \hbox{for some }R, \alpha, \sigma.
\]
It follows that
\begin{equation}
\label{e.url1}
b^{-1}j_{\pm}= j_{\pm}\epsilon_{\infty}^{-1}+j(\varphi_{\pm}, r),
\end{equation}
where $r\in S_{0}$.

Therefore we obtain
\[
\hat{J}= \12\mat{j_{+}-j_{-}}{-(j_{+}+ j_{-})\epsilon_{\infty}^{-1}}{-(j_{+}+ j_{-})\epsilon_{\infty}}{j_{+}-j_{-}}+ R= T+ R,
\]
where the entries of the matrix $R$ are of the form   $j(\varphi_{\pm}, r)$ for $r\in S_{0}.$  We claim that
 \begin{equation}
\label{eurl.2}
\slim_{t\to \pm\infty}R\e^{-\i tB_{\infty}}=0.
\end{equation}
In fact by stationary phase arguments we see that 
\[
\slim_{t\to\pm\infty}\Op r\e^{-\i tL_{\infty}}=0, \ \forall r\in S_{0}.
\]
The same fact is also true for $\e^{- \i tB_{\infty}}= U_{\infty}^{-1}\e^{- \i tL_{\infty}}U_{\infty}$. Therefore we obtain (\ref{eurl.2}), which completes the proof of the Lemma. \qed

\appendix
\section{ A brief summary of Krein space theory}\label{sec0}\init
In this section we recall some classical results on Krein spaces. Proofs can be found in the book \cite{B} or in  the survey paper \cite{La}.
\subsection{Krein spaces}\label{sec0.1}
 If $\cH$ is a topological complex vector space, we denote by $\cH^{\t}$ the space of continuous linear forms on $\cH$ and by $\langle w, u\rangle$, for $u\in \cH$, $w\in \cH^{\t}$ the duality bracket between $\cH$ and $\cH^{\t}$.

\begin{definition}
 A {\em Krein space}  $\cK$ is a hilbertizable vector  space $\cH$ equipped with a bounded hermitian sesquilinear form $[u, v]$ non-degenerate in the sense that if $w\in \cH^{\t}$ there exists a unique $u\in \cH$ such that 
\[
[u, v]= \langle w, v\rangle, \ v\in \cH.
\]
\end{definition}
If we fix a scalar product $(\cdot |\cdot)$ on $\cH$ endowing $\cH$ with its hilbertizable topology, then by  the Riesz theorem there exists a bounded, invertible selfadjoint operator $M$ such that
\[
[u, v]= (u| Mv), \ u, v\in \cH.
\]
If $A\in B(\cH)$,
we will denote by $A^{*}\in B(\cH)$ the adjoint of $A$ on $(\cH,(\cdot | \cdot))$ and by $A^{\dag}\in B(\cH)$ 
the  adjoint of $A$ on $(\cK, [\cdot, \cdot])$ defined by 
\[
[A^{\dag}u, v]:= [u, Av],\ u, v\in \cK.
\] 
Using the polar decomposition of $M$, $M= J|M|$ where $J=J^{*}$, $J^{2}=\one$, one can equip $\cH$ with the equivalent scalar product 
\beq\label{e0.00}
(u| v)_{M}:=(u||M| v),
\eeq so that
\begin{equation}
\label{e0.01}
[u, v]= (u|J v)_{M}, \ u, v\in \cH.
\end{equation}

\begin{definition}\label{pontryag}
 A Krein space $(\cK, [\cdot , \cdot])$ is a {\em Pontryagin space} if either $\one_{\rr^{-}}(J)$ or $\one_{\rr^{+}}(J)$ has  finite rank.
\end{definition}
Replacing $[\cdot , \cdot]$ by $-[\cdot , \cdot]$ we can assume that $\one_{\rr^{-}}(J)$ has finite rank, which is the usual convention for Pontryagin spaces.

\begin{definition}
 If  $(\cK, [\cdot , \cdot])$ is a Krein space, we set:
 \[
C_{\pm}:=\{u\in \cK \ : \ \pm[u, u]\geq 0\}, \ C_{0}:= C_{+}\cap C_{-}.
\]
The vectors in $C_{+}, C_{-}, C_{0}$ are called {\em positive, negative, null} respectively.
\end{definition}
\subsection{Subspaces of a Krein space}\label{sec0.2}
Let $\cK_{1}\subset \cK$ be a subspace of the Krein space $(\cK, [\cdot , \cdot])$.  We denote by $\cK_{1}^{\perp}$ the orthogonal of $\cK_{1}$ in $(\cK, [\cdot , \cdot])$.  Clearly  $\cK_{1}^{\perp}$ is closed and 
$\cK_{1}^{\perp \perp}= \cK_{1}^{\rm cl}$, the closure of $\cK_{1}$ for $(\cdot | \cdot)$.
Non-degeneracy
clearly implies 
\begin{equation}\label{eq:sum}
\cK=\cK_{1}+\cK_{1}^\perp \Rightarrow \cK_{1}\cap\cK_{1}^\perp=\{0\}.
\end{equation} 
A vector subspace $\cK_{1}\subset\cK$ will be called  a \emph{Krein
  subspace} if $\cK_{1}$ is a Krein space when equipped with the induced
topology and  $[\cdot,\cdot]$.

 The following results are well-known (see eg \cite{La}).

\begin{proposition}\label{pr:krein}
  If $\cK_{1}$ is a linear subspace of $\cK$ then the following
  assertions are equivalent: 
  \ben
  \item $\cK_{1}$ is a Krein subspace; 
  \item
  $\cK_{1}^\perp$ is a Krein subspace; 
  \item $\cK=\cK_{1}+\cK_{1}^\perp$. 
  \een
\end{proposition}

\begin{definition}
A projection $P$ on $(\cK, [\cdot, \cdot])$ is {\em orthogonal }  if   $P= P^{\dag}$, {\em positive }  if $[u, Pu]\geq 0$, $u\in \cK$.
\end{definition}
\begin{proposition}\label{pr:oproj}
  A linear subspace $\cK_{1}$ of $\cK$ is a Krein subspace if and only
  if there is an orthogonal projection $P$ such that
  $\cK_{1}=P\cK$. Then $P$ is uniquely determined: it is the projection
  on $\cK_{1}$ along $\cK_{1}^\perp$.
\end{proposition}
\
\begin{definition}\label{df:pos}
  A linear subspace $\cK_{1}\subset\cK$ is called \emph{positive} if $\cK_{1}\subset C_{+}$, 
   \emph{strictly positive} if
  $[u,u]>0$ for all $u\in\cK_{1}, u\neq0$, and \emph{uniformly
    (strictly) positive} if it is strictly positive and the topology
  associated to the norm $[u,u]^{1/2}$ on $\cK_{1}$ coincides with the
  topology induced by $(u|u)^{\12}$.
\end{definition}

 Observe that if $\cK_{1}$ is a positive subspace
then $\|u\|_{\cK_{1}}:=[u,u]^{1/2}$ is a semi-norm on $\cK_{1}$ and
\begin{equation}\label{eq:cauchy}
|[u,v]|\leq \|u\|_{\cK_{1} }\|v\|_{\cK_{1}} \quad \forall \ u,v\in\cK_{1}. 
\end{equation}

\begin{proposition}\label{pr:kp}
  If $\cK_{1}$ is a Krein subspace of $\cK$ then $\cK_{1}$ is positive iff $\cK_{1}$ is strictly positive iff $\cK_{1}$ is uniformly strictly positive.
  \end{proposition}
\section{Operators on Krein spaces}\init\label{secIk.}
Let $A:\Dom A\to \cK$ be a densely defined linear operator on the Krein space $\cK$.

The adjoint $A^{\dag}$ of $A$ on $(\cK, [\cdot, \cdot])$ is defined as
\[
\Dom A^{\dag}:=\{u\in \cK \ : \ \exists  f=: A^{\dag }u\hbox{ such that } [f, v]= [u, Av], \ \forall \ v\in \Dom A\}.
\]
  If we denote by $A^{*}$ the adjoint of $A$ for the scalar product $(\cdot| \cdot)$  then
\[
A^{\dag}= M^{-1} A^{*}M. 
\]
If   $A$ is bounded and $M= \one + K$    the above formula  simplifies to:
\begin{equation}
\label{link}
A^{\dag}= A^{*}+ (\one +K)^{-1}[A^{*}, K].
\end{equation}

A densely defined operator  $A$ is {\em selfadjoint}, resp. {\em unitary }  on $\cK$ if $A= A^{\dag}$,  resp. $A^{\dag}A= AA^{\dag}=\one$.

If $A$ is a closed, densely defined operator on $\cK$ we denote by $\sigma(A)$, $\rho(A)$ the spectrum and resolvent set of $A$.
If $\lambda$ is an isolated point of $\sigma(A)$ we  denote by
\[
E(\lambda, A)=\frac{1}{2\i \pi}\ointctrclockwise_{C}(z-A)^{-1}\d z
\]
where $C$ is a small curve in $\rho(A)$ surrounding $\lambda$, the Riesz spectral projection for $\lambda$. Clearly  
\[
E(\lambda, A)E(\lambda', A)=0, \hbox{ if }\lambda\neq \lambda', \ E(\lambda, A)^{\dag}= E(\overline{\lambda}, A^{\dag}).
\] 
\subsection{Definitizable operators}\label{secIk.1}
\begin{definition}
A selfadjoint operator $A$ is {\em definitizable} if
\ben
\item $\rho(A)\neq \emptyset$;
\item there exists a real polynomial $p(\lambda)$ such that
\[
[u, P(A)u]\geq 0, \ \forall u\in \Dom A^{k}, \ k:={\rm deg}p.
\]
\een
\end{definition}
A real polynomial $p$ satisfying condition (2) above is called  {\em definitizing} for $A$.
\begin{definition}\label{criti}
 Let $A$ a definitizable selfadjoint operator and $p$ a definitizing polynomial for $A$. The set 
 \[
c_{p}(A):= p^{-1}(\{0\})\cap \sigma(A)\cap \rr
\]
is called the set of (finite) {\em critical points } of $A$. 
\end{definition}
The usefulness of the notion of Pontryagin spaces in this context comes from the following theorem.
\begin{theoreme}\label{arsouille}
 A selfadjoint operator $A$ on a Pontryagin space is definitizable with a definitizing polynomial $p$ of even degree.
\end{theoreme}
We sketch the proof below because it contains some information that will be useful later on in Subsect. \ref{secIk.5}.

\proof Set ${\rm dim}\one_{\{-1\}}(J)=:\kappa<\infty$, where $J$ is defined in (\ref{e0.01}). By a theorem of Pontryagin there exists  a $\kappa-$dimensional 
 subspace $L_{-}\subset C_{-}$ such that $L_{-}\subset \Dom A$ and $AL_{-}\subset L_{-}$. Let $p_{0}$ be the minimal polynomial of $A_{|L_{-}}$, $k_{0}= {\rm deg}p_{0}$. Then since $p_{0}(A)L_{-}=\{0\}$ we get that $\overline{p_{0}}(A) \Dom (A^{k_{0}})\subset L_{-}^{\perp}\subset C_{+}$, using the fact that $L_{-}$ is a maximal negative subspace. Therefore $p(z)= \overline{p_{0}}(z)p_{0}(z)$ is a definitizing polynomial for $A$. \qed

 \medskip
 
\begin{remark}
 In the literature the set of critical points of $A$ is defined as the {\em intersection} of the sets $c_{p}(A)$ over all definitizing polynomials. For the applications in this paper it is sufficient to work with a fixed definitizing polynomial $p$, the only important property  of $p$ being that its degree is even. 
\end{remark}
\begin{remark}
 Usually the point $\infty$ of the one-point compactification $\hat{\cc}:= \cc\cup \{\infty\}$ is also considered as a critical point, which corresponds to the possible divergence of  the spectral projectors $\one_{[a, R]}(A)$ when $R\to \pm\infty$. 
 On Pontryagin spaces, one can choose definitizing polynomials of even degree, and  $\infty$ is not a critical point (see e.g. \cite{C}). We will recover this result in Subsect. \ref{secIk.5}.
\end{remark}
 
 The following result is due to Langer \cite{La}.
\begin{proposition}\label{coco}
Let $A$ be a definitizable selfadjoint operator with definitizing polynomial $p$. Then:
 \ben
\item $\sigma(A)\backslash \rr$ is the union of pairs $\{\lambda_{i}, \overline{\lambda}_{i}\}$ of eigenvalues of finite Riesz index; 
\item if $\nu(\lambda)$ is the Riesz index of an isolated eigenvalue $\lambda$, then
\[
\nu(\lambda)\leq \left\{\begin{array}{l}
k(\lambda)\hbox{ if }\lambda\not\in \rr, \\[2mm]
k(\lambda)+ 1 \hbox{ if }\lambda\in \rr,
\end{array}\right.
\]
where $k(\lambda)$ is multiplicity of $\lambda$ as a zero of $p$ (with the convention that $k(\lambda)=0$ if $\lambda$ is not a zero of $p$);
 \item Set now
 \[
E_{0}= \sum_{\lambda\in \sigma(A), \ {\rm Im}\lambda>0}E(\lambda, A)+ E(\overline{\lambda}, A), \ \cK_{0}:= E_{0}\cK.
\]
Then  $E_{0}$ is an orthogonal projector, hence $\cK_{0}$ is a Krein space and
\[
\cK= \cK_{0}\oplus\cK_{0}^{\perp}=: \cK_{0}\oplus\cK_{1}.
\]
\een
\end{proposition}

 In the rest of this section we review the functional calculus for definitizable operators (see \cite{La}, \cite{J}).  

 \subsection{Resolvent estimates.}\label{secIk.3}

In this subsection, we review some resolvent estimates for definitizable operators.

 Let  $A$ be a definitizable selfadjoint operator with definitizing polynomial $p$.  We fix $z_{0}\in \rho(A)$, $z_{0}\not \in \rr$ and
 set
 \beq\label{e0.03}
 q(z)= p(z)(z- z_{0})^{-k}(z- \overline{z_{0}})^{-k}, 
 \eeq
where \[
k= \left\{\begin{array}{l}
{\rm deg} p/2\hbox{ if }{\rm deg} p\hbox{ even},\\
 (\hbox{ deg} p+1)/2\hbox{ if }{\rm deg} p\hbox{ odd}
\end{array}.
\right.
\]
 For $z\in \rho(A)$ and $z\neq z_{0}, \overline{z}_{0}$ one sets:
\beq\label{def-de-q}
Q(z, A):= (q(A)- q(z)\one )(A- z)^{-1}.
\eeq

The following result is  proved in \cite[Sect. II.2]{La}, \cite[Sect. 2.1]{J} . 

\begin{lemma}\label{besson}
 \ben
\item The function $\rho(A)\ni z\mapsto q(A)(A-z)^{-1}\in B(\cH)$ extends holomorphically to  $\cc\backslash \rr$ with:
 \[
\|q(A)(A-z)^{-1}\|\leq C |{\rm Im}z|^{-1}, \ z\in \cc\backslash \rr;
\]
\item  Let   for $R, a, \delta>0$ $U(R, a, \delta):= U_{0}(R, a)\cup U_{\infty}(R, \delta)$ where:
\[
\begin{array}{rl}
U_{0}(R, a)&=\{z\in \cc\ :\   0<| {\rm Im}z|<a, \ |{\rm Re}z|\leq R\},\\[2mm]
U_{\infty}(R, \delta)&=\{z\in \cc\ :\     0<| {\rm Im}z|\leq \delta|{\rm Re}z|, \  |{\rm Re}z|\geq R\}.
\end{array}
\]
where $R, a$ are chosen such that $\sigma(A)\backslash \rr$ does not intersect $U(R, a, \delta)$.
Then there exists $C>0$ such that:
\[
\|(A-z)^{-1}\|\leq \left\{
\begin{array}{l}
C| {\rm Im}z|^{-m-1}, \hbox{ for }z\in U_{0}(R, a),\\[2mm]
C \langle z\rangle^{2k -{\rm  deg}p}|{\rm Im}z|^{-1}, \hbox{ for }z\in U_{\infty}(R, \delta),
 \end{array}
\right.
 \]
 where $m$ is the maximal order of  real zeroes of $p$ .
  \een
\end{lemma}

\subsection{Smooth functional calculus}\label{secIk.4}

The resolvent estimates in Lemma \ref{besson} allows to construct a functional calculus for $A$ using almost analytic extensions (see \cite{HS}, \cite{D}).
Let $S^{\rho}(\rr)$ for $\rho\in \rr$ be the  class of functions
$f$ such that:
\[
|f^{(\alpha)}(\lambda)|\leq C_{m}\langle\lambda\rangle^{\rho-\alpha}, \ \ \alpha\in \nn,
\]
equipped with the semi-norms:
\[
\|f\|_{m}:=\sup_{\lambda\in \rr, \alpha\leq
m}|\langle\lambda\rangle^{-\rho-\alpha}f^{(\alpha)}(\lambda)|.
\]
Let $\chi\in\coinf(\rr)$ with $\chi(s)\equiv 1$ in $|s|\leq \12$,
$\chi(s)\equiv 0$ in $|s|\geq 1$. Set:
\[
\tilde{f}(x+\i y):=(\sum_{r=0}^{N}
f^{(r)}(x)\frac{(\i y)^{r}}{r!})\chi(\frac{y}{\delta\langle x\rangle}),
\]
for $N$ fixed large enough, $\delta>0$. Then if $f\in S^{\rho}(\rr)$:
\beq\label{besson.1}
\begin{array}{l}
\tilde{f}_{|\rr}=f, \\[2mm]
\supp\tilde{f}\subset \{(x+\i y)\ :\  |y|\leq \delta\langle x\rangle,\ \ x\in\supp f\},\\[2mm]
|\frac{\p\tilde{f}(z)}{\p\bar{z}}|\leq C\langle x\rangle^{\rho-N-1}|y|^{N},
\end{array}
\eeq
and:
\beq\label{e6}
f(t)
= \frac{\i}{2\pi}\int_{\cc}\frac{\p \tilde{f}}{\p\overline{z}}(z)(z-t)^{-1}\d z \wedge \d \overline{z},\ t\in\rr.
\eeq
\begin{proposition}\label{irlitu}
 \ben
 \item let $f\in S^{\rho}(\rr)$  for $\rho<0$ if ${\rm deg}p$ is even and $\rho<-1$ if ${\rm deg}p$ is odd.  Then the integral:
 \begin{equation}
\label{e3b}
f(A)
:= \frac{\i}{2\pi}\int_{\cc}\frac{\p \tilde{f}}{\p\overline{z}}(z)(z-A)^{-1}\d z \wedge \d \overline{z}
\end{equation}
is norm convergent in $B(\cH)$ and independent on the choice of the almost analytic extension $\tilde{f}$;
\item
For $\rho$ as in {\rm (1)}, the map $S^{\rho}(\rr)\ni f\mapsto f(A)\in B(\cH)$ is a homomorphism of algebras with:
\[
\begin{array}{l}
f(A)^{\dag}= \overline{f}(A), \\[2mm]
\|f(A)\|\leq \|f\|_{m}, \hbox{ for some }m\in\nn.
\end{array}
\]

 \een
\end{proposition}
\proof The fact that the integral is norm convergent follows from  Lemma \ref{besson} (2) and the estimates (\ref{besson.1}), as is the bound in (2). 
The fact that $f(A)$ is independent on the choice of $\tilde{f}$ and that $f\mapsto f(A)$ is a homomorphism of algebras is proved as in the selfadjoint case. \qed

\subsection{Borel functional calculus}\label{secIk.5}
The  main result on the functional calculus for definitizable operators is the fact that it can be extended to a large class of Borel functions on $\rr$. It relies on the following lemma due to \cite{La}. 
\begin{lemma}\label{coucou}
 There exists  a map $\rr\ni t\mapsto F(t)$ with values in the positive selfadjoint operators  on $(\cH, (\cdot| \cdot))$, left continuous for the weak operator topology such that
 \[
 \begin{array}{rl}
& F(-\infty)=0, \  F(+\infty)= J q(A),\\[2mm]
 &s\leq t\Rightarrow F(s)\leq F(t), 
\end{array}
\]
 \beq\label{e3}
(A-z)^{-1}= q(z)^{-1}J\int_{-\infty}^{+\infty}(t-z)^{-1}\d F(t)- q(z)^{-1}Q(z, A), \ z\in \rho(A).
\eeq
\end{lemma}

 In \cite{La} the functional calculus for definitizable operators is extended to a large class of bounded Borel functions on $\rr$. In particular it is possible to define
  spectral projections  $\one_{I}(A)$ which are selfadjoint on $(\cK, [\cdot, \cdot])$ for all intervals $I$ such that $\partial I$ does not contain critical points.  
  
  We now explain how to recover this result  from Prop. \ref{irlitu}.
 We start by giving more explicit formulas for $f(A)$, $f\in \coinf(\rr)$.
\begin{lemma}
Let $f\in \coinf(\rr)$
 with $\supp f\cap c_{p}(A)= \emptyset$. Then:
 \begin{equation}
\label{e8}
f(A)=J\int_{-\infty}^{+\infty} f(t)q(t)^{-1}\d F(t).
\end{equation}
\end{lemma}

\proof  
Using (\ref{e3}) we get:
\beq\label{e7}
\begin{array}{rl}
f(A)= &J\int_{-\infty}^{+\infty} \left(\frac{\i}{2\pi}\int_{\cc}\frac{\p \tilde{f}}{\p\overline{z}}(z)q(z)^{-1}(z-t)^{-1}\d z \wedge \d \overline{z}\right)\d F(t)\\[2mm]
&+J\frac{\i}{2\pi}\int_{\cc}\frac{\p \tilde{f}}{\p\overline{z}}(z)q(z)^{-1}Q(z, A)\d z \wedge \d \overline{z}.
\end{array}
\eeq
To compute the first term we use that $\tilde{f}$ can be chosen with support close to $\supp f$ so that $q(z)^{-1}$ is holomorphic on $\supp \tilde{f}$. Hence $\tilde{f}(z)q(z)^{-1}$ is an almost analytic extension of $f(\lambda)q(\lambda)^{-1}$ with $\frac{\p \tilde{f}q^{-1}}{\p\overline{z}}(z)= q^{-1}(z)\frac{\p \tilde{f}}{\p\overline{z}}(z)$. We perform the integral in $z, \overline{z}$ using (\ref{e6}) and get that the first term equals:
\[
J\int_{-\infty}^{+\infty} f(t)q(t)^{-1}\d F(t).
\]
The second term is zero by Green's formula, since $q(z)^{-1}Q(z, A)$ is holomorphic in $z$ near $\supp \tilde{f}$, which proves (\ref{e8}).  \qed

 If $\Omega\subset \rr$ is a finite union of  disjoints intervals, we denote by ${\mathcal B}_{c}(\Omega)$ the $*-$algebra of  bounded Borel functions on $\Omega$ which are locally constant near $c_{p}(A)$. 
 \begin{proposition}\label{lati}
 \ben
 \item Let $\Omega\subset \rr$ a finite union of disjoint bounded intervals   $I$ such that $\p I\cap c_{p}(A)=\emptyset$. Then 
 the map $\coinf(\rr)\ni f\mapsto f(A)\in B(\cH)$ can be   extended to an homomorphism of $*-$algebras:
 \[
{\mathcal B}_{c}(\Omega)\ni f\mapsto f(A)\in B(\cH),
\]
with $\overline{f}(A)= f^{\dag}(A)$ for all $f\in {\mathcal B}_{c}(\Omega)$;
\item Let $\lambda_{0}\in \rr\backslash c_{p}(A)$. Then:
\[
\one_{\{\lambda_{0}\}}(A)= \slim_{\epsilon\to 0}\one_{[\lambda_{0}- \epsilon, \lambda_{0}+\epsilon]}(A)
\]
equals the orthogonal projection on ${\rm Ker}(A- \lambda_{0})$;
\item Let $I$ a bounded interval with $I^{\rm cl}\cap c_{p}(A)=\emptyset$. Then there exists $C_{I}\geq 0$ such that
\[
\|f(A)\|\leq C_{I} \|f\|_{\infty}, \  f\in {\mathcal B}_{c}(I);
\]
\item  Assume that  $p$ is of even degree. Then  the above map extends to all $f\in {\mathcal B}_{c}(\rr)$ with the same properties.  In particular 
statement (3) extends to all intervals $I$ with $I^{\rm cl}\cap c_{p}(A)=\emptyset$.
Moreover one has:
\[
1(A)+ E_{0}= \one,
\]
where the projection $E_{0}$ is defined in Prop. \ref{coco}.
\een
\end{proposition}
\begin{remark}\label{roi-des-connards}
Assume for simplicity that $p$ is of even degree. 
The space $C_{0}^{\infty}(\rr)+ {\mathcal B}_{c}(\rr)$ is clearly a $*-$algebra to which the homomorphism $ f\mapsto f(A)$ can be  
uniquely extended. In fact  if $f_{i}\in \coinf(\rr)$, $ g_{i}\in {\mathcal B}_{c}(\rr)$, $ i=1, 2$ with 
$ f_{1}+ g_{1}= f_{2}+ g_{2}$ then $ f_{1}(A)+ g_{1}(A)= f_{2}(A)+ g_{2}(A)$.
\end{remark}
\proof 
Let  first $I\subset \rr$   a bounded interval such that  $I^{\rm cl}\cap c_{p}(A)=\emptyset$.  The function $q(t)$ has constant sign on $I$ and without loss of generality we can assume that $q(t)\geq 0$ on $I$.

Since $t\mapsto F(t)$ is left weakly continuous and increasing, we can using (\ref{e8})  and  the usual monotonicity argument on the Hilbert space $(\cH, (\cdot | \cdot)_{M})$, extend the definition of $f(A)$ to $f\in {\mathcal B}_{c}(I)$.

Let now  $I$ be  a bounded interval with $\p I\cap c_{p}(A)=\emptyset$  and  $I\cap c_{p}(A)= \{\lambda_{0}\}$.  Any function in $f\in{\mathcal B}_{c}(I)$ can be split as $f=f_{1}+ f_{2}$, where $f\in \coinf(\rr)$ is supported close to $\lambda_{0}$ and $\supp f_{2}$ does not intersect $c_{p}(A)$. Therefore we can define $f(A)$ for all $f\in {\mathcal B}_{c}(I)$. This proves (1).

Statement (2) is proved by the same argument as for selfadjoint operators on Hilbert spaces.

To prove  (3)  we note that if $I$ is a bounded  interval with $I^{\rm cl}\cap c_{p}(A)= \emptyset$ then:
\[
\| f(A)\|\leq \|\int_{-\infty}^{+\infty} f(t)q(t)^{-1}\d F(t)\|\leq \| fq^{-1}\|_{L^{\infty}(I)}\| q(A)\|\leq C_{I} \|f\|_{L^{\infty}(I)},
\]
since $q^{-1}$ is uniformly bounded on $I$.

Let us now prove (4).
If $p$ is of even degree  and $I$ is an unbounded  interval with $I^{\rm cl}\cap c_{p}(A)= \emptyset$,
 then $q^{-1}$ is still  uniformly bounded on $I$ so the above estimate extends to $f\in {\mathcal B}(I)$.
 The same monotonicity argument allows to extend  $f(A)$ to all $f\in {\mathcal B}_{c}(\rr)$.

It remains to prove the second statement of (4).   Let us fix $z_{0}\in \rho(A)$ with ${\rm Im }z_{0}>0$.

By the Riesz-Dunford functional calculus  (see e.g. \cite[Sect. VII.9]{DS}) we have:
\begin{equation}
\label{besson.2}
(A- z_{0})^{-1}= \frac{-1}{2\i \pi}\ointctrclockwise_{\gamma_{0}} (z- z_{0})^{-1}(z-A)^{-1}\d z,
\end{equation}
where $\gamma_{0}$ is a circle in $\rho(A)$ surrounding $z_{0}$.

Similarly
\begin{equation}
\label{besson.3}
(A- z_{0})^{-1}E_{0}= \frac{1}{2\i \pi}\ointctrclockwise_{\gamma_{1}\cup \overline{\gamma_{1}}} (z- z_{0})^{-1}(z-A)^{-1}\d z,
\end{equation}
where $\gamma_{1}$ is a circle in $\rho(A)\cap \{{\rm Im }z>0\}$ surrounding  $\sigma(A)\cap \{{\rm Im }z>0\}$.
Let now $r(\lambda)= (\lambda-z_{0})^{-1}\in S^{-1}(\rr)$. An almost analytic extension of $r$ is given by:
\[
\tilde{r}(z)= (z-z_{0})^{-1}\chi(z),
\]
where $\chi\in \cinf(\cc)$ is supported in  some $U(R, a, \delta)$ and equal to $1$ in $U(R, a/2, \delta/2)$, and  the sets $U(R, a, \delta)$ are defined in 
Lemma \ref{besson}.  It follows that for  $C_{\epsilon}= \supp \chi\cap \{|{\rm Im}z|>\epsilon\}$:
\[
\begin{array}{rl}
r(A)= &\frac{\i}{2\pi}\int_{\cc}\frac{\p \tilde{r}}{\p \overline{z}}(z)(z-A)^{-1}\d z\wedge \d \overline{z}\\[2mm]
=&\frac{\i}{2\pi}\int_{\cc}\frac{\p \tilde{\chi}}{\p \overline{z}}(z)(z-z_{0})^{-1}(z-A)^{-1}\d z\wedge \d \overline{z}\\[2mm]
=&\lim_{\epsilon\to 0}\frac{\i}{2\pi}\int_{C_{\epsilon}}\frac{\p \tilde{\chi}}{\p \overline{z}}(z)(z-z_{0})^{-1}(z-A)^{-1}\d z\wedge \d \overline{z}\\[2mm]
=&\lim_{\epsilon\to 0}\frac{\i}{2\pi}\int_{\p C_{\epsilon}}\tilde{\chi}(z)(z-z_{0})^{-1}(z-A)^{-1}\d z\\[2mm]
=&-\lim_{\epsilon\to 0}\frac{1}{2\i \pi}\int_{\rr+\i \epsilon}\tilde{\chi}(z)(z-z_{0})^{-1}(z-A)^{-1}\d z\\[2mm]
&+\lim_{\epsilon\to 0}\frac{1}{2\i \pi}\int_{\rr-\i \epsilon}\tilde{\chi}(z)(z-z_{0})^{-1}(z-A)^{-1}\d z,
\end{array}
\]
where we have used Green's formula. The last two line integrals are independent on $\epsilon$, therefore
\begin{equation}
\label{besson.4}
\begin{array}{rl}
r(A)= &-\frac{1}{2\i \pi}\int_{\rr+\i \epsilon}(z-z_{0})^{-1}(z-A)^{-1}\d z\\[2mm]
&+\frac{1}{2\i \pi}\int_{\rr-\i \epsilon}(z-z_{0})^{-1}(z-A)^{-1}\d z,
\end{array}
\end{equation}
for  $\epsilon>0$ small enough.

We claim that 
\beq
\label{besson.5}
(A-z_{0})^{-1}= (A- z_{0})^{-1}E_{0}+ r(A).
\eeq
In fact this follows from (\ref{besson.2}), (\ref{besson.3}) and (\ref{besson.4}) by deforming the integration contour. 
We have to add integrals of 
$(z-z_{0})^{-1}(z-A)^{-1}$ over two  half-circles  $\Gamma_{\pm}(R)$ in the upper and lower half-plane   of centers $\pm\i \epsilon$ and of radius $R\gg 1$. On $\Gamma_{\pm}(R)$ we obtain as in Lemma \ref{besson} that $\|(A-z)^{-1}\|\in O(|{\rm Im}z|^{-1})$, therefore the contribution of these integrals is  $O(\ln(R)R^{-1})$ and vanishes when $R\to \infty$.

To complete the proof of (4),  note  that 
if $f\in S^{-1}(\rr)$ and $g(\lambda)= (\lambda-z_{0}) f(\lambda)$, then $g\in \coinf(\rr)+ {\mathcal B}_{c}(\rr)$ 
(see Remark \ref{roi-des-connards}) and  $(A- z_{0}) f(A)= g(A)$.
 Therefore multiplying (\ref{besson.5}) by $(A-z_{0})$ we obtain that
\[
\one=  E_{0}+ 1(A),
\]
as claimed. \qed

\section{Abstract propagation estimates for definitizable operators}\init\label{sec4}
In this section we explain how to prove propagation estimates for definitizable operators on a Krein space. The basic framework is parallel to the Hilbert space case, developed in the context of scattering theory for Schr\"{o}dinger operators by Sigal and Soffer \cite{SS}. A detailed exposition can be found in the book \cite[Appendix B.4]{DG}.
\subsection{Propagation estimates}\label{king-of-assholes}
Let $(\cK, \ [\cdot, \cdot])$ be a Krein space. We fix a hilbertian scalar product $(\cdot |\cdot)$ whose associated norm $\|\cdot \|$ defines the topology of $\cK$ such that  $[u, u]= (u|(\one +K)u)$.  

As in Appendix \ref{secIk.} the adjoints of an operator $A$ w.r.t.  $[\cdot , \cdot]$ (resp. $(\cdot |\cdot)$) will be denoted by $A^{\dag}$ (resp. $A^{*}$). Operator inequalities like for example $A\geq 0$ will be understood in the Hilbert sense i.e. $(u|Au)\geq 0$.

We fix a selfadjoint operator $L$ on $(\cK, [\cdot, \cdot])$ and assume that $L$ is definitizable, with definitizing polynomial $p$ of even degree. 
It follows from the functional calculus in  Subsect. \ref{secIk.5} (see Remark \ref{roi-des-connards}) that  $(\e^{-i t L})_{t\in \rr}$ is well defined as a strongly continuous group of bounded operators, unitary on $(\cK, [\cdot, \cdot])$. (Alternatively one could assume  that $\infty$ is not a singular critical point, see \cite{La}).

If $ A\in B(\cK)$ the expression $[L, A]$ can be given different meanings. In our context the most natural one is to define $ [L, A]$ as a quadratic form on $ \Dom L$ by:
\[
 [L, A](u,u):= [Lu, Au]- [u, ALu], \ u \in \Dom L
\] 
If $A\in B(\Dom L)$ then $LA- AL\in B(\Dom L, \cK)$. Denoting by $(LA- AL)_{\rm op}$ this operator one has of course
\[
[L, A](u, u)= [u, (LA- AL)_{\rm op}u], \ u\in \Dom L.
\]
\begin{proposition}\label{abs2}
 Let $\rr\ni t\mapsto \Phi(t)\in B(\cK)$ be strongly differentiable, uniformly  bounded. Let $\chi\in \coinf(\rr)$  such that $\supp \chi\cap p^{-1}(\{0\})=\emptyset$. 
 Set
 \[
{\bf D}\Phi(t)= \p_{t}\Phi(t)+ [L, \i \Phi(t)].
\]
\ben
\item Assume:
\beq
\label{abs.e1}
\begin{array}{rl}
[\chi(L)u, {\bf D}\Phi(t)\chi(L)u]=&(\chi(L)u|C(t)\chi(L)u)+ (\chi(L)u|C_{1}(t)\chi(L)u)\\[2mm]
&+(\chi(L)u| C_{2}(t)\chi(L)u)
\end{array}
\eeq
where:
\[
\begin{array}{l}
i)\ C(t)= C^{*}(t), \ C(t)\geq B(t)^{*}B(t),\\[2mm]
ii)\ |(u|C_{1}(t)u)|\leq \sum_{1\leq j,k\leq N}\|B_{j}(t)u\|\|B_{k}(t)u\|,\\[2mm]
iii)\ \int_{0}^{+\infty}\|B_{j}(t)\e^{- \i tL}\chi(L)u\|^{2}dt\leq C\|u\|^{2}, \ 1\leq j\leq N,\\[2mm]
 iv)\ |(u| C_{2}(t)u)|\leq \delta \|B(t)u\|^{2}, \ 0\leq \delta <1.
\end{array}
\]
Then:
\[
\int_{0}^{+\infty}\|B(t)\e^{-\i tL}\chi(L)u\|^{2}dt\leq C\|u\|^{2}.
\]
\item Assume that (\ref{abs.e1}) holds for $C(t)= C_{2}(t)\equiv 0$. Then
\[
\lim_{t\to +\infty}[\chi(L)\e^{- \i tL}u, \Phi(t)\chi(L)\e^{- \i tL}u]\hbox{ exists}.
\]
\een
\end{proposition}
\begin{corollary}\label{abs4}
 Let $t\mapsto \Phi(t)$ and $\chi$ as above. 
 \ben
 \item Assume that for some $\chi_{1}\in \coinf(\rr)$ with $\chi_{1}\chi= \chi$ one has:
 \beq\label{abs.e2}
\chi_{1}(L){\bf D}\Phi(t)\chi_{1}(L)= \chi_{1}(L)D(t)\chi_{1}(L)+ R_{0}(t),
\eeq
where:
\[
\begin{array}{l}
i) \ \| R_{0}(t)\|\in L^{1}(\rr^{+}), \\[2mm]
ii) \ \|K D(t)\|\in L^{1}(\rr^{+})\\[2mm]
iii) \ D(t)= D^{*}(t), \ D(t)\geq B^{*}(t)B(t)-\sum_{j}B_{j}^{*}(t)B_{j}(t), \\[2mm]
iv)\ \int_{0}^{+\infty}\|B_{j}(t)\chi(L)u_{t}\|^{2}dt\leq C\|u\|^{2},\ 1\leq j\leq N.
\end{array}
\]
Then:
\[
\int_{0}^{+\infty}\|B(t)\e^{-\i tL}\chi(L)u\|^{2}dt\leq C\|u\|^{2}.
\]
\item Assume i), ii), iii') and iv), where:
\[
iii') \ |(u| D(t)v)|\leq \sum_{j} \|B_{j}(t)u\|\|B_{j}(t)v\|.
\]
Then 
\[
\lim_{t\to +\infty}[\chi(L)\e^{- \i tL}u, \Phi(t)\chi(L)\e^{- \i tL}u]\hbox{ exists}.
\]
\een
\end{corollary}

{\bf Proof of Prop. \ref{abs2}:}

 Set $u_{t}= \e^{-\i tL}u$ and
 \[
f(t)= [\chi(L)u_{t}, \Phi(t)\chi(L)u_{t}], 
\]
so that:
\begin{equation}
\label{e.abs5}
f'(t)= [\chi(L)u_{t}, {\bf D}\Phi(t)\chi(L)u_{t}].
\end{equation}
Then:
\[
\begin{array}{rl}
&\int_{t_{1}}^{t_{2}}\|B(t)\chi(L)u_{t}\|^{2}dt\\[2mm]
\leq &\int_{t_{1}}^{t_{2}}(\chi(L)u_{t}\ C(t)\chi(L)u_{t})dt\\[2mm]
=&\int_{t_{1}}^{t_{2}}f'(t)dt - \int_{t_{1}}^{t_{2}} (\chi(L)u_{t}| C_{1}(t)\chi(L)u_{t})dt - \int_{t_{1}}^{t_{2}} (\chi(L)u_{t}| C_{2}(t)\chi(L)u_{t})dt\\[2mm]
\leq &|f(t_{2})- f(t_{1})| + 
\sum_{j,k}\int_{t_{1}}^{t_{2}}\|B_{j}(t)\chi(L)u_{t}\|\|B_{k}(t)\chi(L)u_{t}\| dt\\[2mm]
 &+ \delta \int_{t_{1}}^{t_{2}}\|B(t)\chi(L)u_{t}\|^{2}\\[2mm]
\leq & |f(t_{2})- f(t_{1})|+ N\sum_{j}\int_{t_{1}}^{t_{2}}\|B_{j}(t)\chi(L)u_{t}\|^{2} dt + \delta \int_{t_{1}}^{t_{2}}\|B(t)\chi(L)u_{t}\|^{2}.
\end{array}
\]
It follows that
\[
(1- \delta)\int_{t_{1}}^{t_{2}}\|B(t)\chi(L)u_{t}\|^{2}dt\leq  |f(t_{2})- f(t_{1})|+ N\sum_{j}\int_{t_{1}}^{t_{2}}\|B_{j}(t)\chi(L)u_{t}\|^{2} dt.
\]
Note that $\chi(L)\e^{-\i tL}= g_{t}(L)$ for $g_{t}(\lambda)= \e^{- \i t \lambda}\chi(\lambda)$. Since $g_{t}(\cdot)$ is supported away from critical points, we have $\|g_{t}(L)\|\leq C \|g_{t}(\cdot)\|_{\infty}$ By Prop. \ref{lati} (3), hence $g_{t}(L)$  and $f(t)$ are uniformly bounded.  This completes the proof of (1). The proof of (2) is immediate. \qed

\medskip

{\bf Proof of Corollary \ref{abs4}:}  we have using that $\chi_{1}\chi= \chi$ and (i):
\[
\begin{array}{rl}
[\chi(L)u, {\bf D}\Phi(t)\chi(L)u]=&[\chi(L)u, D(t)\chi(L)u]+ [\chi(Lu, R_{0}(t)\chi(Lu]\\[2mm]
=&(\chi(L)u| D(t)\chi(L)u)+ (\chi(L)u| KD(t)\chi(L)u)\\[2mm]
&+ (\chi(L)u| (\one + K)R_{0}(t)\chi(L)u).
\end{array}
\]
We apply Prop. \ref{abs2} for 
\[
\begin{array}{rl}
 C(t)&= D(t)+\sum_{j}B_{j}^{*}(t)B_{j}(t),\\[2mm]
 C_{1}(t)&= -\sum_{j}B_{j}^{*}(t)B_{j}(t)+ KD(t)+ (\one +K)R_{0}(t),\\[2mm]
 C_{2}(t)&=0.
\end{array}
\]
The proof of (2) is similar. \qed

\subsection{Existence of limits}\label{sec4.1}
In this subsection we describe the analog of the Cook-Kato argument on a Krein space.  As before we fix a definitizable 
selfadjoint operator $L_{1}$ on $(\cK, [\cdot, \cdot])$, and a selfadjoint operator $L_{0}$ on $(\cK, (\cdot | \cdot))$.  For simplicity we assume that
$\Dom L_{1}= \Dom L_{0}$.

The following discussion is similar to the  one at the beginning of  Subsect. \ref{king-of-assholes}.
If $A\in B(\cK)$ then we  define $L_{0}A- AL_{1}$ and $L_{1}A- AL_{0}$  as  quadratics forms on $\Dom L_{0}$ by:
 \[
 \begin{array}{rl}
(L_{0}A- AL_{1})(u, u):=& (L_{0}u| Au)- (u| AL_{1}u),\\[2mm]
 (L_{1}A- AL_{0})(u, u):=&[L_{1}u, Au]- [u, AL_{0}u],
 \end{array}
\]
(see Prop. \ref{abs3}).

Recalling that $ \Dom L_{1}= \Dom L_{0}$, we see again that if  $A\in B(\Dom L_{0})$, 
one can define the operators  $\left(L_{0}A- AL_{1}\right)_{\rm op}$,  $\left(L_{1}A- AL_{0}\right)_{\rm op}$ which belong to $ B(\Dom L_{0}, \cK)$. Then:
 \[
\begin{array}{rl}
(L_{0}A- AL_{1})(u, u):= (u| \left(L_{0}A- AL_{1}\right)_{\rm op}u),\\[2mm]
 (L_{1}A- AL_{0})(u, u):=[u|\left(L_{1}A- AL_{0}\right)_{\rm op}u].
\end{array}
\]

Let now $\rr^{+}\ni t\mapsto M(t)\in B(\cH)$ a strongly differentiable map.   
We define the asymmetric Heisenberg derivatives:
\[
\begin{array}{rl}
{}_{0}{\rm D}_{1}M(t)= \p_{t}M(t)+ \i (L_{0}M(t)- M(t)L_{1}), \\[2mm]
{}_{1}{\rm D}_{0}M(t)= \p_{t}M(t)+ \i (L_{1}M(t)- M(t)L_{0}).
\end{array}
\]
\begin{proposition}\label{abs3}
Let $\chi\in \coinf(\rr)$ such that $\supp \chi\cap c_{p}(L_{1})= \emptyset$. 
\ben
\item Assume that:
\[
\begin{array}{l}
|(u_{0}| \chi(L_{0})\left({}_{0}{\bf D}_{1}M(t)\right)\chi(L_{1})u_{1})|\leq \sum_{i=1}^{N}\|B_{0i}(t)\chi(L_{0})u_{0}\|\|B_{1i}(t)\chi(L_{1})u_{1}\|,\\[2mm]
\int_{1}^{+\infty}\|B_{0i}(t)\chi(L_{0})\e^{\i t L_{0}}v\|^{2}dt\leq C\|v\|^{2},\\[2mm]
\int_{1}^{+\infty}\|B_{1i}(t)\chi(L_{1})\e^{\i t L_{1}}v\|^{2}dt\leq C\|v\|^{2}.
\end{array}
\]
Then there exists
\[
\slim_{t\to +\infty}\e^{\i t L_{0}}\chi(L_{0})M(t)\chi(L)\e^{- \i tL_{1}}.
\]
\item Assume that:
\[
\begin{array}{l}
|[u_{1}, \chi(L_{1})\left({}_{1}{\bf D}_{0}M(t)\right)\chi(L_{0})u_{0}]|\leq \sum_{i=1}^{N}\|B_{0i}(t)\chi(L_{0})u_{0}\|\|B_{1i}(t)\chi(L_{1})u_{1}\|,\\[2mm]
\int_{1}^{+\infty}\|B_{0i}(t)\chi(L_{0})\e^{-\i tL_{0}}v\|^{2}dt\leq C\|v\|^{2},\\[2mm]
\int_{1}^{+\infty}\|B_{1i}(t)\chi(L_{1})\e^{- \i tL_{1}}v\|^{2}dt\leq C\|v\|^{2}.
\end{array}
\]
Then there exists
\[
\slim_{t\to +\infty}\e^{\i t L_{1}}\chi(L_{1})M(t)\chi(L_{0})\e^{- \i tL_{0}}.
\]
\een
\end{proposition}
 \proof
 We only prove (2), the proof of (1) being simpler.   We fix an interval $I$ such that $\supp \chi\subset I$ and$I^{\rm cl}\cap c_{p}(L_{1})=\emptyset$. Without loss of generality we can assume that $[\cdot , \cdot]\geq 0$ on $\one_{I}(L_{1})\cK$.  Let $u, v\in \cK$  and set
 \[
 u_{t}:= \e^{\i t L_{1}}\chi(L_{1})M(t)\chi(L_{0})\e^{- \i tL_{0}}u.
\]
Then 
 \[
\begin{array}{rl}
&|[v, u_{t}]- [v,u_{s}]|\\[2mm]
\leq &\int_{s}^{t}|[\e^{- \i \sigma L_{1}}v, \chi(L_{1})\left({}_{1}{\bf D}_{0}M(\sigma)\right)\chi(L_{0})\e^{- \i \sigma L_{0}}u]|\d \sigma\\[2mm]
\leq &\sum_{i=1}^{N}\left(\int_{s}^{t}\|B_{1i}(\sigma)\chi(L_{1})\e^{-\i tL_{1}}v\|^{2} d\sigma\right)^{\12}
\left(\int_{s}^{t}\|B_{0i}(\sigma)\chi(L_{0})\e^{-\i tL_{0}}u\|^{2} d\sigma\right)^{\12}.
\end{array}
\]
It follows that
\begin{equation}
\label{paulgen}
|[v, u_{t}]- [v,u_{s}]|\leq C \|v\|m(t,s), \hbox{where }m(t,s)\to 0\hbox{ when }t, s\to+\infty.
\end{equation}
We use now that $u_{t}= \one_{I}(L_{1})u_{t}$ hence  $u_{t}\in \cK_{1}:= \one_{I}(L_{1})\cK$. Since $\cK_{1}$ is positive, we have by Prop. \ref{pr:kp}:
\[
\begin{array}{rl}
\|u_{t}- u_{s}\|\leq &C [u_{t}- u_{s}, u_{t}- u_{s}]^{\12}\\[2mm]
=& C \sup_{v\in \cK, \ [v, \one_{I}(L_{1})v]=1}|[v, u_{t}- u_{s}]|\\[2mm]
\leq& C\sup_{v\in \cK, \ \|v\|\leq C'}|[v, u_{t}- u_{s}]|.
\end{array}
\]
Applying (\ref{paulgen}) we obtain the existence of $\slim_{t\to +\infty}u_{t}$.\qed
\section{Pseudo-differential calculus}\init\label{pdocalc}
In this section we recall some well-known results about pseudo-differential and Fourier integral operators.
\subsection{Pseudo-differential operators}
We denote by $\cS(\rr^{d})$ the Schwartz class of functions on $\rr^{d}$ and by
$\cS'(\rr^{d})$ the Schwartz class of tempered distributions on $\rr^{d}$. 

We set as usual  $\langle s\rangle=(s^{2}+1)^{\12}$, $s\in \rr^{n}$.

For $p,m\in \rr$ we denote by $S^{p, m}(\rr^{2d})$  or simply $S^{p,m}$ the class of symbols $a\in
\cinf(\rr^{2d})$ such that
\[
|\p^{\alpha}_{x}\p^{\beta}_{\xi}a(x,\xi)|\leq C_{\alpha,
\beta}\langle \xi\rangle^{p-|\beta|}\langle
x\rangle^{m-|\alpha|}, \ \ \alpha,
\beta\in \nn^{d}.
\]
 The symbol classes above are equipped with the topology
given by the semi-norms equal to the best constants in the
estimates above.

For $a\in S^{p,m}$, we denote by  $\Op(a)$ the  Weyl quantization of $a$
defined by:
\[
\Op(a)(x,D)u(x):=(2\pi)^{-1}\int\int\e^{\i
(x-y)\xi}a(\frac{x+y}{2}, \xi)u(y)\d y\d \xi,
\]
which is well defined as a continuous map from $\cS(\rr^{d})$ to $\cS'(\rr^{d})$.
We recall  the following facts (see \cite[Thm. 18.5.4]{Ho}):
\beq\label{calc}
\begin{array}{l}
\Op(b)^{*}= \Op(\overline{b}),\\[2mm]
[\Op(b_{1}), \i \Op(b_{2})]= \Op(\{b_{1}, b_{2}\})+ \Op(S^{p_{1}+
p_{2}-3, m_{1}+ m_{2})}),\\[2mm]
\Op(b_{1})\Op(b_{2})+ \Op(b_{2})\Op(b_{1})=
2\Op(b_{1}b_{2})+ \Op(S^{p_{1}+
p_{2}-2, m_{1}+ m_{2}}),
\end{array}
\eeq
if  $b_{i}\in S^{p_{i}, m_{i}}$ and $\{ \: , \:\}$ denotes
the Poisson bracket.

We will need the following fact.
\begin{lemma}\label{urla}
 Let  $[c^{jk}](x)$,  $b(x)=(b_{1}(x), \dots, b_{d}(x))$, $c(x)$ such that
 \[
\begin{array}{l}
c_{0}\one\leq [c^{jk}](x)\leq c_{1}\one, \hbox{ for some }c_{0}>0, \\[2mm]
[c^{jk}](x)-\one, \ b(x), c(x)\in S^{0, -\mu}, \mu>0. 
\end{array}
\]
Set
\[
p(x, \xi)= \sum_{jk}\xi_{j}c^{jk}(x)\xi_{k}+ \sum b_{j}(x)\xi_{j}+ c(x),
\]
and let $C>0$ such that $\Op b+C\geq c_{2}$ for some $c_{2}>0$. Then:
\[
(\Op p +C)^{\12}= \Op ((p+C)^{\12})+ \Op S^{0, -1-\mu},
\]
\[
(\Op p +C)^{-\12}= \Op ((p+C)^{-\12})+ \Op S^{-2, -1-\mu},
\]
\end{lemma}
\proof   A proof in the case $b(x)\equiv 0$ can be found in \cite[Props. 2.10, 2.11]{GP}.  The extension  to the case $b(x)\neq 0$ is straightforward. \qed
\subsection{Fourier integral operators}\label{astruc}
 We now recall the definition of some Fourier integral operators. 
\begin{definition}\label{urlc}
Let  Let  $\varphi(x, \xi)\in C^{\infty}(\rr^{2d})$ such that 
  \[
 \varphi(x, \xi)- x\cdot \xi\in S^{-1, 1-\rho} \hbox{ for some }\rho>0.
 \]
 Then for $a\in S^{m, p}$ we define:
 \[
\begin{array}{rl}
j(\varphi, a): &{\mathcal S}(\rr^{d})\to {\mathcal S}(\rr^{d})\\[2mm]
j(\varphi, a)u(x):= &(2\pi)^{-d}\int\int\e^{\i\varphi(x, \xi)- \i y\cdot \xi}a(x, \xi)u(y)\d y \d \xi
\end{array}
\]
\end{definition}
It is well known that if $a\in S^{0, 0}$ then $j(\varphi, a)$ extends as a bounded operator on $L^{2}(\rr^{d})$.  We state  a version of Egorov theorem (see eg \cite{IK}).
\begin{lemma}\label{urlb}
 Let  $\varphi(x, \xi)$ as above. Then  for all $d\in S^{m,p}$ one has:
\[
(\Op d) j(\varphi, 1)= j(\varphi, \tilde{d}), \hbox{ for }\tilde{d}(x, \xi)= d(x, \p_{x}\varphi(x, \xi))+ S^{m-1, p-1-\rho}.
\]
\end{lemma}

\section{Some technical results}\init
\subsection{Proof of Prop. \ref{3.1}.}\label{ap1}

Set $\mu=\min(\mu_{0}, \mu_{\l})$. An easy computation shows that
\[
\begin{array}{rl}
\epsilon^{2}=& \sum_{jk}D_{j}c^{jk}(x)D_{k} - 2\sum_{j}d_{j}(x)D_{j}+ r(x)+ m^{2}\\[2mm]
=&\Op(\tilde{\epsilon}^{2})+ \Op S^{0, -\infty},
\end{array}
\] 
for
\[
\tilde{\epsilon}^{2}(x, \xi)= \sum_{jk}\xi_{j}c^{jk}(x)\xi_{k}- 2\sum_{j} d_{j}(x)\xi_{j}+ r_{1}(x)+ m^{2}, 
\]
and
\[
c^{jk}(x)= c^{-2}(x)a^{jk}(x), \ d_{j}(x)= c^{-2}(x)b_{j}(x),\ [c^{jk}](x)-\one, d_{j}(x), r_{1}(x)\in S^{-\mu}(\rr^{d}).
\]
It follows from Lemma \ref{urla} that
\[
b=\Op( \tilde{b}) + \Op S^{0, -1- \mu}, \ b^{-1}= \Op(\tilde{b}^{-1})+ \Op S^{-2, -1- \mu},
\]
for
 \[
\tilde{b}(x, \xi)= (\tilde{\epsilon}^{2}(x, \xi)- v_{\l}^{2}(x))^{\12}= (\xi^{2}+ m^{2})^{\12}+ S^{1, -\mu}\in S^{1, 0}.
\]
 Similarly
\[
a= \Op(\tilde{a}), \ \tilde{a}(x, \xi)= (\xi^{2}+m^{2})^{-1}x\cdot \xi + S^{-2, 0}\in S^{-1, 1}.
\]
Statements (1), (2), (4), (5) follow then directly from pseudo-differential calculus. To prove (3) we write
\[
[b, \i a]= \Op (\{\tilde{b}, \tilde{a}\})+ \Op S^{-2, -1}= \Op (\{(\xi^{2}+ m^{2})^{\12}, \tilde{a}\})+ S^{-1, -\mu},
\]
where $\{\cdot, \cdot \}$ is the Poisson bracket and use that 
\[
\{(\xi^{2}+ m^{2})^{\12}, \tilde{a}\}= \xi^{2}(\xi^{2}+ m^{2})^{-3/2}=\tilde{b}^{-3}(\tilde{b}^{2}-m^{2})+ S^{-1, -\mu}.
\]
Using once again pseudo-differential calculus we obtain (3). \qed

\subsection{Proof of Prop. \ref{3.1bis}.}
 (1): follows from pseudo-differential calculus.

(2):  We fix a function $G\in \cinf(\rr^{d})$ with  $0\not\in \supp G$ and $G\equiv 1$ on $\supp F$. Then $F(\xt, D_{x})v_{\s}b^{-1}=F(\xt, D_{x})G(\xt)v_{\s}b^{-1}\in O(t^{-\mu_{s}})$. The other term is
\[
v_{\s}F(\xt, D_{x})b^{-1}=G(\xt)v_{\s}b^{-1}b F(\xt, D_{x})b^{-1}\in O(t^{-\mu_{s}}).
\]
(3): Similarly  $F(\xt, D_{x}) r b^{-2}= F(\xt, D_{x})G(\xt)rb^{-2}\in O(t^{-\mu_{s}})$.  To handle the other term we write 
$rb^{-1}Fb^{-1}= rb^{-1}G(\xt)Fb^{-1}$. From Prop. \ref{3.1} (5), we get that
\[
b^{-1}G(\xt)= G_{1}(\xt)b^{-1}G(\xt)+ R(t), \hbox{ where }b^{-2}R(t)b^{-1}\in O(t^{-2}).
\]
Hence:
\[
\begin{array}{rl}
rb^{-1}GFb^{-1}=&rG_{1}b^{-1}GFb^{-1}+ r R(t)Fb^{-1}\\[2mm]
=& rG_{1}b^{-2}bGF b^{-1}+ rb^{-2} b^{2}R(t)b^{-1}b Fb^{-1}.
\end{array}
\]
Since  $b GFb^{-1}$  and  $bFb^{-1}\in O(1)$, $rb^{-2}$ is bounded, $rG_{1}(\xt)b^{-2}\in O(t^{-\mu_{s}})$ and $ b^{2}R(t)b^{-1}\in O(t^{-2})$ we obtain (3). \qed

\end{document}